\documentclass[%
preprint,
 amsmath,amssymb,
 aps,
 prb,
]{revtex4}

\usepackage{graphicx}% Include figure files
\usepackage{dcolumn}% Align table columns on decimal point
\usepackage{bm}% bold math
\usepackage{hyperref}% add hypertext capabilities

\usepackage{color}
\usepackage{xspace}
\usepackage{setspace}
\newcommand{\ea}{~\textit{et~al.}\xspace} %% et alii
\newcommand{\etc}{~\textit{etc.}\xspace}
\newcommand{\eg}{\textit{e.g.}\xspace}     %% id est
\newcommand{\md}[1]{\mathrm{#1}}           %% in math mode, roman typeset "Droit"

\graphicspath{{figures/}}
\begin{document}

\preprint{}%npj Computational Materials?}

\title{Supervised deep learning prediction of the formation enthalpy \\of the full set of configurations in complex phases:\\ the $\sigma-$phase as an example}
%Predicting enthalpy of every ordered atomic configurations of a crystal phase using machine learning: the $\sigma-$phase as an example\\or\\Using the only binary configurations to predict multicomponent crystal structure properties with supervised machine learning

\author{Jean-Claude Crivello}
 \email{crivello@icmpe.cnrs.fr}
 \affiliation{%
 Univ Paris Est Creteil, CNRS, ICMPE, UMR 7182, 2 rue Henri Dunant, 94320 Thiais, France
}%
\author{Nataliya Sokolovska}%
 \affiliation{%
 NutriOmics, INSERM, Sorbonne University Paris, France
}%
\author{Jean-Marc Joubert}
 \affiliation{%
 Univ Paris Est Creteil, CNRS, ICMPE, UMR 7182, 2 rue Henri Dunant, 94320 Thiais, France
}%

\date{\today}% 

\begin{abstract}
Machine learning (ML) methods are becoming integral to scientific inquiry in numerous disciplines, such as material sciences.
In this manuscript, we demonstrate how ML can be used to predict several properties in solid-state chemistry, in particular the heat of formation of a given complex crystallographic phase (here the $\sigma-$phase, $tP30$, $D8_{b}$).
Based on an independent and unprecedented large first principles dataset containing about 10,000 $\sigma-$compounds with $n=14$ different elements, we used a supervised learning approach, to predict all the $\sim$500,000 possible configurations within a mean absolute error of 23\,meV/at ($\sim$2\,kJ.mol$^{-1}$) on the heat of formation and $\sim$0.06\,\AA\ on the tetragonal cell parameters. 
We showed that neural network regression algorithms provide a significant improvement in accuracy of the predicted output compared to traditional regression techniques. 
Adding descriptors having physical nature (atomic radius, number of valence electrons) improves the learning precision. 
Based on our analysis, the training database composed of the only binary-compositions plays a major role in predicting the higher degree system configurations.
Our result opens a broad avenue to efficient high-throughput investigations of the combinatorial binary calculation for multicomponent prediction of a complex phase.

\end{abstract}
\maketitle

%---------------------------------------------------------------------
\section{Introduction}
The statistical machine learning (ML) is the art to construct statistical models from observational data. 
Machine learning, being a domain of artificial intelligence, revolutionised research in many fields (image processing, natural language, speech processing, biology and medicine,\etc)\cite{Lee18,Seg18,Rua19}, and the number of publications introducing new statistical methods has exploded within the last decades. 
Strange but true that the applications of the machine learning methods to the material sciences, although more and more visible,\cite{Sch19} are still falling behind compared to other applications. 
Among the successful applications of the ML to the materials science, is automatic extraction of predictive models from existing materials data,\cite{Tshi19,War16} and discovery of new class of promising materials or composition, such as the fashionable high entropy alloys (HEA)\cite{Zho19,Kos19,Pei20,Pen20}. 
So far, materials scientists have used ML to build predictive models for a few applications such as to predict heat capacity,\cite{Kau18} semiconducting band gap,\cite{Pil16,Gla20}\etc, but also the heat of formation of inorganic compounds.\cite{Mer14,Kir15,Uba17,Bar20}
All these recent studies have emerged with the powerful use of high-throughput calculations, such as density functional theory (DFT) impulsed for large projects in the last decade (AFLOW\cite{Ose18}, OQMD\cite{Ose18}, NOMAD\cite{Dra18}\etc).
In fact, the increasing availability of DFT data, combined with modern data mining and ML techniques, has enabled the construction of a predictive model to replace DFT calculations and accelerate data generation.\cite{Dem16}
Prediction of crystal structure is still the holy Grail in inorganic chemistry, while the component prediction is one promising approach.\cite{Liu17}.
However, in recent work of Kim\ea,\cite{Kim20} the effect of the space quality has been investigated and illustrated that ML performance can be rather poor if there are several bad (noisy) training data quite close to good candidates.\cite{Jai13} %, such as the energy data in the some Materials Project dataset

The contribution of the current work is two-fold: ($i$) we introduce a general high performance ML-based framework for predicting the heat of formation, corresponding to the energy scale which measures the strength of chemical bonds in a compound, where an input to the ML methods are the combinations of the elements in a given crystal structure; 
and ($ii$) we present and explore new original data set that we constructed and managed. 

Note that heat of formation prediction was the aim of several studies and attempts, like the semi-empirical Miedema's model.\cite{Mie75}
In fact, this fundamental value, called also enthalpy of formation ($\Delta_fH$ in meV) is the key parameter in thermodynamics modelling, such as in the calculation of phase diagrams (Calphad method).\cite{Kauf70}
We have applied the $\Delta_fH$ determination to a large combinatorial challenge, yielded by the distribution of every atom from a given space ($n$-base) into every $s$ non-equivalent crystallographic sites of a given phase.
This kind of description is well known in thermodynamic modelling for addressing the energy of a multicomponent and non-stochiometric phases and is called the Compound Energy Formalism (CEF).\cite{Sun81}
In CEF, each crystal site is considered as a sublattice and the distribution of every atom generates $n^s$ unique configurations, called end-members, the $\Delta_fH$ of which has to be given to use this model.

To illustrate the efficiency of our ML framework, we investigated an important phase in the field of metallurgy: the $\sigma-$phase ($D8_{b}$).
Its complex crystallographic structure is composed of 30 atoms in its tetragonal cell, occupying $s=5$ distinct sites ($i$, $j$, $k$, $l$, $m$),
as summarized in Supplementary Informations SI-\ref{a:cry}.
Its features has been discussed in details.\cite{Jou08}
This phase appears in many types of engineering alloys and its formation prediction requires reliable thermodynamical description. 
Properly, it is shown that it is important to keep a 5-sublattice model in CEF to properly describe the configuration of the $\sigma-$phase in multicomponent alloys.\cite{Mat13, Dup18}
The difficulty lies in the large number of end members which must be considered in multicomponent systems. 
In fact, the description for a binary system (2 elements) leads to the generation of $2^5=32$ different ordered configurations to express, but this number rapidly increases with the degree of the system: ternary with $3^5=243$, quaternary with with $4^5$=1024, ... up to a real alloys with $\sim14$ different elements and its $14^5=537,824$ configurations.
Since the corresponding huge number of $\Delta_fH$ cannot be calculated by classical DFT, their prediction using ML is computationally tractable and, therefore, looks attractive and is one of the major contributions of this paper.

%---------------------------------------------------------------------
\section{Results}
The results of this study -- our new and original data set, ML metrics used to test the statistical methods, and the corresponding results -- are described below. 
First, we discuss the architecture of our method, detailed later in the Methods section. Second, 
 we demonstrate its application to the heat of formation prediction, and, finally, we report the prediction performance.

%------------------------------------
\subsection{General-purpose method}
\begin{figure}[htb]
\includegraphics[width=15cm]{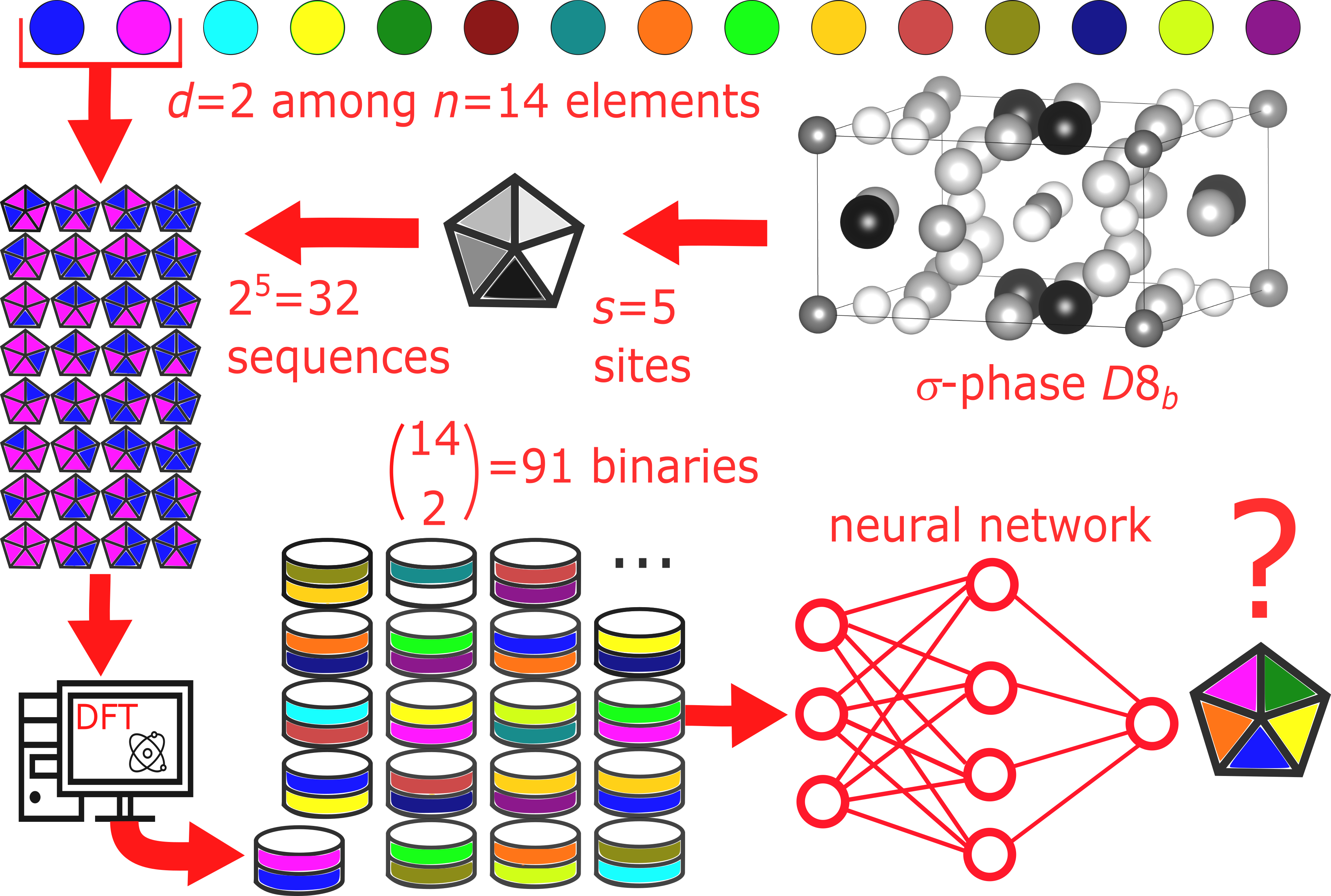}% 
\caption{\label{fig:chartflow} Chart-flow of the methodology presented in the paper.
$(i)$ The crystal structure is summarized as a $s$ non-equivalent sites figure;
$(ii)$ from $n$ available elements, a given system of system degree~$d$ is selected (\eg $d=2$ for binary);
$(iii)$ the permutation leads to $d^5$ unique configurations ;
$(iv)$ every configuration is calculated by DFT, forming a unit of data;
$(v)$ the stack of all $\binom{n}{d}$ units forms a learning database;
then $(vi)$ a supervised machine learning is used to predict multicomponent configurations. 
}
\end{figure}

The originality of current work is in construction of our learning database.
Instead of a mishmash of massive data coming from several independent phases and from various high-throughput sources (calculated using different parameters),
we built a single $\sigma$-phase oriented database with our own consistent massive DFT calculations.
In addition to some additional physical parameters, the main descriptors are the combination of $n=14$ different elements on the different $s=5$ crystallographic sites that described the $\sigma$-phase.
Among the combinatorial set (described in SI-\ref{a:comb}), a selection of  data in the training set includes all the possible binary compositions (system degree $d=2$), which represents only 0.5\% of the all possible configurations.
A graphical chartflow of the present methodology is given on Figure~\ref{fig:chartflow}.\\

%------------------------------------
\subsection{Prediction of the heat of formation}
The observations are the independent variables (from configurations $X_{ijklm}$ of the training database with $N\simeq$ 10,000 data, detailed in SI-\ref{a:ldb}), and
the aim of the regression analysis is to generate a statistical model that can predict a dependent variable, $y_{ijklm}$ (the heat of formation in our case).
Several regression algorithms have been investigated by choosing the best parameter to produce the most accurate generalising results.
The evaluation of the prediction accuracy of a model is characterized using the coefficient of determination $R^2$, the mean absolute error MAE, and the root mean squared error RMSE, given as:
\begin{eqnarray}
R^2 & = & 1 - \frac{\sum \left( y_{ijklm} - \hat{y}_{ijklm}\right)^2}
                   {\sum \left( y_{ijklm} - \bar{y}\right)^2},\\
\md{MAE} & = & \frac{1}{N}\sum_{\{i,j,k,l,m\}}^N \left| y_{ijklm} - \hat{y} \right|, \\
\md{RMSE} & = & \sqrt{\frac{1}{N}\sum_{\{i,j,k,l,m\}}^N \left( y_{ijklm} - \hat{y}_{ijklm}\right)^2},
\end{eqnarray}
where $\hat{y}$ is the predicted value based on the learned model, and $\bar{y}$ the average value.

First, we have estimated a number of regression models from our learning database. 
Namely, we tested the Ridge Linear Regression, Elastic Net Linear Regression, Random Forest Regression (RFR), Multi-Layer Perceptron Regression (MPR), Gradient Boosting Machine (BBM), Support Vector Machine (SVM), K-Nearest Neighbours, Bayesian Ridge Regression, and Gaussian Process Regression (GPR).  
Our averaged results from 10-fold cross validation are summarized in Table~\ref{tab:CV}, and also shown in SI-\ref{a:CV}. 
Classical regression with various regularization such as LASSO (Least Absolute Shrinkage ans Selection Operator), Ridge regression, or their combination, also known as Elastic Net, are not accurate enough, since the number of observations in the database is not very big.
Moreover, the sparsity inducing penalties (LASSO and Elastic Net) are not very relevant to our case, since the number of parameters is quite small.

On the other hand, non-linear supervised learning methods achieve very reasonable performance.
The $R^2$ closest to 1 are obtained with RFR, MPR, GBM and SVM regression algorithms.
The associated best MAE (average from 10-fold CV) are obtained for the MPR method with 13\,meV ($\sim$1\,kJ/mol) using 3 hidden layers, each containing 500 units.

\begin{table}[htb]
\caption{\label{tab:CV}%
Cross validation scores on the complete data set (average values from 10-fold)  using various machine learning methods. MAE, MSE and  RMSE in meV/at (1\,meV$\sim$0.0965\,kJ/mol), illustrated in SI-\ref{a:CV}.
}
\begin{ruledtabular}
\begin{tabular}{lrrr}
Algorithm                        & $R^2$ & MAE &  RMSE \\
\colrule
Ridge Linear Regression                      & 0.45 & 73 &  110 \\
Elastic Net Linear Regression                            & 0.46 & 73 &  109 \\
Random Forest Regressor (RFR)          & 0.89 & 31 &  56 \\ %
Multi-layer Perceptron Regressor (MPR) & 0.96 & 13 &  31 \\ % 
Gradient Boosting Machine (GBM)        & 0.95 & 19 &  37 \\ %
Support Vector Machine Reg (SVM)       & 0.91 & 26 &  54 \\ % 
K-nearest neighbors                    & 0.61 & 55 &  93 \\
Bayesian Ridge Regression              & 0.45 & 73 & 110 \\
Gaussian Process Regression (GPR)      & 0.87 & 30 &  66 \\
\end{tabular}
\end{ruledtabular}
\end{table}

%-------------------------------
\subsection{Validation of learned models on an independent validation set}

\begin{figure}[htb]
\includegraphics[height=6cm]{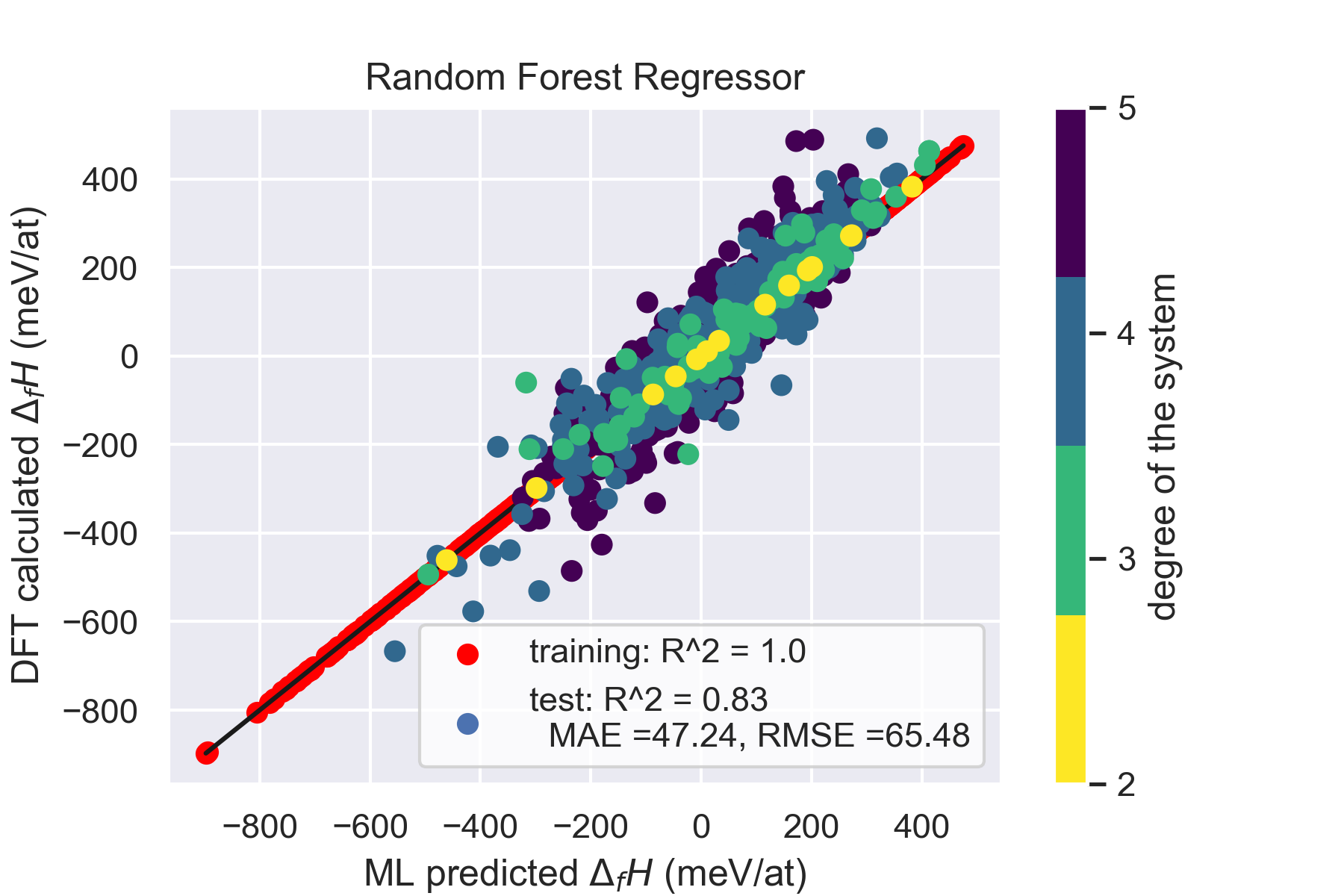}% 
\includegraphics[height=6cm]{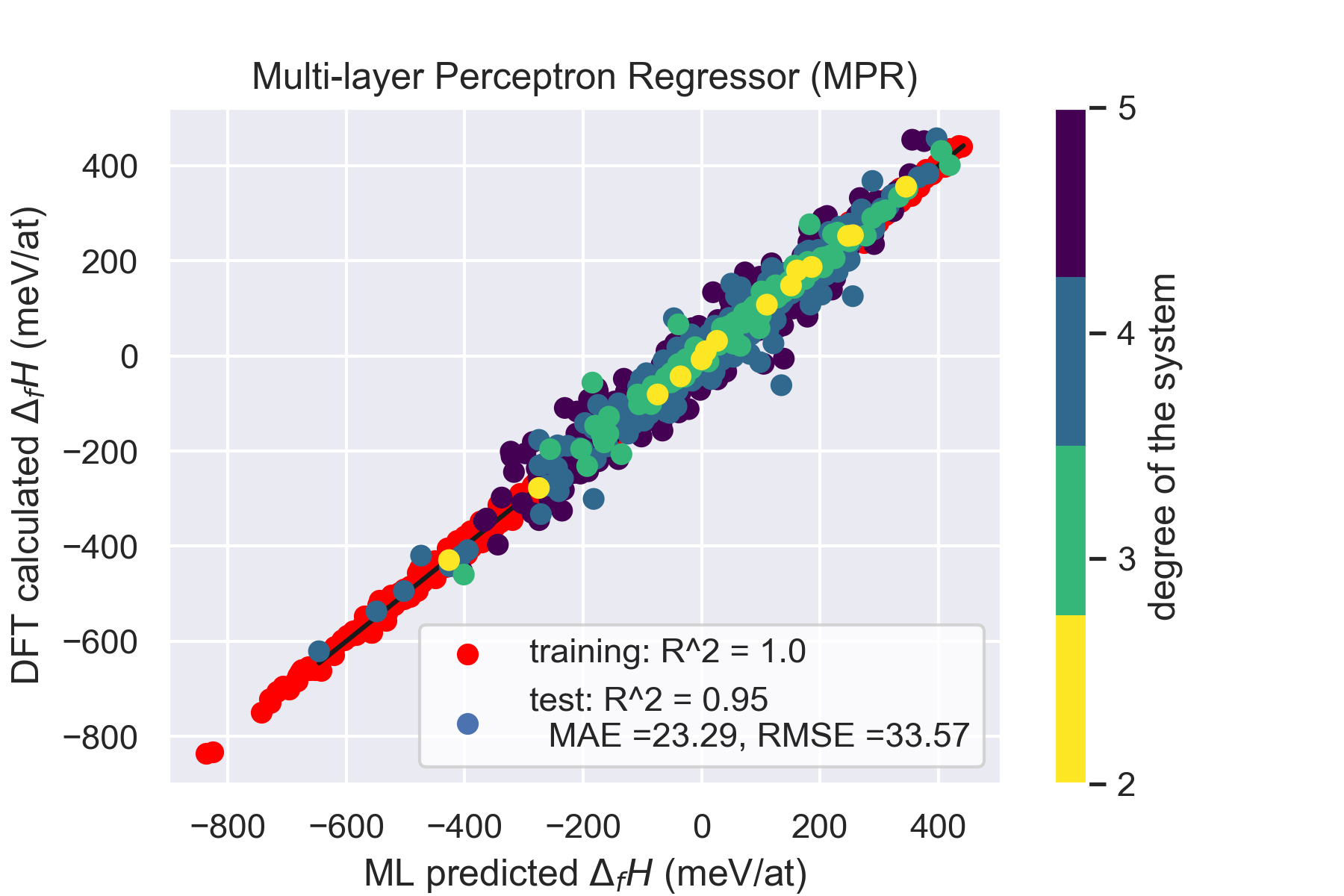}\\ 
\includegraphics[height=6cm]{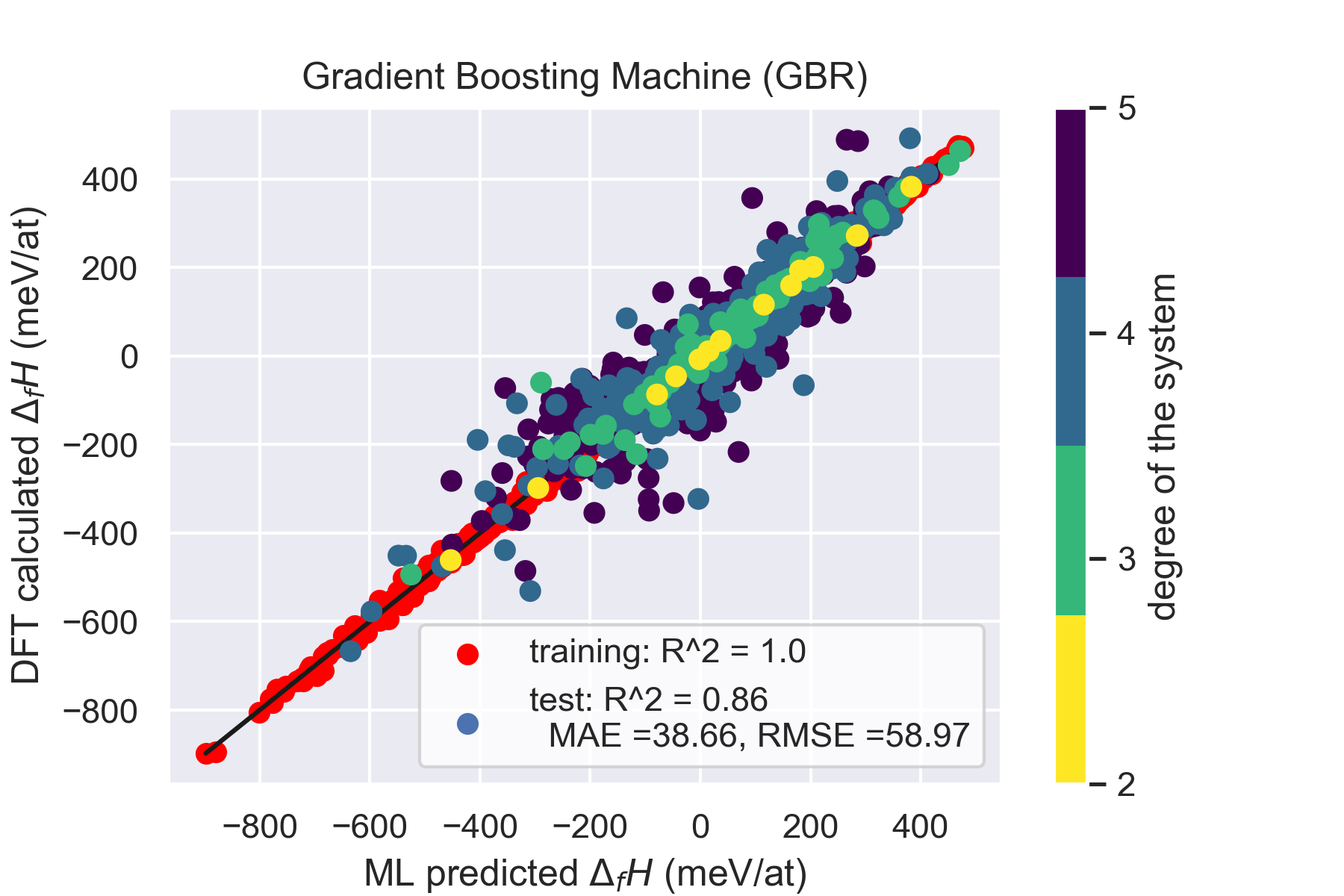}% 
\includegraphics[height=6cm]{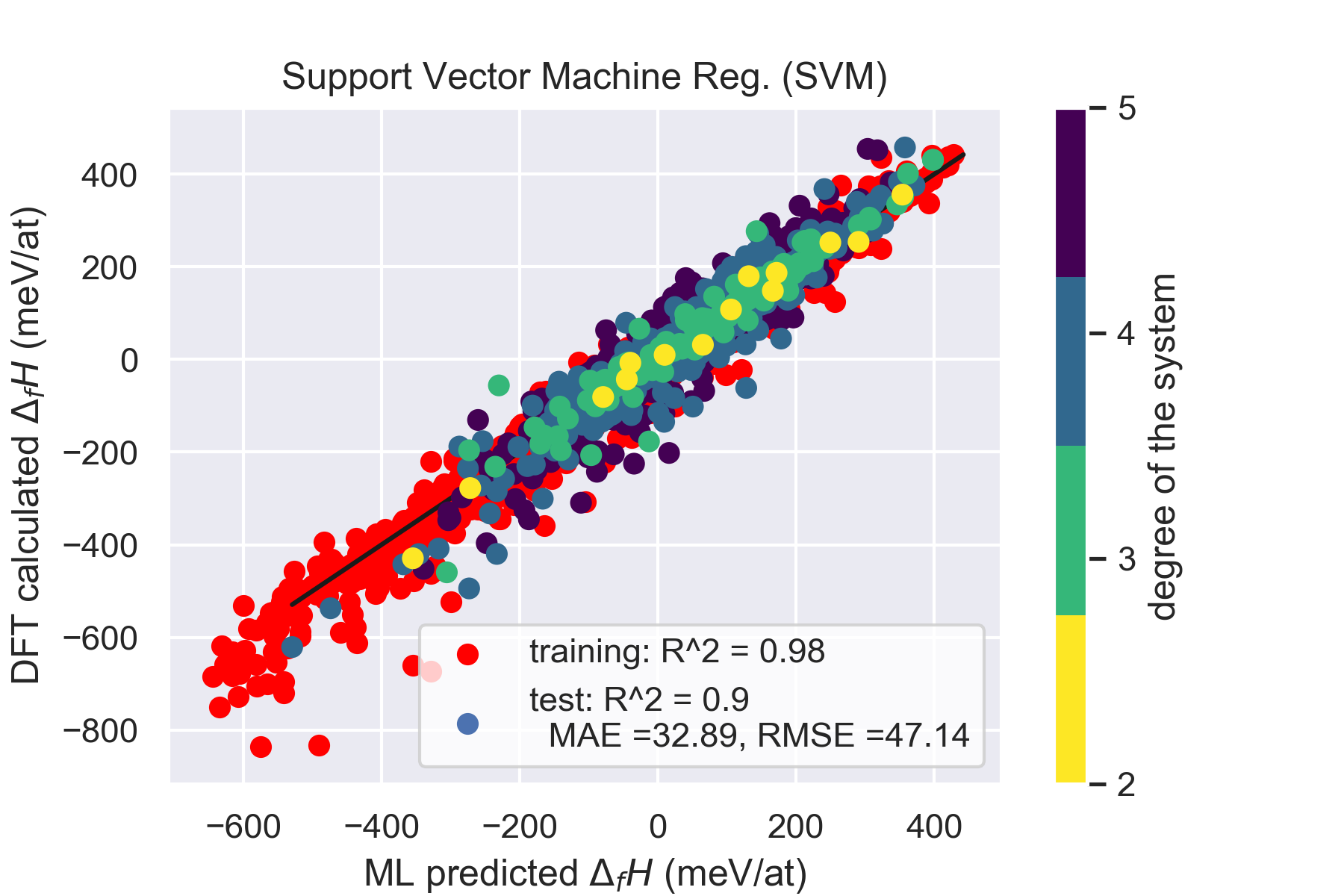}% 
\caption{\label{fig:predi} Prediction of randomized 1001 configurations among the 537,824 ones from the learning of the training database (9974 data in red).
The tested 1001 configurations are reported in colors (blue to yellow) corresponding to the degree~$d$ of their system (right side legend).
The diagonal line indicates the perfect agreement between DFT calculated and ML predicted values.}
\end{figure}

Random Forest Regression, Multi-Layer Perceptron Regression, Gradient Boosting and Support Vector Machines have shown the best performance on the data set containing 9974 inputs (our training database, SI-\ref{a:ldb}). 
We tested these regressors on a new, previously unobserved during the training procedure, set of configurations: 1001 randomized observations among the 537,824 possible ones (testing set given in SI-\ref{a:test}).
Using MPR, the achieved accuracy of total energy, and therefore heat of formation, with a MAE of about 23\,meV/at (Fig.~\ref{fig:predi}) provides a very reasonable accuracy compared to other ML methods of the literature, where the standards are usually around MAE$\sim$50\,meV/at for systems higher than binaries.\cite{Mer14,Kir15,Dem16,Jha18,Zha20}

%---------------------------------------------------------------------
\newpage
\section{Discussion}

%------------------------------------
\subsection{Influence of the system degree}
For the best method, here the regressor-type neural network MPR, the accuracy of prediction depends of the degree of the system of the tested configuration, illustrated by the color code of Fig.~\ref{fig:predi}.
As an example, the MAE$\sim$23\,meV for the whole testing set could be decomposed as contribution depending of the system degree~$d$.
It increases with~$d$: 7\,meV ($d=2$), 22 ($d=3$), 28 ($d=4$) and 38 ($d=5$).
This result illustrates obviously that multi-component systems are more difficult to predict. 

\begin{figure}[htb]
\includegraphics[height=5.5cm]{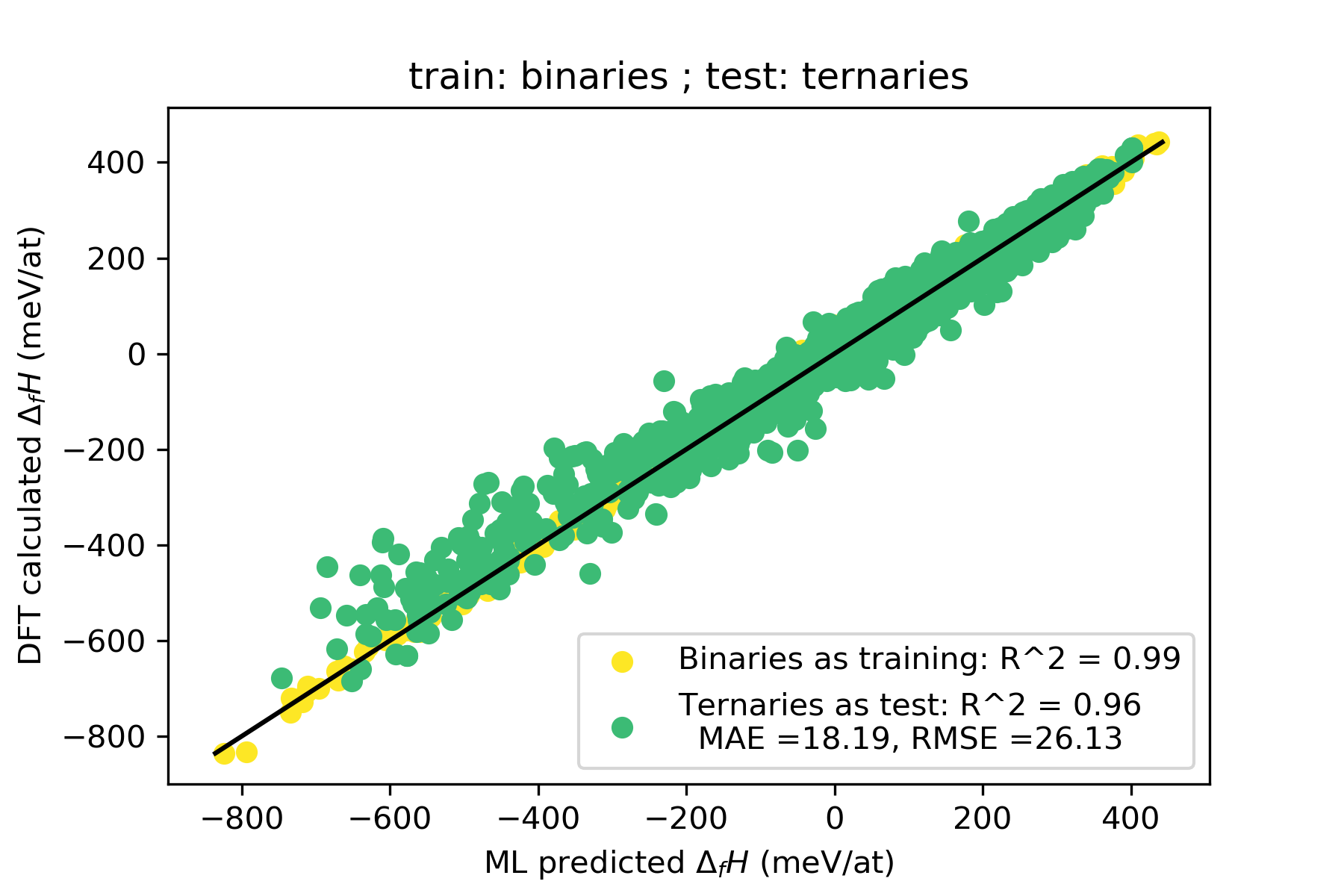}% 
\includegraphics[height=5.5cm]{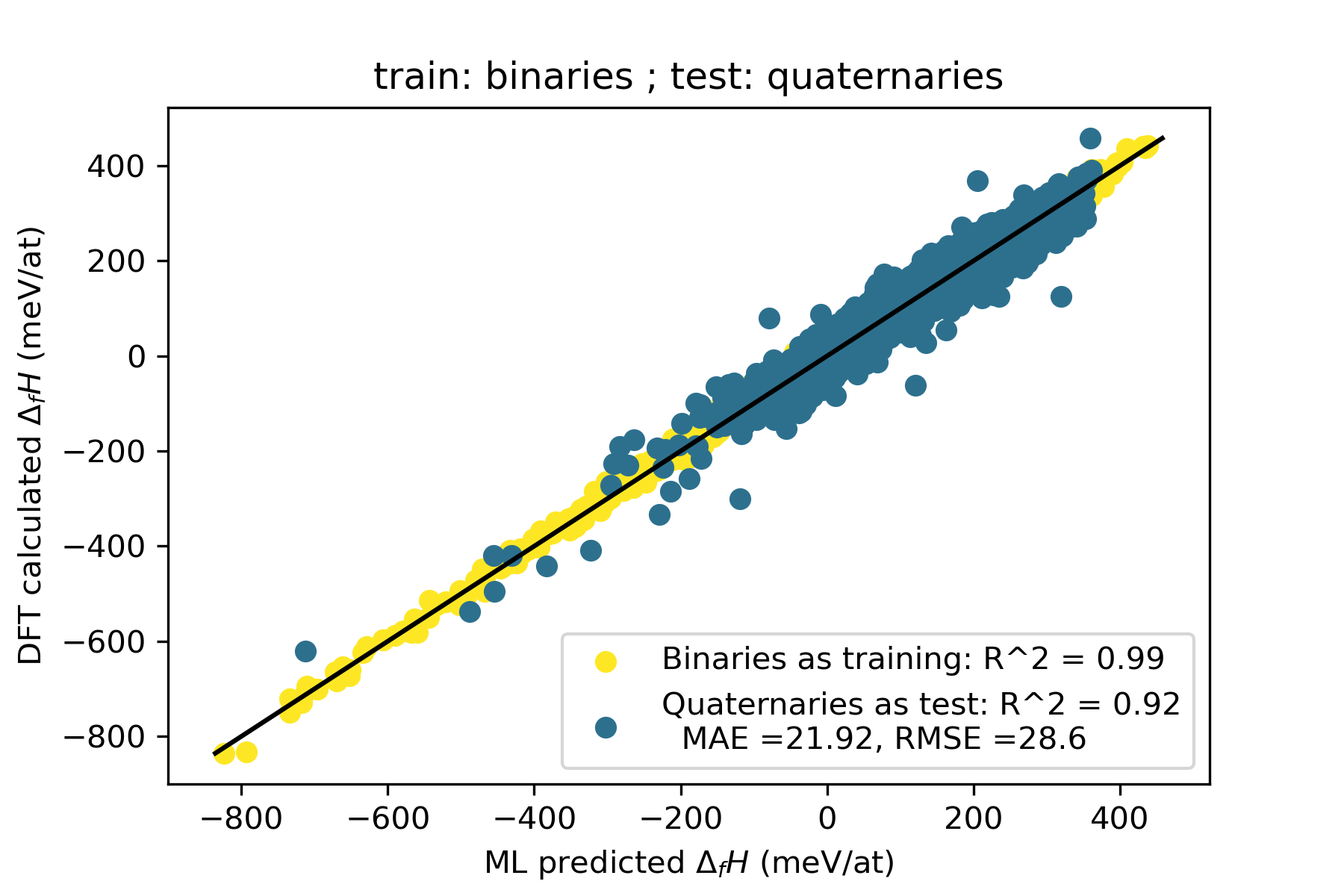}\\%
\includegraphics[height=5.5cm]{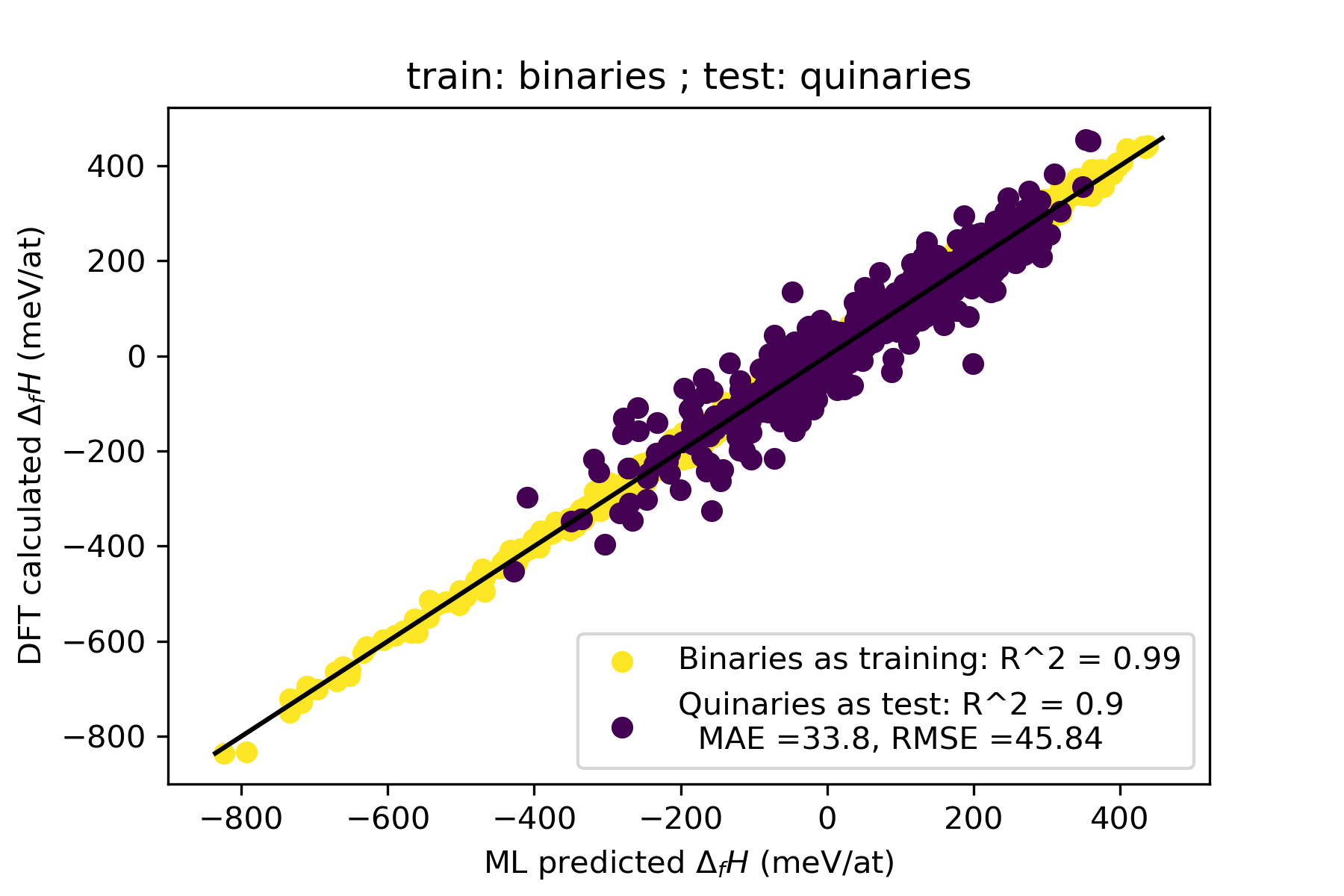}% 
\includegraphics[height=5.5cm]{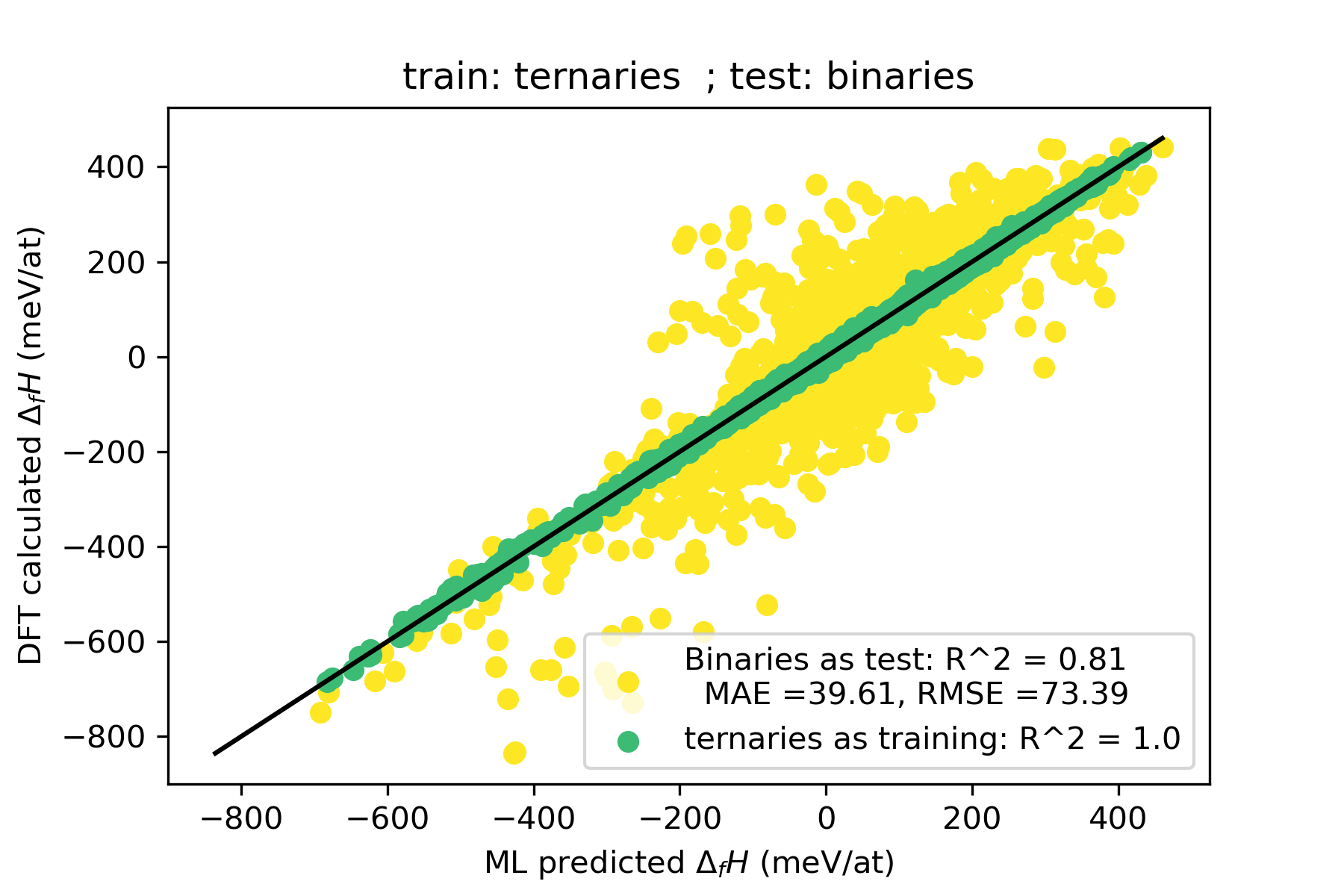}% 
\caption{\label{fig:sim} From training on only binaries, prediction of ternaries, quaternaries or quinaries. 
The diagonal line indicates the perfect agreement between predicted and real values.}
\end{figure}

Another question could be addressed to the learning weight of binaries: does the whole $d=1$ (14 elements) and $d=2$ systems (here the 91 different sub-systems $\times30$ configurations) are sufficient to predict higher degree systems?
In other words, is it possible to predict accurately the whole possible $14^5$ combinations only from all unary and binary configurations (2744 unique data), which is representative of only 0.5\% of the total set?
In order to answer, we merged our training and testing sets leading to 10941 unique configurations and split them in the 5 sub-systems: 14 ``$d=1$" , 2730 ``$d=2$", 5051 ``$d=3$", 2571 ``$d=4$" and 575 ``$d=5$" configurations.
Then, from the all unique unary and binary configurations, we have tested the predictive behaviour for higher degree systems as shown in Fig.~\ref{fig:sim}.
Whereas the ternary and quaternary systems are well predicted with MAE$\sim$18 and 22\,meV respectively, the quinaries still present surprisingly reasonable results with MAE$\sim$34\,meV respectively.
However, from higher to lower degree systems, a learning from a portion of the ternary configurations (5051 among the  54,600 possible) gives larger dispersion on prediction on the binaries with higher MAE$\sim$40\,meV and RMSE$\sim$73\,meV.
Other combinations of training/testing subsystems are illustrated in SI-\ref{a:simu}.

%------------------------------------
\subsection{Contribution of additional descriptors}
The consideration of additional physical descriptors improves slightly the learning scores. 
As shown in SI-\ref{a:wwu}, the prediction with neither atomic radius nor the number of valence electron is slightly worse (MAE$\sim$24 with MPR).
This result might seem unexpected at first glance.
In fact, it is well known  that topologically close packed (TCP) structures, as $\sigma-$phase, are driven by geometric arguments: 
since atoms are in the center of a coordination sphere, the atomic radius reflects the capability to occupy small or large coordination number (CN) sphere. 
The number of valence electron is also known to be important.
In fact, for similar radius, a study has shown that the degeneracies of electronic levels plays a role on the site preference.\cite{Slu09}
To summarize this point, it looks that the total energy calculated by DFT contains these additional properties and does not need to be given in separated descriptors especially using a versatile deep learning approach like the MPR.

%------------------------------------
\subsection{Prediction of crystal properties}
The $\Delta_fH$ is only one predictive variable among many describing the crystal structure of the $\sigma$ phase.
Considering the only crystal definition, 9 variables are necessary to describe a configuration: 2 cell parameters ($a$, $c$) and 7 internal parameters ($x^{4f}$, $x^{8i_1}$,	$y^{8i_1}$, $x^{8i_2}$,	$y^{8i_2}$, $x^{8j}$,	$z^{8j}$).
In the present work, the supervised learning is optimized for predicting the $\Delta_fH$ but was also applied for every other variables and helps to initialize new DFT input files. 
In fact, it has been used to predict the starting structure of a large part of our learning database, and helped to reduce $\sim$10\,times the CPU consumption for the DFT relaxation steps.
From our best model (optimized MPR) and from all available learning sets (9974+1001=10975 data), the predictive  $\Delta_fH$ and the 9 other crystal variables are given for the every 537,824 configurations in SI-\ref{a:predi}.
As an example, the prediction of both $a$ \& $c$ tetragonal cell parameters presents a MAE$\sim$0.06 \& 0.07\,\AA\ and a RMSE$\sim$0.08 \& 0.10 respectively (SI-\ref{a:cell}).

%------------------------------------
\newpage
\subsection{Conclusions and outlook}
This work addresses the issue of the crystal phase stability from the machine learning viewpoint.
Because the $\Delta_fH$ is the key descriptor to model the formation of compounds, we have investigated the prediction of this variable using a supervised approach, using a complex crystallographic structure as an example: the $\sigma$-phase.
Based on an unprecedented large first principles dataset containing about 10,000  compounds with $n=14$ different elements, we optimized several supervised learning approaches, where the Multi-layers Perceptron Regressor presents best results to predict all the $\sim$500,000 possible configurations within a mean absolute error of 23\,meV ($\sim$2\,kJ.mol$^{-1}$). 
Additional descriptors with roots in the physical nature of the problem are minor contribution to the learning score in comparison with the only combinatorial DFT set.
It is shown that the training database from the only binary-compositions (0.5\% occurrence of whole set) are able to predict multicomponent configurations with a high accuracy.
This result suggests that several complex  phases including non-equivalent sites could be easily determined from the only binary contribution.

As an outlook, the learning database size will be increased up to twenty elements, including Ta and  Si.
%Moreover, the magnetism will be considered in future.
This work will be extended to other complex TCP phases with more than 2 sites to demonstrate the efficiency of our approach, such as $A13$, $C14$,\etc
Indeed, it opens broad avenues in the study of complex structures with the only binary configurations as a learning set, this could be efficient even with low number of data.

%---------------------------------------------------------------------
\newpage
\section{Methods}

%------------------------------------
\subsection{Training database from DFT calculations}
First, a database from DFT calculations has been compiled.
Since 2008, many active groups have calculated $\sigma$-phase configurations in binary\cite{Ber01,Kor05,Slu09,Cie10,Pal10}, ternary\cite{Chv04,Cri10b,Pal11,Yaq12} and quaternary systems.\cite{Cri15}
Since all these sparse studies were calculated with different methods and parameters, our present original database includes only new calculations obtained under the same conditions, required millions of hours of CPU time to construct.
The DFT methodology details are explained in SI-\ref{a:dft}. 
The learning database includes $n=14$ different elements: Al, Co, Cr, Fe, Mn, Mo, Nb, Ni, Pt, Re, Ru, V, W, Zr, and contains 9974 unique configurations embracing all the $\binom{14}{2}=91$ binaries (degree $d=2$), 33 on the $\binom{14}{3}=364$ ternaries ($d=3$), 9 on the 1001 quaternaries ($d=4$) and only 1 on the 2002 possible quinaries ($d=5$, see SI-\ref{a:comb} for the combinatorial descriptions and SI-\ref{a:ldb} for the detailed list of systems in our training database).
The elemental distribution is not uniform because of some chemical reasons explaining that we wanted to have more data for pertinent systems (\eg Zr-based $\sigma$-phase is not frequent).
This analysis could be shown from SI-\ref{a:stat}.
In addition, an independent testing set for 1001 randomized configuration were calculated (detailed in SI-\ref{a:test}).

%------------------------------------
\subsection{Database construction format}
In a second step, data was arranged as a learning database, $X_{ijklm}\in\mathbb{R}^{(n+2)s\times N}$. 
The $n=14$ elements are categorical variables but need to be treated with analytical methods that require numbers.
Thus, each of the 5 crystal sites ($i$, $j$, $k$, $l$, $m$) has been considered as a 14-dim vector of dummies (spin variables: 0 or 1) by the one hot encoding method. 
In addition, because it is well known that the stability in this kind of compounds is driven by the two geometric and electronic constraints\cite{Jou08} (\eg large electropositive atoms have a preference on high coordination sites), atom size and electron concentration have been used as additional descriptors.
In total, we use a set of $p=(n+2)s=80$ attributes corresponding to the following features for each configuration $X_{ijklm}$, with ($i$, $j$, $k$, $l$, $m$) $\in \{$ Al, Co, Cr, \dots , V, W, Zr $\}$:
\begin{itemize}
\item Ordering configuration of atoms in the crystal ($14\times5$ vectors of dummies)
\item Atomic radius (5 normalized values, related to the 5 atoms in $ijklm$ configuration)
\item Number of valence electrons (5 normalized values)
\end{itemize}
leading to a $9974\times80$ matrix as the learning database, associated to the target $y_{ijklm}$ vector, here the heat of formation, $\Delta_fH(ijklm)$, but could be any crystallographic properties such as cell parameters.

At last, based on the ML best results, the learning on the whole database (9974 configurations) was done and a final prediction of 1001 random configurations among the 537,824 was estimated.

%------------------------------------
\subsection{Estimation of the machine learning models}
\begin{figure}
\includegraphics[scale=0.45]{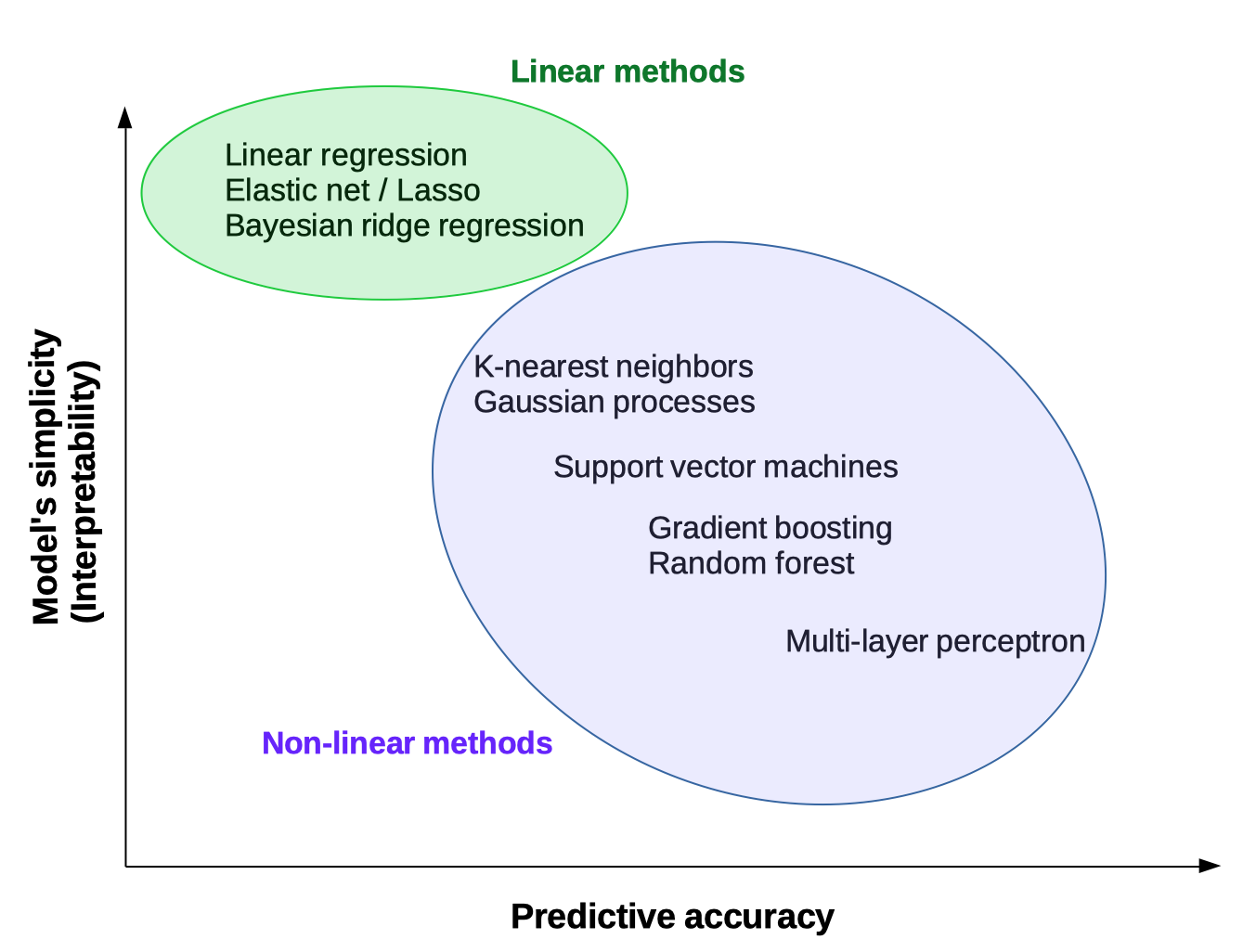}
\caption{A schematic representation of the models' simplicity and the generalising accuracy for the tested Machine Learning approaches.}
\label{fig:MLmethods}
\end{figure}

The machine learning models mentioned above are estimated using Scikit-Learn 0.23 library in Python 3.5.
Each approach offers different advantages, such as speed or interpretability, but our main goal was the high accuracy. 
Figure~\ref{fig:MLmethods} illustrates the relationship between the model's simplicity (interpretability) and the generalising performance for the machine learning methods we considered. 
Note that this scheme (Figure~\ref{fig:MLmethods}) is approximate. 
What is true for our results is that the linear methods perform worse than the non-linear regressions. 
We also noticed that the grid search for an optimal Multi-Layer Perceptron architecture (number of hidden layers and units), as well as the number of trees and their maximal depth in the Random Forest and Gradient Boosting, is important, and the search for an optimal configuration can be computationally expensive.

%We have taken steps to automate optimization of models with the aim to make the better results, such as number of branches in random forest, and layer number map for neural networks.
For each method, the corresponding hyper-parameters were fixed using the grid search module and the cross-validation error rate.
The generalising performance is the test accuracy using 10-fold cross validation procedure: the database is randomly split into  10 subsets (folds), and the model is trained on 9 parts, and tested on 1 part. 
The procedure is repeated 10 times. 
The average test accuracy is the mean value over performances on the test data.

Below, we provide a short description of the considered machine learning methods.
The data consists in $N$ observations $\{X_i, y_i\}_{i=1}^N$, where $y \in \mathbb{R}$. 
Each observation is a vector of length $p$, where $p$ is the number of features (parameters), $p=80$ in our case. 
\begin{itemize}
\item \textit{Ordinary least squared linear regression.} The predicted value is modelled as follows: 
\begin{align}
y_i = \beta_1 x_{i1} +  \beta_2 x_{i2} + \dots + \beta_p x_{ip} + \epsilon_i,
\end{align}
where $\epsilon_i$ is unobserved noise. 
To avoid overfitting, the loss function is penalised by the $L_2$ penalty term, also known as Ridge regression:
\begin{align}
\hat{\beta} = \arg \min_{\beta} \| y - X \beta \|_2^2 + \lambda \| \beta \|_2^2,
\end{align}  
where $\lambda$ is associated with the penalty term, ans is usually is fixed using the grid search.
\item \textit{Elastic net linear regression.} The elastic net is a combination of the $L_1$ and $L_2$ penalty terms:
\begin{align}
\hat{\beta} = \arg \min_{\beta} \| y - X \beta \|_2^2 + \lambda_2 \| \beta \|_2^2 + \lambda_1 \| \beta \|_1.
\end{align}
In case where $\lambda_1 = 0$, we get the above mentioned ridge regression, and if $\lambda_2 = 0$, the problem boils down to the Lasso. 
An optimal choice of $\lambda_1$ and $\lambda_2$ leads to a sparse but accurate model. 
The number of parameters is quite small in our database, and the feature selection is not relevant for our case.
\item \textit{Random forest regression (RFR).} The random forest approach is an ensemble non-linear machine learning method that fits a number of decision trees (number of estimators) that vote for the final decision. 
A number of hyper-parameters, such as the number of trees and their maximal depth has to be fixed (usually via the cross validation) to achieve a reasonable trade off between the model's complexity and the generalising performance.  
\item \textit{Gradient boosting regression (GBR).} The gradient boosting method constructs an additive model. 
At each iteration of the learning procedure, a new regression tree which corrects the current model, is added. 
Similarly to the random forest, here also, the number of trees (or the number of estimators) and their maximal depth are important hyper-parameters that control the model's complexity and accuracy (very deep decision trees are likely to lead to overfitting).
\item \textit{Multi-layer Perceptron Regression (MPR).} A multi-layer perceptron (or artificial neural network) includes an input layer, several (or at least one) hidden layers, and an output layer. 
In our experiments, we use three hidden layers, and we get the following transformations:
\begin{align}
f(X) = h\Big(b^3 + W^3 h \big(b^2 + W^2 h (b^1 + W^1 X)\big)\Big),
\end{align}
where $h$ is the activation function, and in our experiments tanh is chosen $h(a) = (e^a - e^{-a})/(e^a + e^{-a})$. The parameters to estimate are $\{b^1, W^1, b^2, W^2, b^2, W^3\}$. The optimised function is the squared loss function. We use a stochastic gradient descent. 
\item \textit{Support vector regression (SVM).} We use the $\epsilon$-support vector regression with the rbf (radial basis function) kernel that leads to a non-linear separator. The loss function is penalised by the $L_2$ norm, and the hyper-parameter $C$ which is the regularisation parameter, is to be fixed. The choice of the non-linear kernel is motivated by our results with both linear (linear regression) and non-linear (MPR, boosting, random forest) methods, illustrating that the non-linear approaches are more efficient for our task.
\item \textit{K-nearest neighbours.} The algorithm uses similarity between points to predict the values of new instances. We use the Manhattan distance, since it is more robust against the outliers. First, the distance between a new point and all available data is computed. Then, the $k$ nearest neighbours are selected, and the average of these points is the predicted value. For our experiments, we fixed $k=3$.
\item \textit{Bayesian ridge regression.} The method is a ridge regression in the Bayesian viewpoint, and it uses probability distributions rather than point estimates. The response $y$ is assumed to be drawn from a probability distribution
\begin{align}
y = \mathcal{N} (\beta^T X, \sigma^T I), 
\end{align}
where $\sigma$ is the standard deviation, $\mathcal{N}(\cdot)$ is a normal (Gaussian) distribution, $T$ stands for \textit{transpose}, and $I$ is the identity. The goal of the Bayesian approach is rather to determine the posterior distribution for the parameters of the model:
\begin{align}
\mathbb{P} (\beta | X, y) = \frac{\mathbb{P}(y | \beta,X) \mathbb{P}(\beta | X)}{\mathbb{P}(y | X)}.
\end{align}
\item \textit{Gaussian process regression (GPR).} The approach is non-parametric and Bayesian. A Gaussian process is like a multivariate Gaussian distribution in an infinite dimension, and any collection of labels of a data set are joint Gaussian distributed:
\begin{align}
y \sim \text{GP} \big( m(X), k(X, X') \big),
\end{align}
where $m(\cdot)$ is a mean function, and $k(\cdot)$ is a covariance function.
\end{itemize}

%--------------------------------------------------------------------------
\section*{Data and code availability}
All data generated in this study are available from the authors on request, and will be included in the regular publication version.

%---------------------------------------------------------------------
\section*{Acknowledgements}
%JCC thanks the Covid19 pandemic and its lock-down which allow him to focus only on this work at home supported by its two fanatic children.
We wish to acknowledge the support of the GENCI-CINES:  DFT calculations were performed using their HPC resources (Grant A0060906175).
In addition, we acknowledge the financial support from the CNRS (programs MaLeFHYCe, PEPS, Cellule Energie CNRS and MALEpHYq, Emergence@INC)

%---------------------------------------------------------------------
\section*{Author contributions}
JCC designed the project, performed the calculation.
JMJ analysed the results as Calphad output.
NS performed the machine learning optimization.
All the authors together finalized the manuscript.
%---------------------------------------------------------------------
\section*{Competing Interests}
The authors declare no competing interests

%---------------------------------------------------------------------
\section*{Additional Information}
Supplementary information is available for this paper in several Appendix.

%--------------------------------------------------------------------------
\section*{References}
%\bibliographystyle{apsrev4-1}
%\bibliographystyle{apsrev}
%\bibliographystyle{unsrt}
\bibliographystyle{ieeetr}
\bibliography{bibliojc}

%---------------------------------------------------------------------
\newpage
\appendix
\section{Crystal details of the $\sigma$-phase}
\label{a:cry}
The crystal structure of the $\sigma$-phase is tetragonal, described by the $P4_2/mnm$ space group (no.~136), with five non-equivalent positions ($s=5$).
Depending on the system, the cell parameters $a$ and $c$ take experimental values from 8.78 to 10.06\,\AA\ and 4.55 to 5.23\,\AA\ respectively.
The $\sigma-$phase belongs to the Frank–Kasper or topologically close packed phases, characterized
by the unique presence of tetrahedral interstices, and a limited number of coordination polyhedra.\cite{Jou08}

\begin{figure}[htb]
\includegraphics[width=7cm]{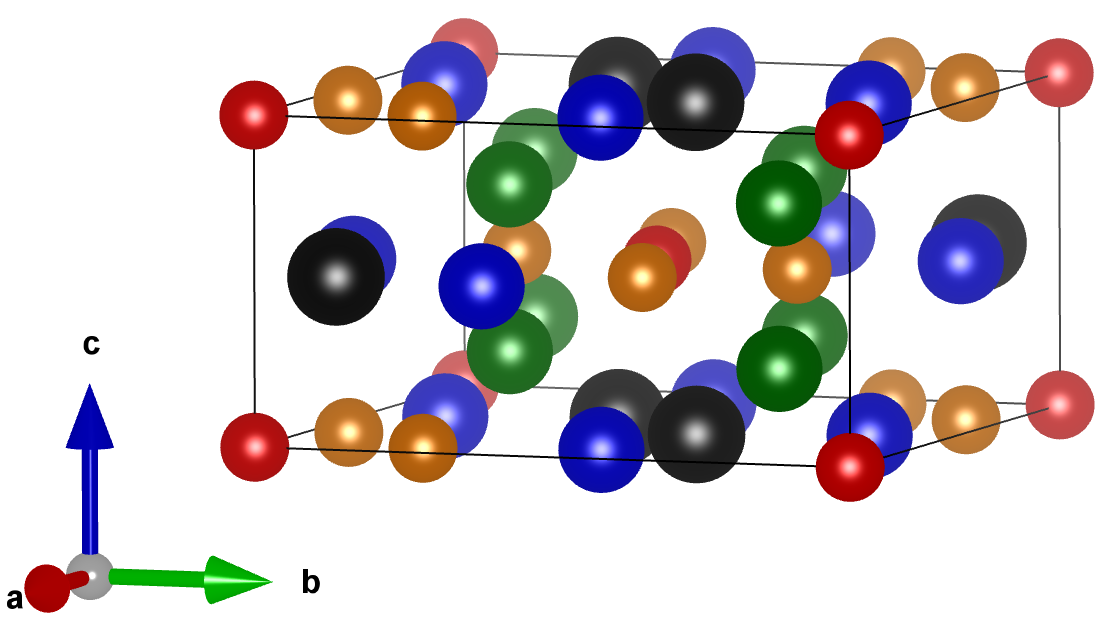}
\caption{\label{fig:cry} Representation of the primitive cell of the crystal phase $\sigma-$phase ($D8_b$), with its 5 non-equivalent sites: $2a$ (red balls), $4f$ (black), $8i_1$ (blue), $8i_1$ (orange) and $8j$ (green).}
\end{figure}

\begin{table}[htb]
\caption{\label{tab:cry}%
Crystal structure of the $\sigma-$phase ($D8_b$): sites, Wyckoff positions, atomic positions (average values) in the $P4_2/mnm$ space group (no.~136). Details of the first neighbors number and their ratio, and coordination number (CN).}
\begin{ruledtabular}
\begin{tabular}{lcccc|cccccc|r}
Site & Wyc. & $x$  & $y$  & $z$ & $i$ & $j$ & $k$ & $l$  & $m$ & ratio & CN\\
\hline
 $i$  & $2a$  & 0  & 0  & 0 & 0 & 4 & 0 & 4 & 4 & 0 & 12 \\
 $j$ & $4f$  & $\simeq$0.399  & $x$  & 0 & 2 & 1 & 2 & 4 & 6 & 0.07 & 15 \\
 $k$& $8i_1$& $\simeq$0.464  & $\simeq$0.131  & 0 & 0 & 1 & 5 & 4 & 4 & 0.36 & 14 \\
 $l$ & $8i_2$& $\simeq$0.741  & $\simeq$0.066  & 0 & 1 & 2 & 4 & 1 & 4 & 0.08 & 12 \\
 $m$  & $8j$  & $\simeq$0.187  & $x$  & $\simeq$0.251 & 1 & 3 & 4 & 4 & 2 & 0.14 & 14 \\
\end{tabular}
\end{ruledtabular}
\end{table}

%--------------------------------------------------------------------------
\newpage
\section{Analysis of the combinatorial descriptions}
\label{a:comb}

Considering a set of $n=14$ different elements that could be arranged in a crystallographic phase~$\varphi$ of $s=5$ inequivalent sites, there is a total of $N=$ 14$^5=$ 537,824 unique configurations. 
The system of $n$ elements is composed of sub-systems of degree $d\leq s$ that could describe a unique $\varphi$ configuration.
Each subsystem owns $d^5$ configurations, that could be decomposed in a number of sub-systems of lower degree as detailed in Table bellow.
In addition, the numbers of subsystems is given by a binomial coefficient $\binom{n}{d}=\frac{n!}{d!(n-d)!}$, \eg there is $\binom{14}{4}=1001$ quaternary systems for $n=14$.
A summary of the combinatorial descriptions is given in the following Table and figure.

\begin{table}[htb]
\caption{\label{tab:comb}%
For a system with $n=14$ elements distributed in a phase~$\varphi$ of $s=5$ sites, are given:
degree~$d$ of a subsystem, number of configuration $d^5$, number of configuration in each subsystem of lower degree $d\simeq s$,
number of different subsystems, 
total and proportion of the number of configurations.
}
\begin{ruledtabular}
\begin{tabular}{lc|ccccc|c|ll}
$d$ & $d^5$ & $N(d=1)$ & $N(d=2)$ & $N(d=3)$ & $N(d=4)$ & $N(d=5)$ & $\binom{n}{d}$ & \multicolumn{2}{c}{$N$}\\
\hline
1 &   1 & 1 &   0 & 0 & 0 & 0 & 14 & 14 & (0.0\%)\\
2 &  32 & 2 &  30 & 0 & 0 & 0 & 91 & 2,730 & (0.51\%)\\
3 & 243 & 3 & 3$\times$30 & 150 & 0 & 0 & 364 & 54,600 & (10.15\%)\\
4 &1024 & 4 & 6$\times$30 & 4$\times$150 & 240 & 0 & 1,001 & 240,240 & (44.67\%)\\
5 &3125 & 5 & 10$\times$30 & 10$\times$150 & 5$\times$240 & 120 & 1,001 & 240,240 & (44.67\%)\\
\end{tabular}
\end{ruledtabular}
\end{table}

\begin{figure}[htb]%left bottom right top
\includegraphics[trim=110 20 30 50,clip, height=5cm]{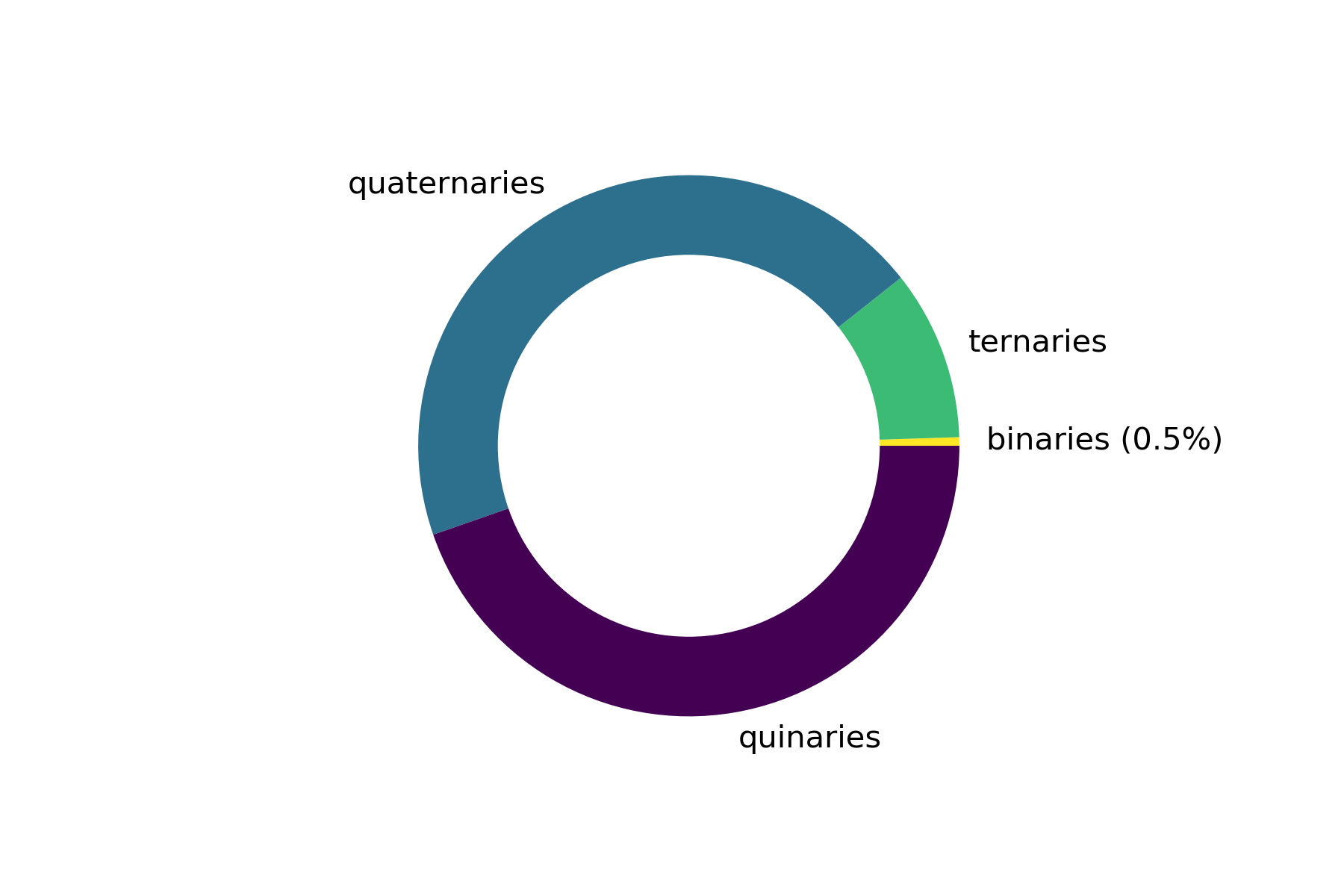}
\includegraphics[trim=75 20 10 50,clip, height=5cm]{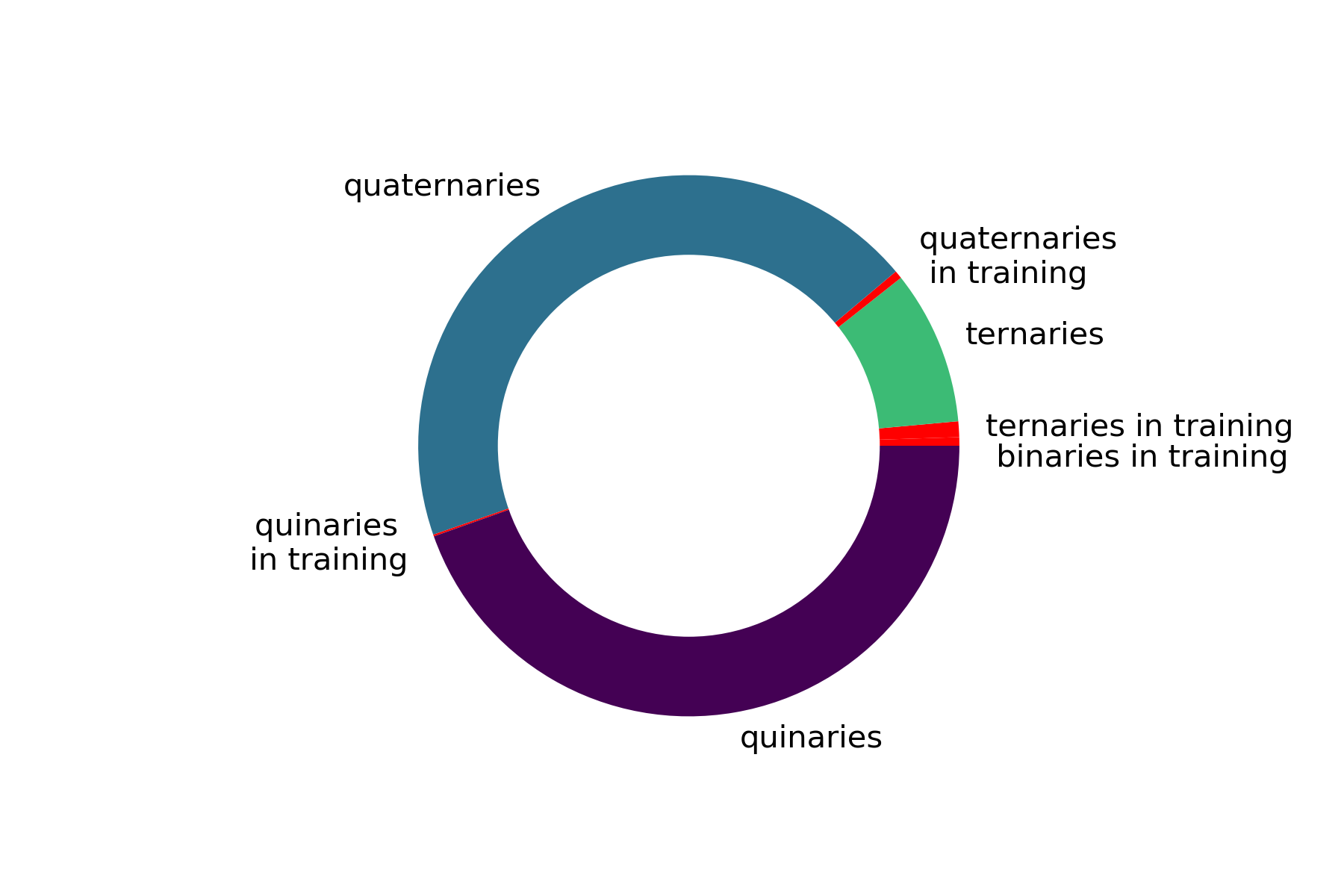}
\caption{\label{f:dis_ele} Proportion of sub-systems in the whole set for $s=5$ and $n=14$ (537,824 unique configurations) (left) and calculated sub-systems included in the learning database in red (right).}
\end{figure}

%--------------------------------------------------------------------------
\newpage
\section{The training database, 9974 configurations}
\label{a:ldb}
\subsection{List of systems included in the training database}
91 binary systems with all the corresponding $2^5=32$ configurations, 91 among $\binom{14}{2}=91$:
\begin{small}
\begin{verbatim}
['Al', 'Co'], ['Al', 'Cr'], ['Co', 'Cr'], ['Al', 'Fe'], ['Co', 'Fe'], 
['Cr', 'Fe'], ['Al', 'Mn'], ['Co', 'Mn'], ['Cr', 'Mn'], ['Fe', 'Mn'], 
['Al', 'Mo'], ['Co', 'Mo'], ['Cr', 'Mo'], ['Fe', 'Mo'], ['Mn', 'Mo'],
['Al', 'Nb'], ['Co', 'Nb'], ['Cr', 'Nb'], ['Fe', 'Nb'], ['Mn', 'Nb'],
['Mo', 'Nb'], ['Al', 'Ni'], ['Co', 'Ni'], ['Cr', 'Ni'], ['Fe', 'Ni'],
['Mn', 'Ni'], ['Mo', 'Ni'], ['Nb', 'Ni'], ['Al', 'Pt'], ['Co', 'Pt'],
['Cr', 'Pt'], ['Fe', 'Pt'], ['Mn', 'Pt'], ['Mo', 'Pt'], ['Nb', 'Pt'],
['Ni', 'Pt'], ['Al', 'Re'], ['Co', 'Re'], ['Cr', 'Re'], ['Fe', 'Re'],
['Mn', 'Re'], ['Mo', 'Re'], ['Nb', 'Re'], ['Ni', 'Re'], ['Pt', 'Re'],
['Al', 'Ru'], ['Co', 'Ru'], ['Cr', 'Ru'], ['Fe', 'Ru'], ['Mn', 'Ru'],
['Mo', 'Ru'], ['Nb', 'Ru'], ['Ni', 'Ru'], ['Pt', 'Ru'], ['Re', 'Ru'],
['Al', 'V'],  ['Co', 'V'],  ['Cr', 'V'],  ['Fe', 'V'],  ['Mn', 'V'],
['Mo', 'V'],  ['Nb', 'V'],  ['Ni', 'V'],  ['Pt', 'V'],  ['Re', 'V'],
['Ru', 'V'],  ['Al', 'W'],  ['Co', 'W'],  ['Cr', 'W'],  ['Fe', 'W'],
['Mn', 'W'],  ['Mo', 'W'],  ['Nb', 'W'],  ['Ni', 'W'],  ['Pt', 'W'],
['Re', 'W'],  ['Ru', 'W'],  ['V', 'W'],   ['Al', 'Zr'], ['Co', 'Zr'],
['Cr', 'Zr'], ['Fe', 'Zr'], ['Mn', 'Zr'], ['Mo', 'Zr'], ['Nb', 'Zr'],
['Ni', 'Zr'], ['Pt', 'Zr'], ['Re', 'Zr'], ['Ru', 'Zr'], ['V', 'Zr'],
['W', 'Zr'].
\end{verbatim}
\end{small}

33 ternary systems with all the corresponding $3^5=243$ configurations, 33 among $\binom{14}{3}=364$:
\begin{small}
\begin{verbatim}
['Cr', 'Fe', 'Mn'], ['Cr', 'Fe', 'Mo'], ['Co', 'Fe', 'Ni'], ['Cr', 'Fe', 'Ni'],
['Cr', 'Mn', 'Ni'], ['Fe', 'Mn', 'Ni'], ['Cr', 'Mo', 'Ni'], ['Al', 'Nb', 'Ni'],
['Al', 'Nb', 'Pt'], ['Co', 'Cr', 'Re'], ['Cr', 'Mo', 'Re'], ['Mo', 'Nb', 'Re'],
['Cr', 'Ni', 'Re'], ['Mo', 'Ni', 'Re'], ['Mo', 'Pt', 'Ru'], ['Al', 'Fe', 'V'],
['Co', 'Fe', 'V'],  ['Cr', 'Fe', 'V'],  ['Cr', 'Mn', 'V'],  ['Cr', 'Mo', 'V'],
['Fe', 'Mo', 'V'],  ['Co', 'Ni', 'V'],  ['Fe', 'Ni', 'V'],  ['Cr', 'Fe', 'W'], 
['Cr', 'Mn', 'W'],  ['Fe', 'Mn', 'W'],  ['Cr', 'Ni', 'W'],  ['Fe', 'Ni', 'W'],
['Mn', 'Ni', 'W'],  ['Mo', 'Pt', 'W'],  ['Mo', 'Ru', 'W'],  ['Pt', 'Ru', 'W'],
['Re', 'W', 'Zr'].
\end{verbatim}
\end{small}

9 quaternary systems with all the corresponding $4^5=1024$ configurations, 9 among $\binom{14}{4}=1001$:
\begin{small}
\begin{verbatim}
['Cr', 'Fe', 'Mn', 'Ni'],
['Cr', 'Mo', 'Ni', 'Re'],
['Cr', 'Fe', 'Mo', 'V'],
['Co', 'Fe', 'Ni', 'V'],
['Cr', 'Fe', 'Mn', 'W'],
['Cr', 'Fe', 'Ni', 'W'],
['Cr', 'Mn', 'Ni', 'W'],
['Fe', 'Mn', 'Ni', 'W'],
['Mo', 'Pt', 'Ru', 'W'].
\end{verbatim}
\end{small}

1 quinary system with all the corresponding $5^5=3125$ configurations, 1 among $\binom{14}{5}=2002$:
\begin{small}
\begin{verbatim}
['Cr', 'Fe', 'Mn', 'Ni', 'W'].
\end{verbatim}
\end{small}

%-----------------------------------------------------------
\newpage
\subsection{Statistical analysis of the training database}
\label{a:stat}

\begin{figure}[htb]
\includegraphics[width=10cm]{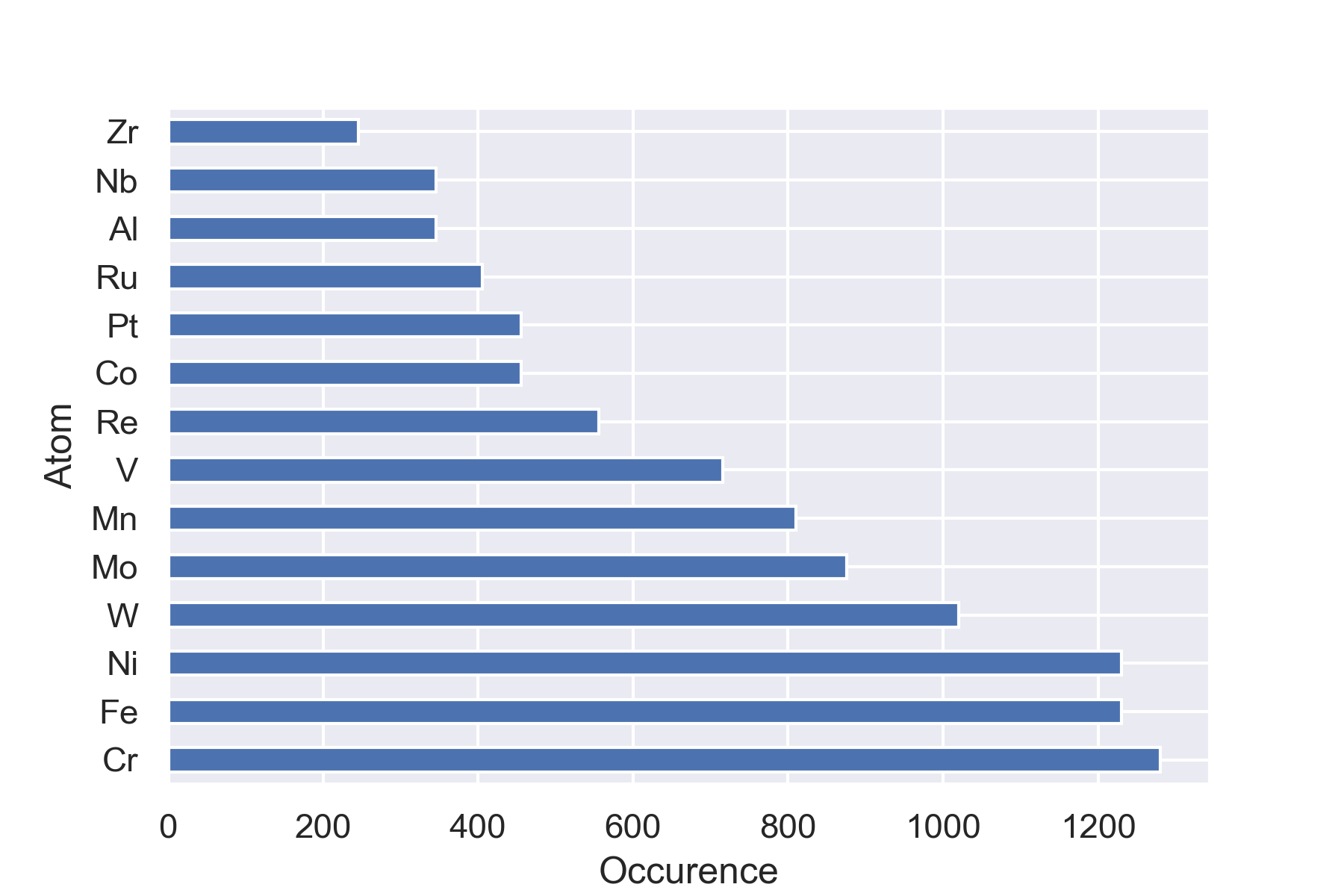}
\caption{\label{f:occu_ele} Occurrence of the $n=14$ elements in the training database (9974 entries).}
\end{figure}

\begin{figure}[htb]
\includegraphics[width=10cm]{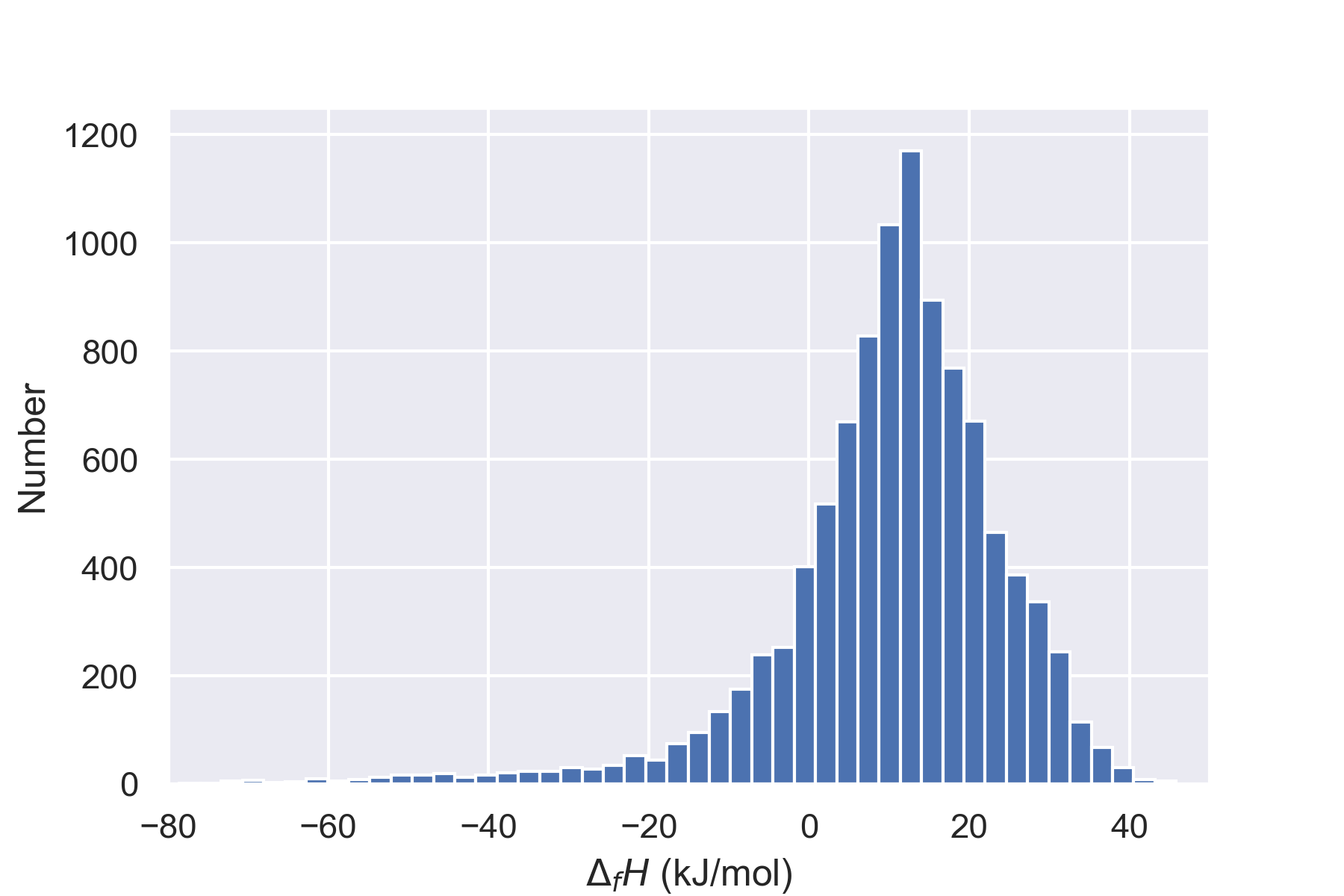}
\caption{\label{f:dis_hf} Distribution of the heat of formation of the 9974 entries of the training database.}
\end{figure}

\clearpage
\begin{figure}[htb]
\includegraphics[width=16cm]{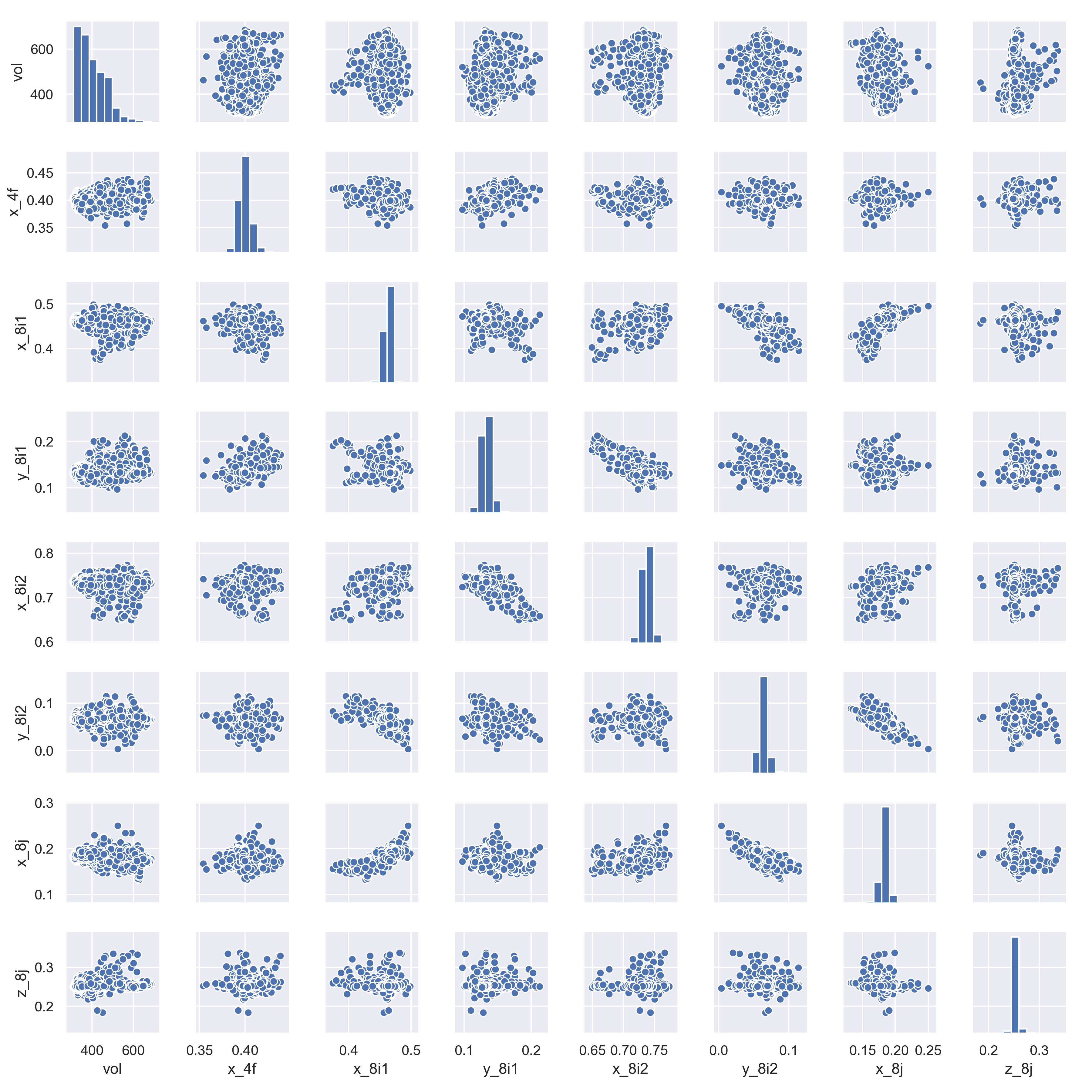}
\caption{\label{f:corre} Correlation matrix figures of cell parameters and internal positions of compounds in the training database.}
\end{figure}

%--------------------------------------------------------------------------
\newpage
\section{Cross validation results from the only training database}
\label{a:CV}

\begin{figure}[htb]
\includegraphics[width=5.5cm]{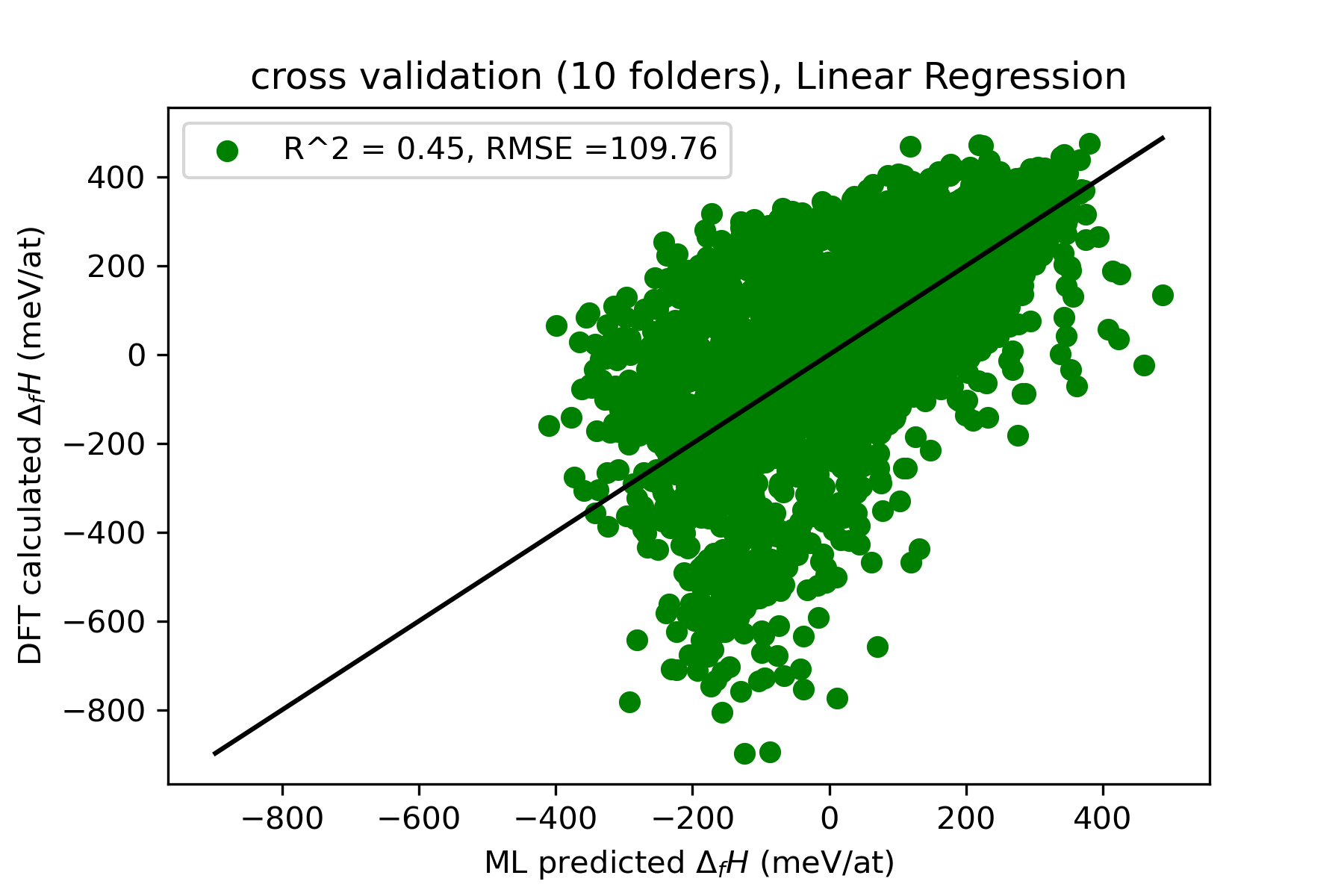}% 
\includegraphics[width=5.5cm]{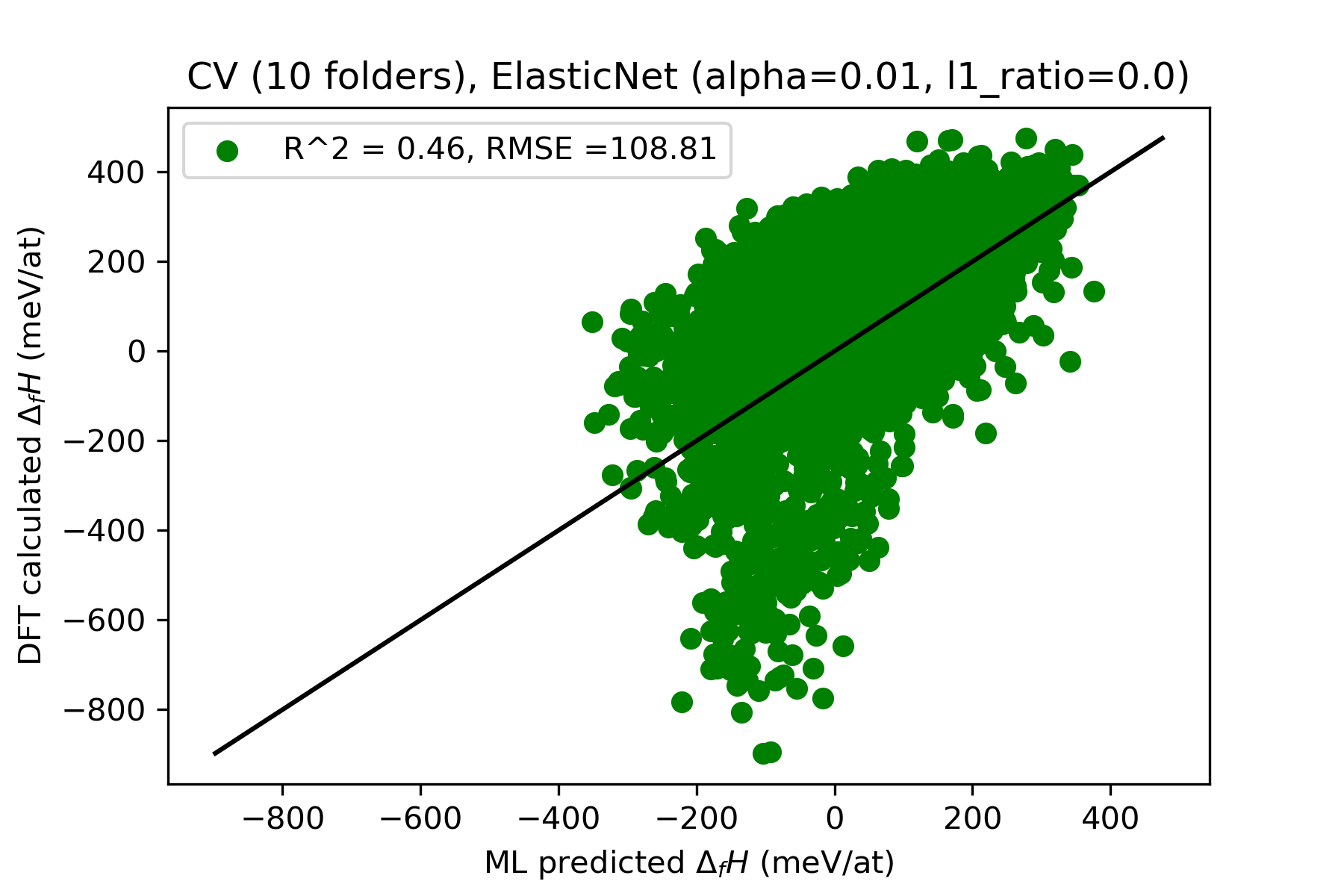}% 
\includegraphics[width=5.5cm]{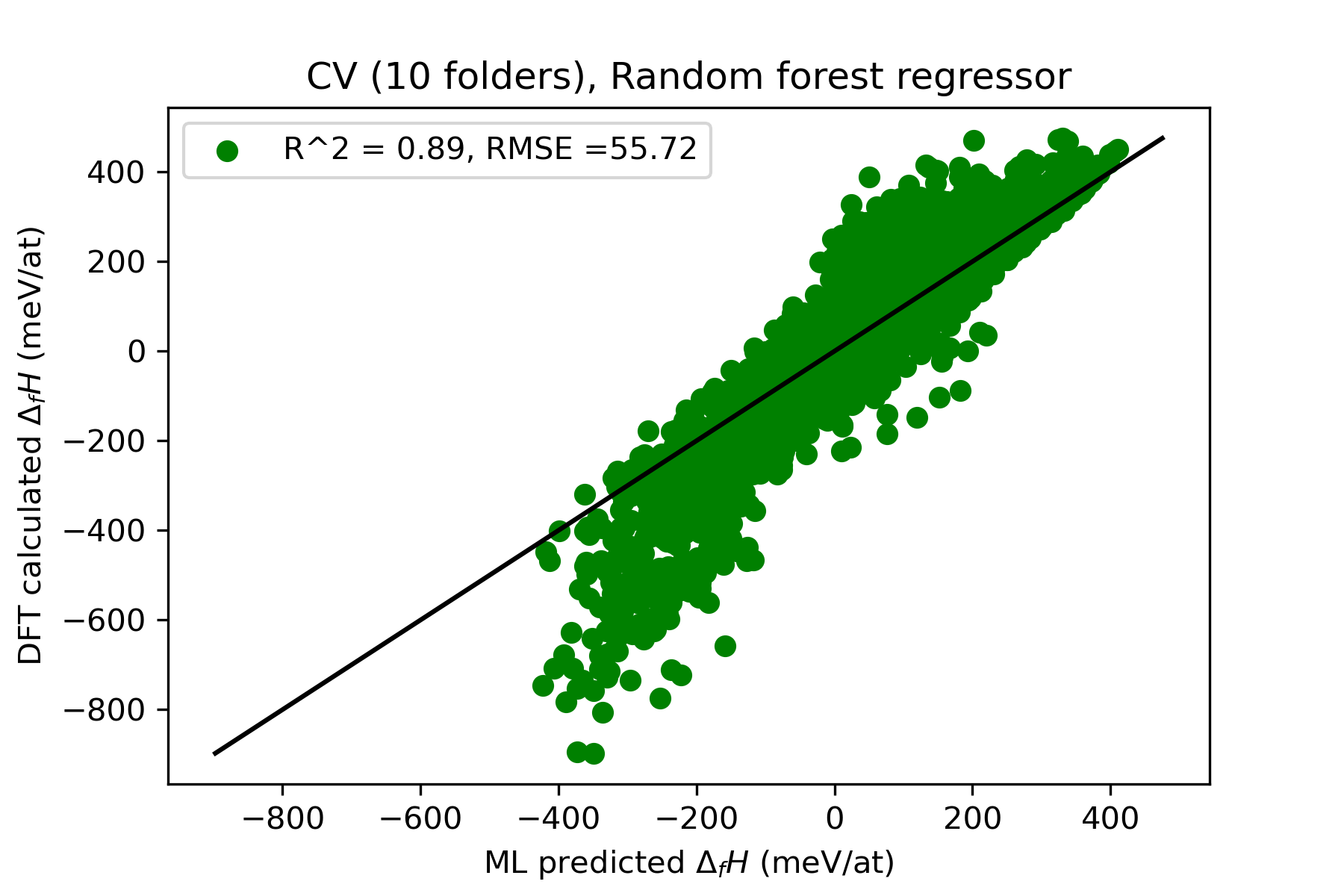}\\% 
\includegraphics[width=5.5cm]{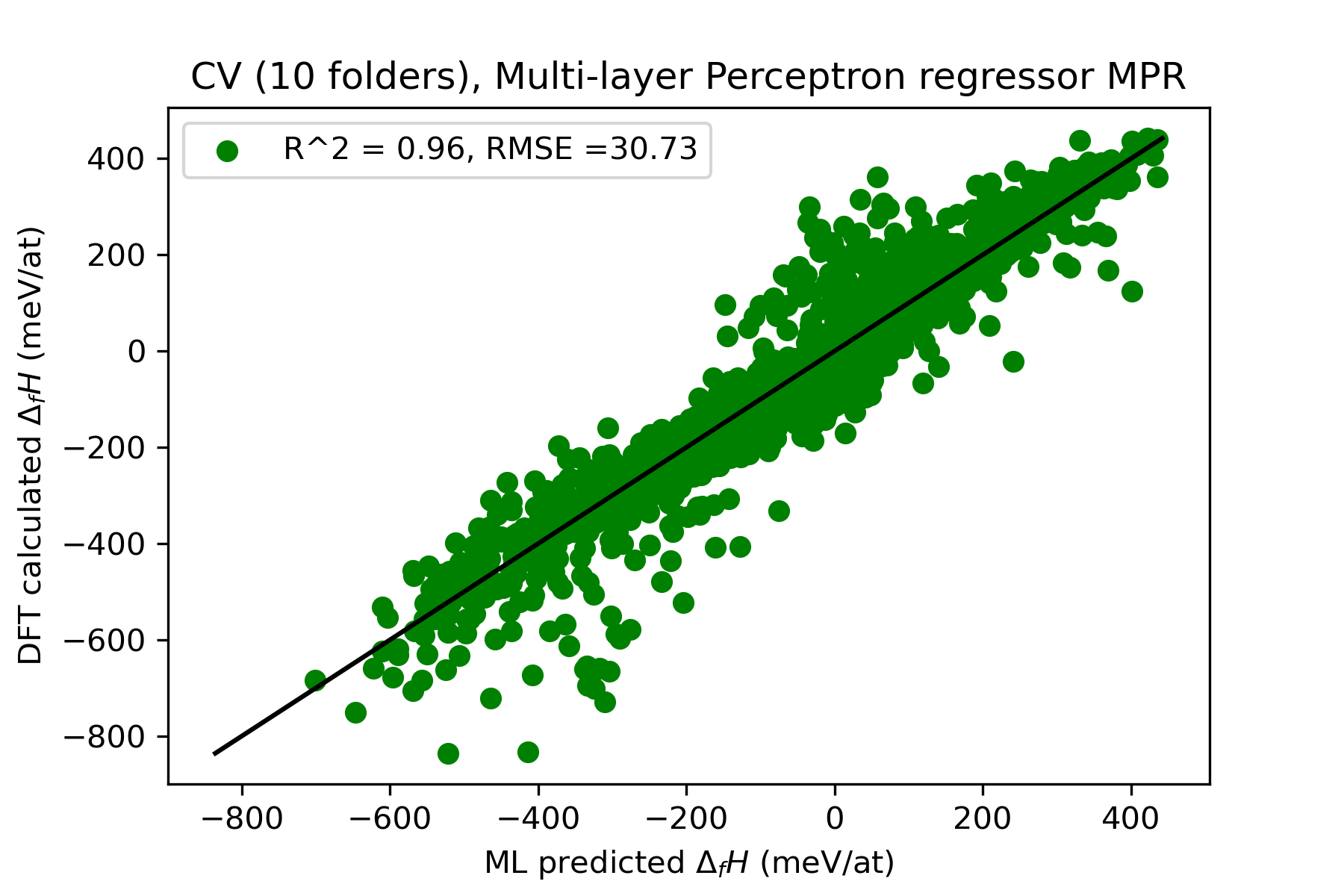}% 
\includegraphics[width=5.5cm]{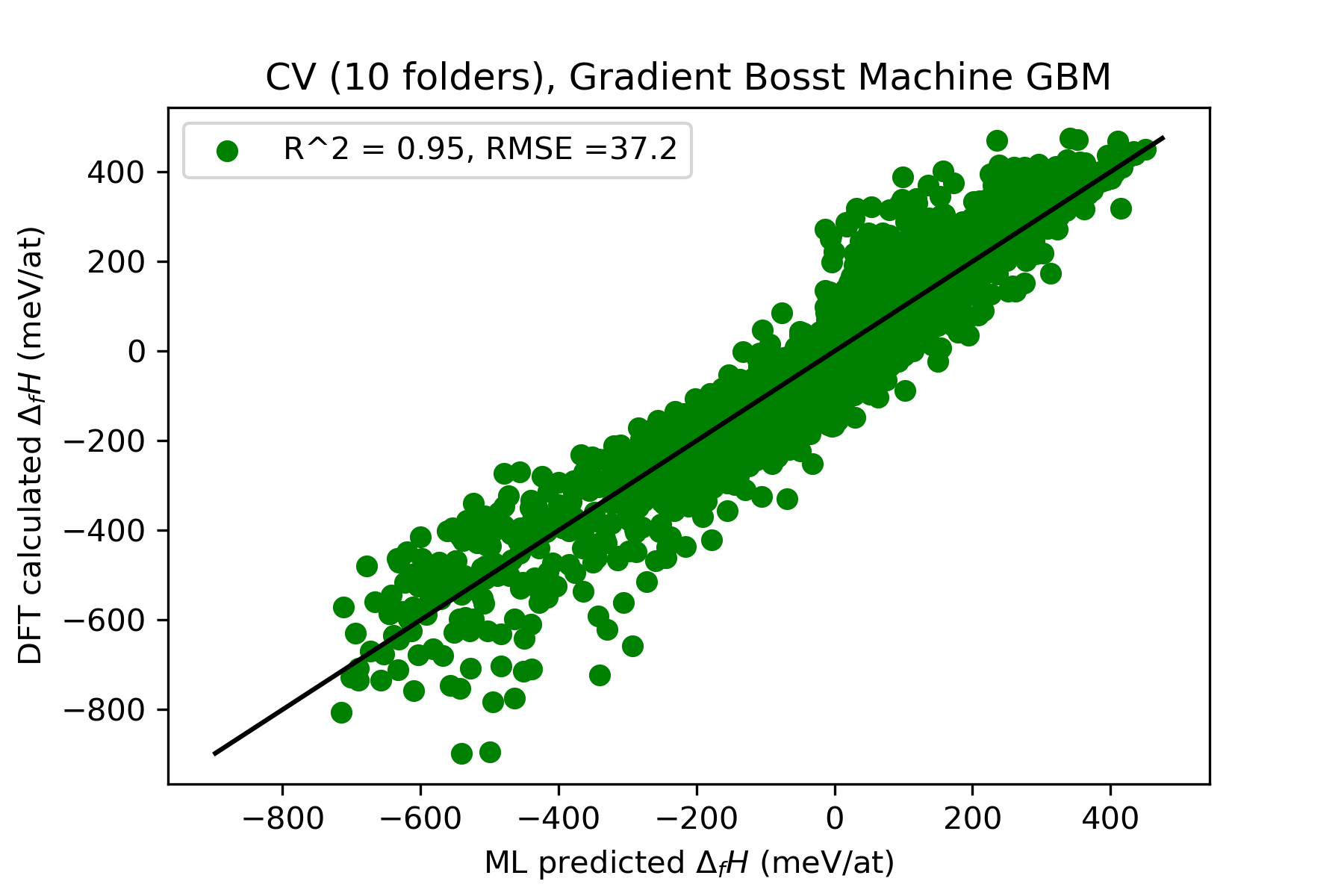}% 
\includegraphics[width=5.5cm]{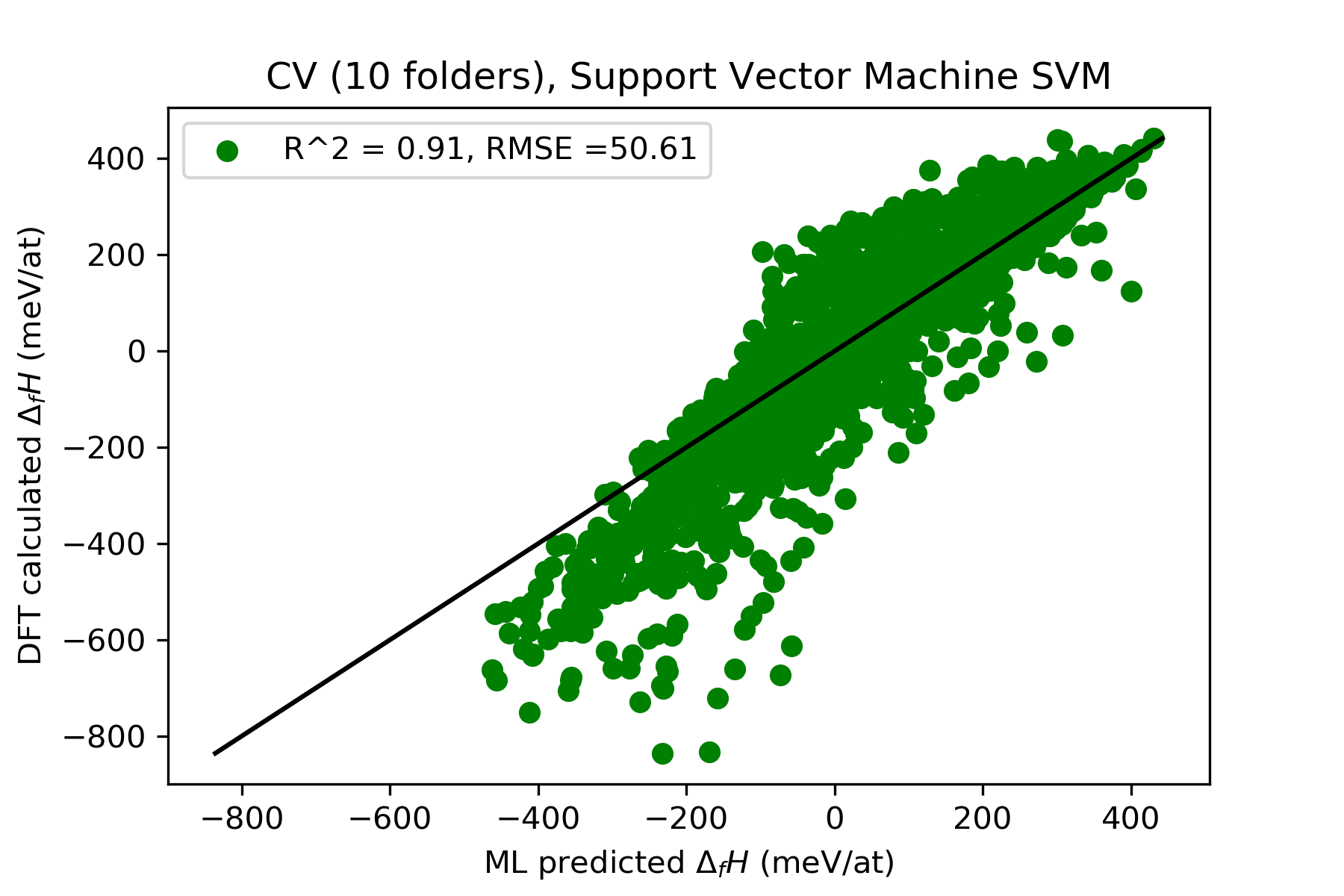}\\% 
\includegraphics[width=5.5cm]{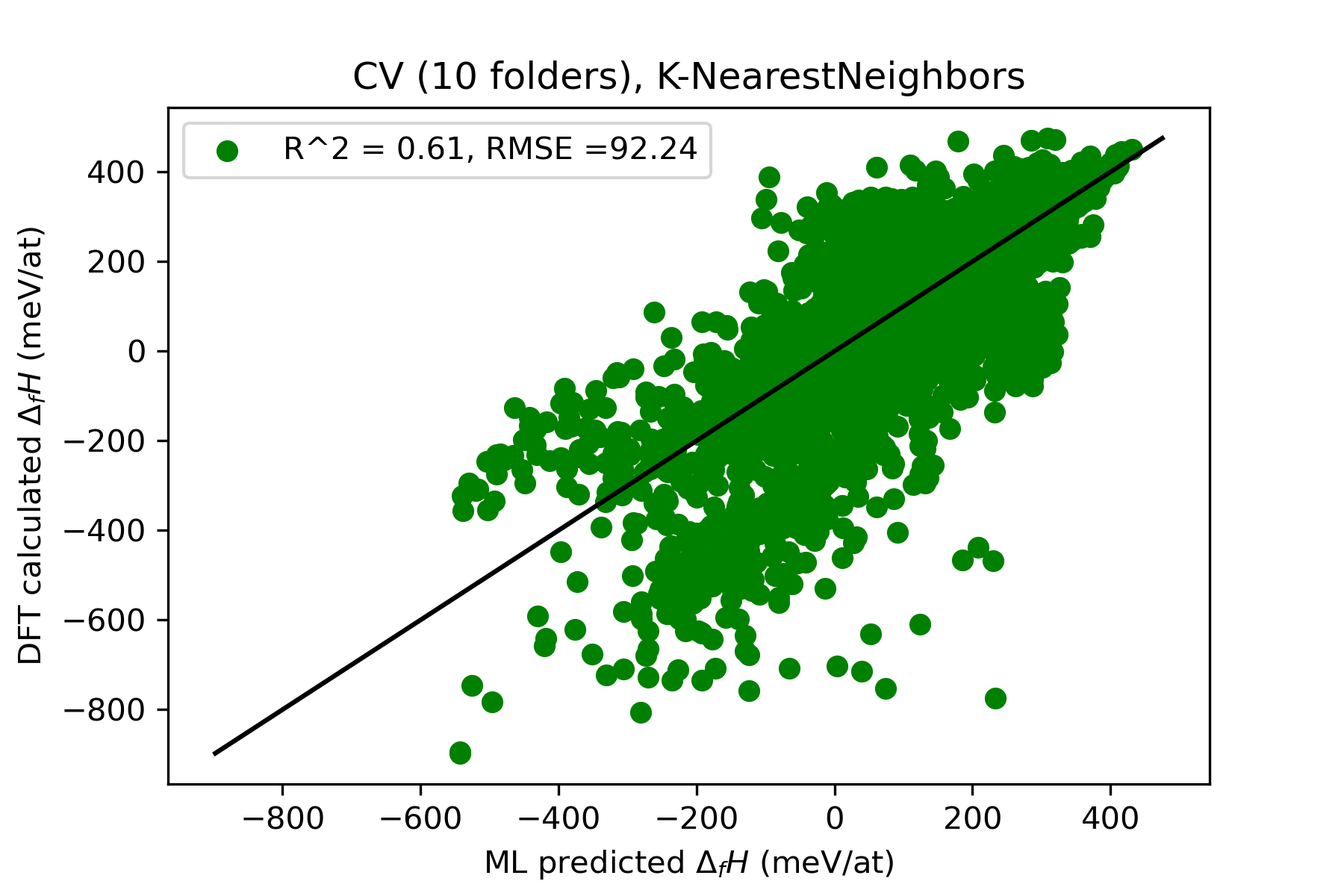}% 
\includegraphics[width=5.5cm]{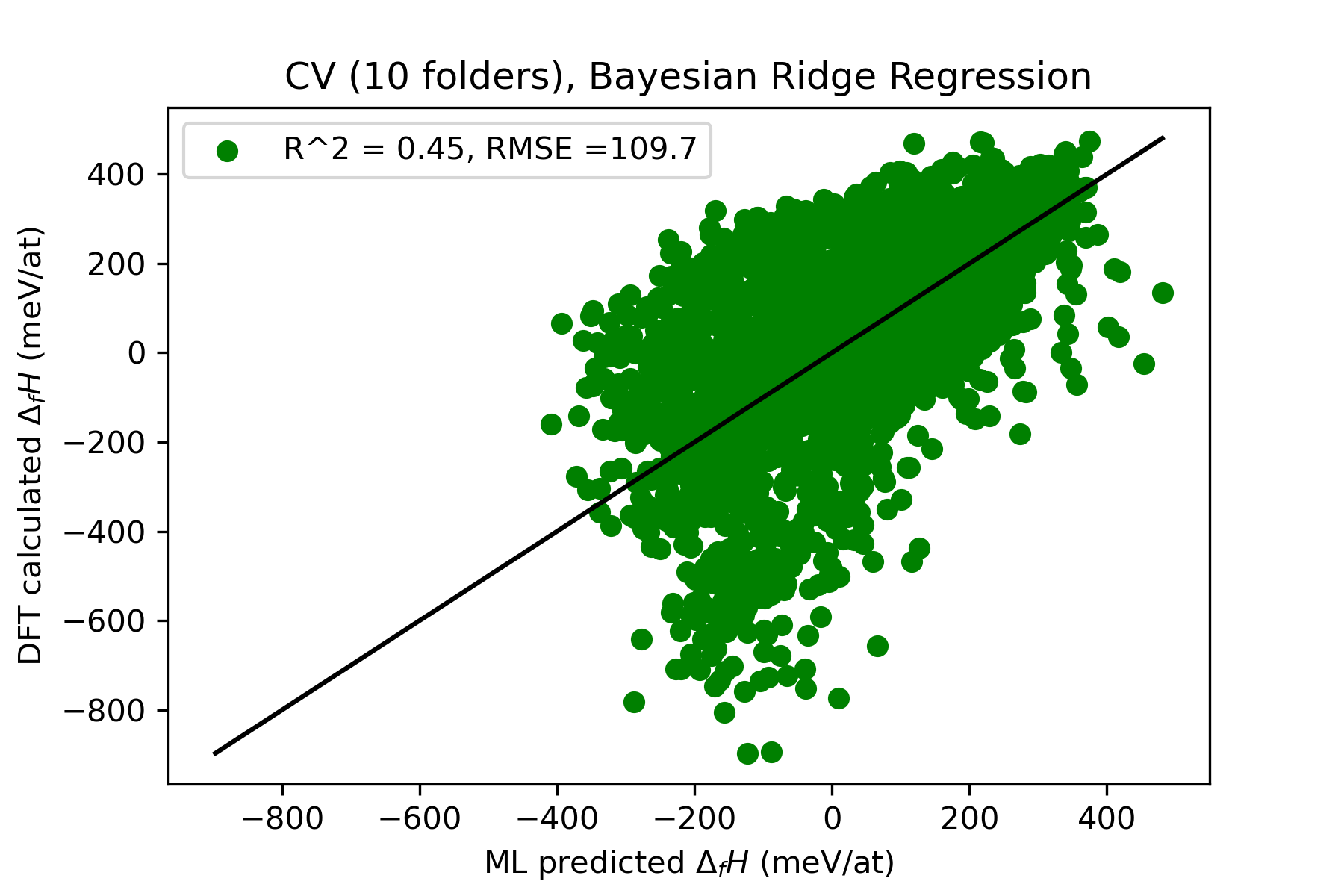}% 
\includegraphics[width=5.5cm]{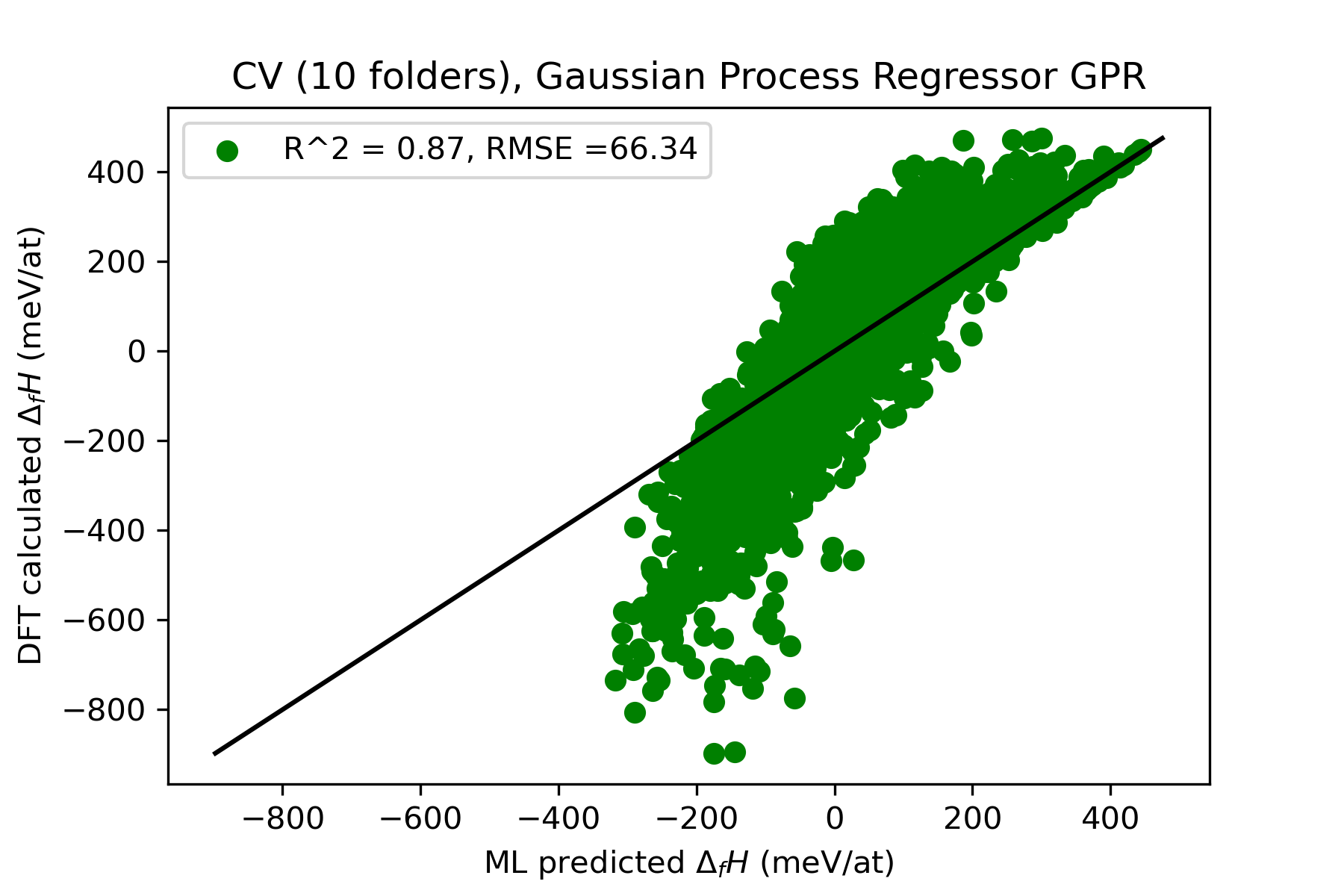}\\% 
\caption{\label{fig:CV} Averaged results of prediction from various learning methods of$\sim$10,000 compounds from cross validation under 10-folders.
The diagonal line indicates the perfect agreement between DFT calculated and predicted values.}
\end{figure}

%--------------------------------------------------------------------------
\newpage
\section{List of the 1001 testing set configurations}
\label{a:test}
{\setstretch{1.0}
\begin{footnotesize}
\begin{verbatim}
'Al:Co:Al:Co:Al', 'Al:Mo:Al:Mo:Al', 'Al:Mo:Mn:Re:Al', 'Al:Mo:Pt:Ru:Al', 'Al:Nb:Ni:Re:Re', 
'Al:Ni:Co:Mo:Ru', 'Al:Re:Fe:Al:Re', 'Co:Cr:V:Re:Co', 'Co:Fe:Cr:Co:Fe', 'Co:Fe:Ni:V:Mo', 
'Co:Fe:Ni:V:Ru', 'Co:Pt:Zr:Mo:Re', 'Co:Re:Re:V:Co', 'Cr:Cr:Cr:Nb:Zr', 'Cr:Fe:Mo:V:Mn', 
'Cr:Fe:Re:Re:Fe', 'Cr:Mn:Ni:W:Nb', 'Cr:Mn:Re:Re:Mn', 'Cr:Mo:Ni:Re:V', 'Cr:Mo:Re:Ni:V', 
'Cr:Pt:Al:Ni:Re', 'Cr:Re:Mn:Re:Pt', 'Cr:W:Cr:Mn:Pt', 'Cr:W:Mo:Cr:W', 'Cr:W:Mo:Ni:Cr', 
'Fe:Al:Re:Fe:Al', 'Fe:Cr:Co:Fe:Cr', 'Fe:Fe:Re:Mo:Re', 'Fe:Mn:Ni:W:Zr', 'Fe:Nb:Fe:Nb:Fe', 
'Fe:Re:Cr:Cr:Re', 'Fe:Re:Mn:Mn:Re', 'Fe:Re:V:Fe:Re', 'Fe:W:Fe:Zr:Mo', 'Mn:Mo:Re:Mn:Mo', 
'Mn:Ni:W:V:V', 'Mn:Re:Cr:Mn:Re', 'Mn:Re:V:Mn:Re', 'Mn:W:Cr:Ni:Fe', 'Mo:Co:Mo:Co:Mo', 
'Mo:Cr:W:Mo:Cr', 'Mo:Mn:Mo:Re:Re', 'Mo:Nb:Mn:V:Mn', 'Mo:Nb:Re:W:W', 'Mo:Pt:Ru:Al:Mo', 
'Mo:Pt:Ru:W:Re', 'Mo:V:Mn:Cr:Fe', 'Nb:Zr:Ru:Al:Mo', 'Ni:Al:Ni:Al:Ni', 'Ni:Mo:Nb:Re:W', 
'Ni:Nb:Cr:Mn:W', 'Ni:Zr:Fe:Mn:Ni', 'Pt:Re:Fe:Zr:Nb', 'Re:Al:Re:Al:Re', 'Re:Cr:Al:W:Al', 
'Re:Mo:Mo:Re:Fe', 'Re:Pt:Re:Pt:Re', 'Re:V:Co:Re:V', 'Re:V:Mn:Re:Mn', 'Ru:Co:Fe:Ni:V', 
'Ru:Co:Zr:Al:Nb', 'Ru:Pt:Re:Ru:Pt', 'V:Al:Ni:V:Mo', 'V:Fe:Fe:Re:V', 'V:Re:Cr:Co:V', 
'V:V:Cr:W:Re', 'V:Zr:W:Ni:Mo', 'W:Co:Co:W:Co', 'W:Cr:Ni:Mo:W', 'Zr:Ru:Mo:Co:Mo', 
'Al:Al:Co:Ni:Nb', 'Al:Al:Fe:Pt:Mo', 'Al:Al:Mn:Ni:Ru', 'Al:Al:Mn:Re:Co', 'Al:Al:Mo:V:Ni', 
'Al:Al:V:Ru:Mo', 'Al:Co:Cr:Mo:Zr', 'Al:Co:Cr:Ni:Ru', 'Al:Co:Re:Ru:Pt', 'Al:Co:V:Co:Re', 
'Al:Co:V:Nb:W', 'Al:Cr:Nb:Nb:Mn', 'Al:Cr:Ni:Mo:Ru', 'Al:Cr:Pt:Nb:Zr', 'Al:Cr:W:Co:Pt', 
'Al:Cr:W:Co:V', 'Al:Cr:Zr:Ru:Pt', 'Al:Fe:Al:Re:Ni', 'Al:Fe:Cr:Al:Ru', 'Al:Fe:Cr:Re:Pt', 
'Al:Fe:Fe:W:V', 'Al:Fe:Mn:Mo:Pt', 'Al:Fe:Mn:Mo:Re', 'Al:Fe:Mn:Re:Ru', 'Al:Fe:Re:Co:V', 
'Al:Fe:Ru:Al:Re', 'Al:Fe:Ru:Mo:Mo', 'Al:Fe:Zr:Al:Zr', 'Al:Mn:Cr:Fe:Fe', 'Al:Mn:Fe:Ni:Zr', 
'Al:Mn:Mo:Re:Ru', 'Al:Mn:Nb:Zr:Co', 'Al:Mn:Ru:Ni:V', 'Al:Mo:Co:Co:V', 'Al:Mo:Fe:V:Re', 
'Al:Mo:Ni:Mo:W', 'Al:Mo:Ni:Zr:W', 'Al:Mo:Zr:W:Mo', 'Al:Nb:Co:Mn:Zr', 'Al:Nb:Cr:Nb:Zr', 
'Al:Nb:Cr:W:Mo', 'Al:Nb:Fe:Al:Al', 'Al:Nb:Nb:Ni:Nb', 'Al:Nb:Ni:Zr:Re', 'Al:Nb:Pt:Zr:Fe', 
'Al:Nb:Ru:Re:Ni', 'Al:Nb:Ru:Zr:V', 'Al:Nb:Zr:Zr:V', 'Al:Ni:Mn:Cr:Al', 'Al:Ni:Mo:Re:Pt', 
'Al:Ni:Nb:Pt:Ru', 'Al:Ni:Ni:Co:Ni', 'Al:Ni:Re:Cr:Nb', 'Al:Ni:Ru:Nb:V', 'Al:Ni:W:Pt:Co', 
'Al:Pt:Al:V:Mn', 'Al:Pt:Mn:V:Mn', 'Al:Pt:V:Al:Ru', 'Al:Re:Cr:W:W', 'Al:Re:Re:Cr:Ru', 
'Al:Re:Ru:Ru:Ni', 'Al:Re:Ru:V:Cr', 'Al:Re:W:Pt:Cr', 'Al:Ru:Fe:Fe:Co', 'Al:Ru:Mn:Mo:Zr', 
'Al:Ru:Re:Re:Nb', 'Al:Ru:Ru:W:Cr', 'Al:Ru:V:Al:Fe', 'Al:Ru:V:Co:Mo', 'Al:Ru:V:Re:Nb', 
'Al:Ru:Zr:Mn:Nb', 'Al:V:Fe:W:Cr', 'Al:V:Mo:Mn:Mn', 'Al:V:Ni:V:Nb', 'Al:V:Pt:Re:Fe', 
'Al:W:Mo:Mo:Re', 'Al:W:Pt:Al:W', 'Al:W:Pt:Ni:Pt', 'Al:W:V:Fe:Mo', 'Al:Zr:Re:Pt:Cr', 
'Al:Zr:W:Co:Mo', 'Co:Al:Mo:Zr:Al', 'Co:Al:Nb:Mn:Pt', 'Co:Al:Pt:Re:Cr', 'Co:Al:Re:Re:Al', 
'Co:Al:Zr:V:Fe', 'Co:Co:Al:Zr:Cr', 'Co:Co:Cr:V:Mo', 'Co:Co:V:Co:Pt', 'Co:Cr:Al:Mo:Fe', 
'Co:Cr:Al:Nb:Cr', 'Co:Cr:Al:Zr:Ni', 'Co:Cr:Cr:Nb:Zr', 'Co:Cr:Nb:Re:V', 'Co:Cr:V:Ni:V', 
'Co:Cr:W:Mo:Re', 'Co:Cr:Zr:Cr:Re', 'Co:Cr:Zr:Fe:Al', 'Co:Fe:Fe:Mo:Ru', 'Co:Fe:Fe:W:Ni', 
'Co:Fe:Mo:Fe:Mo', 'Co:Fe:Re:Zr:Fe', 'Co:Fe:V:W:Ni', 'Co:Mn:Fe:Ru:Zr', 'Co:Mn:Ru:Mo:Ru', 
'Co:Mn:V:Mo:Mn', 'Co:Mo:Ni:Cr:Cr', 'Co:Mo:V:Re:Al', 'Co:Mo:W:Zr:Cr', 'Co:Mo:Zr:Nb:Ni', 
'Co:Nb:Al:Ru:Re', 'Co:Nb:Ni:Zr:Nb', 'Co:Nb:Ru:Pt:Mo', 'Co:Ni:Co:Mn:Al', 'Co:Ni:Mn:Al:V', 
'Co:Ni:Mo:Co:Re', 'Co:Ni:Ni:Mo:Pt', 'Co:Ni:Pt:Al:Ru', 'Co:Ni:Pt:Zr:Mn', 'Co:Ni:Re:Co:Re', 
'Co:Pt:Al:Pt:Pt', 'Co:Pt:Al:Ru:Fe', 'Co:Pt:Nb:Mn:Al', 'Co:Pt:Ni:Ni:Nb', 'Co:Pt:Pt:Nb:Cr', 
'Co:Pt:Ru:W:Zr', 'Co:Pt:V:Ni:Fe', 'Co:Re:Al:Mo:Cr', 'Co:Re:Co:Mo:Al', 'Co:Re:Nb:Zr:Al', 
'Co:Re:Zr:V:Co', 'Co:Ru:Cr:V:Mo', 'Co:Ru:Mn:Pt:V', 'Co:Ru:Ni:Ni:Zr', 'Co:Ru:Pt:Al:Al', 
'Co:V:Al:Fe:Mo', 'Co:V:Co:Mo:Mo', 'Co:V:Mo:Zr:Co', 'Co:V:Ni:Re:Zr', 'Co:V:Pt:W:Zr', 
'Co:V:W:Nb:W', 'Co:V:Zr:Fe:Ru', 'Co:W:Mo:Ru:Ni', 'Co:W:Re:Pt:Al', 'Co:W:Ru:Fe:W', 
'Co:W:W:Zr:Al', 'Co:Zr:Co:Ru:Mo', 'Co:Zr:Nb:Al:Al', 'Co:Zr:V:Mo:V', 'Cr:Al:Ni:Cr:Re', 
'Cr:Al:Pt:Al:V', 'Cr:Al:Pt:Ni:Ru', 'Cr:Al:Zr:Nb:W', 'Cr:Co:Co:Mn:Ni', 'Cr:Co:Cr:Ni:Pt', 
'Cr:Co:Fe:Co:Ru', 'Cr:Co:Re:Mn:Pt', 'Cr:Co:Re:Ru:Al', 'Cr:Co:W:Zr:V', 'Cr:Co:Zr:Pt:Fe', 
'Cr:Cr:Cr:Nb:Zr', 'Cr:Cr:Re:Mo:Nb', 'Cr:Cr:V:Ru:Mn', 'Cr:Fe:Al:Co:Fe', 'Cr:Fe:Al:Cr:Pt', 
'Cr:Fe:Cr:Mn:Al', 'Cr:Fe:Ni:Al:Co', 'Cr:Fe:Ni:Mn:Al', 'Cr:Fe:Ru:Re:W', 'Cr:Fe:Ru:Ru:Nb', 
'Cr:Mn:Co:W:Pt', 'Cr:Mn:Fe:Re:Nb', 'Cr:Mn:Mo:W:Pt', 'Cr:Mo:Cr:Nb:Co', 'Cr:Mo:Fe:V:Zr', 
'Cr:Mo:Mn:Co:Cr', 'Cr:Mo:Pt:V:W', 'Cr:Mo:V:W:Co', 'Cr:Mo:W:Zr:Al', 'Cr:Nb:Fe:Cr:W', 
'Cr:Nb:Mo:Mn:V', 'Cr:Nb:Pt:Nb:Mo', 'Cr:Nb:V:Al:Fe', 'Cr:Nb:W:V:Al', 'Cr:Ni:Co:V:Co', 
'Cr:Ni:Nb:Mo:Ni', 'Cr:Ni:Pt:Zr:Mo', 'Cr:Ni:V:Ni:Ru', 'Cr:Pt:Co:Cr:Ru', 'Cr:Pt:Fe:Mo:Zr', 
'Cr:Pt:Mo:Ru:Zr', 'Cr:Pt:Ni:Mn:V', 'Cr:Pt:Ni:V:V', 'Cr:Pt:Pt:Mo:Re', 'Cr:Re:Co:Cr:Re', 
'Cr:Re:Fe:Al:Ru', 'Cr:Re:Mn:Re:Pt', 'Cr:Re:Ni:Re:Mo', 'Cr:Ru:Co:Re:Ru', 'Cr:Ru:Cr:Mo:Ru', 
'Cr:Ru:Cr:Nb:Re', 'Cr:Ru:Ni:Co:Mn', 'Cr:Ru:Ni:Re:Fe', 'Cr:V:Mo:Al:Ni', 'Cr:V:Nb:V:Zr', 
'Cr:V:Ni:W:Fe', 'Cr:V:Ru:Nb:W', 'Cr:V:Zr:Ni:Pt', 'Cr:W:Nb:Mn:Ni', 'Cr:W:Pt:Pt:Fe', 
'Cr:W:V:Zr:Ni', 'Cr:Zr:Ni:Cr:Ru', 'Cr:Zr:Ru:V:Co', 'Fe:Al:Al:W:W', 'Fe:Al:Co:Ni:Al', 
'Fe:Al:Cr:Mo:Ru', 'Fe:Al:Cr:Ni:Ru', 'Fe:Al:Fe:V:Fe', 'Fe:Al:Mn:Al:Co', 'Fe:Co:Fe:Al:Al', 
'Fe:Co:Mo:Ru:Pt', 'Fe:Co:Ni:Al:Zr', 'Fe:Cr:Fe:V:Co', 'Fe:Cr:Mo:V:Re', 'Fe:Cr:Nb:Mo:Pt', 
'Fe:Cr:Ni:V:Al', 'Fe:Cr:Re:Al:Ru', 'Fe:Cr:Ru:Al:Cr', 'Fe:Cr:V:Nb:Mo', 'Fe:Cr:V:Re:Co', 
'Fe:Fe:Al:Co:Mn', 'Fe:Fe:V:Mn:Re', 'Fe:Fe:Zr:Mo:Al', 'Fe:Mn:Fe:Al:Fe', 'Fe:Mn:Mn:Re:Ni', 
'Fe:Mn:Nb:Al:Zr', 'Fe:Mn:Nb:Mo:Al', 'Fe:Mn:W:Ru:V', 'Fe:Mo:Mn:Ni:V', 'Fe:Mo:Mo:V:Ru', 
'Fe:Mo:Ru:Al:Nb', 'Fe:Mo:W:Zr:W', 'Fe:Nb:Cr:Pt:Al', 'Fe:Nb:Fe:Mn:V', 'Fe:Nb:Mo:Mn:Cr', 
'Fe:Nb:Mo:Nb:Ni', 'Fe:Nb:Mo:Re:Ru', 'Fe:Nb:Re:Al:Zr', 'Fe:Nb:Re:V:Mo', 'Fe:Nb:Ru:Nb:Zr', 
'Fe:Ni:Mo:Ni:Fe', 'Fe:Ni:Nb:Al:Pt', 'Fe:Ni:Nb:Ru:Co', 'Fe:Ni:W:Nb:V', 'Fe:Pt:Cr:Al:Mo', 
'Fe:Pt:Fe:Cr:Mn', 'Fe:Pt:V:Fe:Cr', 'Fe:Re:Mn:W:W', 'Fe:Re:Pt:Fe:Al', 'Fe:Ru:Cr:Nb:Re', 
'Fe:Ru:Re:Nb:Mn', 'Fe:Ru:Re:Pt:W', 'Fe:V:Nb:Co:Ru', 'Fe:V:Zr:Fe:W', 'Fe:W:Ni:Mn:Cr', 
'Fe:W:Ni:Zr:Pt', 'Fe:W:Re:Re:Mo', 'Fe:W:W:Re:V', 'Fe:Zr:Nb:Pt:V', 'Mn:Al:Cr:W:Cr', 
'Mn:Al:Ni:Re:Pt', 'Mn:Co:Co:V:Nb', 'Mn:Co:Fe:Al:Mo', 'Mn:Co:Ni:Ru:Co', 'Mn:Co:Pt:Co:W', 
'Mn:Co:V:Pt:Pt', 'Mn:Co:V:V:Fe', 'Mn:Co:W:Fe:Nb', 'Mn:Co:Zr:Mn:Al', 'Mn:Cr:Mn:Fe:V', 
'Mn:Cr:Nb:Ru:Fe', 'Mn:Cr:Ru:Fe:Mo', 'Mn:Cr:V:Re:V', 'Mn:Fe:Cr:Cr:Ru', 'Mn:Fe:Cr:Mn:Al', 
'Mn:Fe:Cr:Zr:Ru', 'Mn:Fe:Mn:Ru:Mn', 'Mn:Fe:Mo:Nb:Re', 'Mn:Fe:Pt:Al:Fe', 'Mn:Fe:Pt:Cr:Fe', 
'Mn:Fe:Re:Al:Al', 'Mn:Mn:Al:Pt:Cr', 'Mn:Mn:Mo:Ru:Fe', 'Mn:Mn:Ni:Ru:Mo', 'Mn:Mn:Ru:Al:Mn', 
'Mn:Mn:Ru:Pt:Ni', 'Mn:Mn:W:Nb:V', 'Mn:Mn:Zr:Re:Ni', 'Mn:Mo:Ru:Pt:Zr', 'Mn:Mo:Ru:Re:Fe', 
'Mn:Mo:Ru:W:Mn', 'Mn:Mo:W:Mn:Mo', 'Mn:Mo:W:Ni:Mo', 'Mn:Nb:Fe:Mn:Re', 'Mn:Nb:Mn:Pt:Fe', 
'Mn:Nb:Mo:Fe:Mo', 'Mn:Nb:V:Fe:Fe', 'Mn:Ni:Al:W:V', 'Mn:Re:Al:Ni:Mn', 'Mn:Re:Mn:Co:Re', 
'Mn:Re:Ni:Mo:Nb', 'Mn:Re:Re:Pt:W', 'Mn:Re:Re:W:Nb', 'Mn:Re:Ru:Co:Ni', 'Mn:Ru:Co:Mo:Ni', 
'Mn:Ru:Nb:Al:Fe', 'Mn:Ru:Re:Cr:Pt', 'Mn:V:Cr:Nb:Al', 'Mn:V:Nb:Pt:W', 'Mn:V:Ru:Ru:Cr', 
'Mn:W:Al:V:Co', 'Mn:W:Co:Pt:Zr', 'Mn:W:Mn:Re:V', 'Mn:W:Ni:Al:Co', 'Mn:W:Zr:Zr:V', 
'Mn:Zr:Al:Mn:V', 'Mn:Zr:Re:V:Nb', 'Mn:Zr:V:Al:Ru', 'Mo:Al:Mn:W:Mn', 'Mo:Al:Nb:Mo:W', 
'Mo:Al:Ni:Co:Ru', 'Mo:Al:Ni:Re:Zr', 'Mo:Al:Ru:Nb:Re', 'Mo:Al:Ru:Nb:Zr', 'Mo:Co:Cr:Mn:Fe', 
'Mo:Co:Nb:Co:Fe', 'Mo:Co:Ru:Al:Zr', 'Mo:Co:Ru:Fe:Ni', 'Mo:Cr:Al:Mn:Zr', 'Mo:Cr:Cr:Ni:W', 
'Mo:Cr:Fe:Re:Mn', 'Mo:Cr:Mn:Fe:Co', 'Mo:Cr:Mo:Mo:Nb', 'Mo:Cr:W:Co:Ni', 'Mo:Fe:Co:Nb:V', 
'Mo:Fe:Cr:Mo:Ni', 'Mo:Fe:Nb:Al:Al', 'Mo:Fe:Ni:Ni:Ni', 'Mo:Fe:Pt:Zr:Nb', 'Mo:Fe:Pt:Zr:Ni', 
'Mo:Mn:Co:Ni:Mo', 'Mo:Mn:Co:W:Nb', 'Mo:Mn:Mn:Fe:Cr', 'Mo:Mn:Ni:Mo:Re', 'Mo:Mn:Re:Pt:Ni', 
'Mo:Mn:Zr:V:W', 'Mo:Mo:Al:Co:Fe', 'Mo:Mo:Cr:Mo:Fe', 'Mo:Mo:Cr:V:Mn', 'Mo:Mo:Ni:Nb:Al', 
'Mo:Mo:Ni:W:Mo', 'Mo:Mo:Re:Al:Zr', 'Mo:Mo:W:Al:Pt', 'Mo:Nb:Nb:Fe:Co', 'Mo:Ni:Cr:Fe:Mn', 
'Mo:Ni:Fe:Ni:Mn', 'Mo:Ni:Nb:Nb:Cr', 'Mo:Ni:Nb:Re:Pt', 'Mo:Ni:W:Ni:Nb', 'Mo:Pt:Co:W:Nb', 
'Mo:Pt:Pt:Re:Zr', 'Mo:Re:Cr:Zr:Re', 'Mo:Re:Mn:Co:Al', 'Mo:Re:Pt:W:V', 'Mo:Re:Ru:Al:Re', 
'Mo:Re:Ru:V:Fe', 'Mo:Re:V:Fe:Fe', 'Mo:Re:V:Re:Pt', 'Mo:Re:V:W:Nb', 'Mo:Ru:Al:Fe:Al', 
'Mo:Ru:Al:Re:V', 'Mo:Ru:Co:Re:Ni', 'Mo:Ru:Cr:Pt:Ni', 'Mo:Ru:Ni:Ni:Fe', 'Mo:Ru:Ru:W:Zr', 
'Mo:V:Pt:Ni:Nb', 'Mo:V:V:Ru:Ru', 'Mo:V:W:Re:Al', 'Mo:V:W:Zr:Co', 'Mo:W:Cr:Ru:Nb', 
'Mo:W:Re:Zr:Cr', 'Mo:W:V:Ni:Fe', 'Mo:Zr:Al:Nb:Co', 'Mo:Zr:Fe:Nb:Pt', 'Mo:Zr:Fe:Ni:Pt', 
'Mo:Zr:Mo:V:W', 'Mo:Zr:Ni:Pt:Fe', 'Mo:Zr:Pt:Co:Ru', 'Mo:Zr:Re:Fe:Ni', 'Mo:Zr:Re:Mn:Nb', 
'Mo:Zr:V:Pt:Cr', 'Mo:Zr:V:V:Fe', 'Nb:Al:Co:Ni:Cr', 'Nb:Al:Re:Ni:Cr', 'Nb:Al:Ru:Mo:Re', 
'Nb:Co:Ni:Co:Pt', 'Nb:Co:Ni:Fe:Fe', 'Nb:Co:Zr:Co:Cr', 'Nb:Cr:Cr:Zr:Pt', 'Nb:Cr:Mn:Pt:Pt', 
'Nb:Cr:Nb:Ni:Co', 'Nb:Cr:Re:Pt:Ni', 'Nb:Fe:Co:Al:Re', 'Nb:Fe:Mn:Mo:Mo', 'Nb:Fe:Nb:Pt:W', 
'Nb:Fe:V:Pt:V', 'Nb:Fe:Zr:Ni:Co', 'Nb:Mn:Cr:Pt:Al', 'Nb:Mn:Fe:V:Pt', 'Nb:Mn:Re:Nb:Nb', 
'Nb:Mn:Zr:Nb:V', 'Nb:Mo:Fe:Zr:Mo', 'Nb:Mo:Ru:W:Fe', 'Nb:Mo:Zr:Co:Co', 'Nb:Nb:Co:Ru:Al', 
'Nb:Nb:Nb:Mn:Re', 'Nb:Nb:Re:Ni:V', 'Nb:Nb:Zr:Re:Nb', 'Nb:Ni:Cr:Ru:Fe', 'Nb:Ni:Fe:Mo:Pt', 
'Nb:Ni:Mn:Zr:Mo', 'Nb:Ni:V:Ni:Co', 'Nb:Pt:Ni:Co:Re', 'Nb:Pt:Ru:Al:Fe', 'Nb:Re:Al:Fe:Fe', 
'Nb:Re:Co:Nb:Al', 'Nb:Re:Co:Pt:Ru', 'Nb:Re:Co:W:Mn', 'Nb:Re:Nb:W:Ru', 'Nb:Re:Pt:Nb:Nb', 
'Nb:Re:W:Nb:Mn', 'Nb:Re:W:Nb:Pt', 'Nb:Ru:Co:Mn:Ru', 'Nb:Ru:Cr:Nb:Re', 'Nb:Ru:Nb:Ni:V', 
'Nb:Ru:W:Re:Pt', 'Nb:V:Mn:Pt:Fe', 'Nb:V:Nb:Zr:Al', 'Nb:V:Re:Mo:V', 'Nb:V:Ru:Nb:Nb', 
'Nb:W:Ni:Zr:Nb', 'Nb:W:Re:Ru:Al', 'Nb:W:V:V:Mn', 'Nb:Zr:Co:Ru:W', 'Nb:Zr:Mo:Ni:Ni', 
'Nb:Zr:Mo:Pt:Fe', 'Nb:Zr:Mo:Re:Fe', 'Nb:Zr:Ni:Ni:Pt', 'Nb:Zr:Re:Mn:Mo', 'Nb:Zr:Ru:W:Re', 
'Nb:Zr:V:Nb:Co', 'Nb:Zr:W:Ni:Zr', 'Nb:Zr:Zr:Ru:W', 'Ni:Al:Cr:Ni:Zr', 'Ni:Al:Cr:Re:Pt', 
'Ni:Al:Cr:W:Nb', 'Ni:Co:Co:Co:Ru', 'Ni:Co:Co:Cr:Pt', 'Ni:Co:Cr:Cr:Cr', 'Ni:Co:Fe:Al:V', 
'Ni:Co:Nb:W:Mn', 'Ni:Co:Ni:Al:V', 'Ni:Co:Pt:Zr:Nb', 'Ni:Co:Ru:Zr:Nb', 'Ni:Co:V:Ni:Ru', 
'Ni:Co:Zr:Ni:Ni', 'Ni:Co:Zr:Re:Ru', 'Ni:Cr:Co:W:W', 'Ni:Cr:Nb:Nb:Fe', 'Ni:Cr:Ni:Nb:Pt', 
'Ni:Cr:Re:Re:Ni', 'Ni:Cr:Ru:Cr:Re', 'Ni:Cr:Zr:Nb:Ru', 'Ni:Fe:Co:W:Pt', 'Ni:Fe:Mo:Cr:Zr', 
'Ni:Fe:Ni:Pt:Fe', 'Ni:Fe:Pt:Cr:Pt', 'Ni:Fe:W:Cr:Ru', 'Ni:Fe:W:Mn:Mo', 'Ni:Fe:Zr:Mn:Nb', 
'Ni:Mn:Re:Cr:W', 'Ni:Mn:V:Zr:Nb', 'Ni:Mo:Mn:W:W', 'Ni:Mo:Nb:W:V', 'Ni:Mo:Ni:Mo:Mn', 
'Ni:Mo:Ni:Ni:Ni', 'Ni:Mo:Zr:Co:W', 'Ni:Nb:Mo:Pt:Mn', 'Ni:Nb:Re:Co:V', 'Ni:Nb:Re:Mn:Mn', 
'Ni:Nb:V:Pt:W', 'Ni:Ni:Ni:V:Re', 'Ni:Ni:V:Ru:Nb', 'Ni:Pt:Al:Mn:Zr', 'Ni:Pt:Co:Al:V', 
'Ni:Pt:Fe:Ni:Mn', 'Ni:Pt:Mn:Cr:Nb', 'Ni:Pt:Ru:Al:Ru', 'Ni:Pt:V:Mn:Co', 'Ni:Re:Fe:Pt:Ni', 
'Ni:Re:Fe:Ru:Fe', 'Ni:Ru:Al:Co:Mo', 'Ni:Ru:Co:Ru:Ru', 'Ni:Ru:Mn:Re:Zr', 'Ni:Ru:Mo:Pt:Ni', 
'Ni:Ru:Ru:Zr:Fe', 'Ni:Ru:V:Ni:Mo', 'Ni:Ru:Zr:Fe:Cr', 'Ni:V:Mn:Re:Mo', 'Ni:V:Mn:V:Pt', 
'Ni:V:Pt:Cr:Pt', 'Ni:V:Pt:Ru:W', 'Ni:V:W:Re:Re', 'Ni:W:Al:Cr:Mo', 'Ni:W:Cr:Ru:Zr', 
'Ni:W:Fe:V:V', 'Ni:W:Ni:Zr:Zr', 'Ni:W:Pt:V:Fe', 'Ni:W:Re:Mn:Mn', 'Ni:W:W:Mn:Mo', 
'Ni:W:W:Mo:V', 'Ni:Zr:Cr:Cr:Zr', 'Ni:Zr:Mo:Cr:Zr', 'Ni:Zr:W:Ni:Co', 'Pt:Al:Al:Re:Pt', 
'Pt:Al:Fe:Mn:Cr', 'Pt:Al:Ni:Mo:V', 'Pt:Al:Pt:Ru:Zr', 'Pt:Al:Ru:Fe:Ru', 'Pt:Al:V:Nb:Nb', 
'Pt:Co:Mn:Nb:Mn', 'Pt:Co:Nb:Co:Re', 'Pt:Co:V:W:Al', 'Pt:Co:W:Nb:Nb', 'Pt:Co:Zr:Mo:V', 
'Pt:Cr:Al:Nb:Re', 'Pt:Cr:Fe:Cr:Nb', 'Pt:Cr:Mn:Mo:Co', 'Pt:Cr:Nb:Mo:Pt', 'Pt:Cr:Pt:Ru:Cr', 
'Pt:Cr:Re:Mn:Fe', 'Pt:Cr:W:Re:Al', 'Pt:Cr:Zr:W:Al', 'Pt:Fe:Fe:Cr:Cr', 'Pt:Fe:Fe:Fe:Pt', 
'Pt:Fe:Mo:V:Fe', 'Pt:Mn:Cr:Ni:Ni', 'Pt:Mn:Mo:W:Mo', 'Pt:Mn:Nb:Co:Pt', 'Pt:Mn:Nb:W:V', 
'Pt:Mn:Re:Pt:Cr', 'Pt:Mn:W:Al:Nb', 'Pt:Mo:Nb:Co:Al', 'Pt:Mo:Nb:Re:Pt', 'Pt:Mo:Ru:W:Ru', 
'Pt:Mo:Zr:Co:W', 'Pt:Nb:Al:Nb:Ni', 'Pt:Nb:Pt:Cr:Mo', 'Pt:Nb:Pt:Ni:Al', 'Pt:Nb:Ru:Cr:Ni', 
'Pt:Ni:Ni:Mn:W', 'Pt:Ni:Ni:Ni:Ru', 'Pt:Pt:Re:Mn:Nb', 'Pt:Re:Mn:Fe:Al', 'Pt:Re:Re:Zr:V', 
'Pt:Re:Ru:Co:Re', 'Pt:Re:V:Mn:Re', 'Pt:Re:W:V:W', 'Pt:Re:Zr:Mo:Co', 'Pt:Ru:Al:Ni:Mn', 
'Pt:Ru:Cr:Al:Cr', 'Pt:Ru:Cr:Mn:Co', 'Pt:Ru:Mn:Fe:Re', 'Pt:Ru:Mn:V:Nb', 'Pt:V:Al:Pt:Zr', 
'Pt:V:Fe:Pt:Pt', 'Pt:V:Mn:Re:Ni', 'Pt:V:Re:W:Pt', 'Pt:V:V:Mn:Mo', 'Pt:W:Al:Cr:V', 
'Pt:W:Mo:Ni:Re', 'Pt:W:Re:Fe:Mn', 'Pt:Zr:Al:Mo:Nb', 'Pt:Zr:Co:Mn:Fe', 'Pt:Zr:Fe:Fe:V', 
'Pt:Zr:Nb:W:Pt', 'Pt:Zr:Re:Nb:Pt', 'Pt:Zr:Ru:W:Ru', 'Re:Al:Co:Fe:Cr', 'Re:Al:Cr:Zr:Ru', 
'Re:Al:Fe:Mo:Co', 'Re:Al:Mn:Ru:Cr', 'Re:Al:Nb:Pt:V', 'Re:Al:V:Mn:W', 'Re:Co:Cr:V:Co', 
'Re:Co:Mo:Re:Co', 'Re:Co:Mo:Ru:Nb', 'Re:Co:Pt:Co:Nb', 'Re:Co:Re:Re:Pt', 'Re:Co:V:Ru:Ni', 
'Re:Cr:Al:Co:Ni', 'Re:Cr:Al:W:Al', 'Re:Cr:Co:Cr:Cr', 'Re:Cr:Ni:Mn:Re', 'Re:Fe:Mn:Mn:Al', 
'Re:Fe:Ni:Ru:Nb', 'Re:Fe:Pt:V:Re', 'Re:Fe:Re:Co:Zr', 'Re:Fe:Ru:Mn:Ni', 'Re:Fe:Zr:Fe:Mn', 
'Re:Mn:Mn:Mo:Nb', 'Re:Mn:Re:Cr:Pt', 'Re:Mn:Re:Pt:Pt', 'Re:Mo:Al:Re:Nb', 'Re:Mo:Cr:Mn:Cr', 
'Re:Mo:Nb:Co:Ru', 'Re:Mo:Pt:Ru:Nb', 'Re:Mo:Pt:Ru:Ni', 'Re:Mo:Ru:W:Pt', 'Re:Nb:Mn:Ru:Fe', 
'Re:Nb:Nb:Re:Al', 'Re:Nb:Pt:W:Pt', 'Re:Ni:Mo:W:Re', 'Re:Ni:Ni:Mn:Nb', 'Re:Ni:Pt:Nb:Re', 
'Re:Ni:Pt:V:Fe', 'Re:Ni:Pt:V:Ru', 'Re:Ni:Re:Fe:Zr', 'Re:Ni:W:Cr:Cr', 'Re:Pt:Cr:Nb:Fe', 
'Re:Pt:Mn:Ni:Co', 'Re:Pt:Nb:Nb:Pt', 'Re:Pt:Ni:Re:Co', 'Re:Pt:V:Re:Zr', 'Re:Pt:W:Pt:Nb', 
'Re:Re:Fe:Re:Cr', 'Re:Re:Pt:Mo:Re', 'Re:Ru:Cr:Fe:Al', 'Re:Ru:Mo:Cr:Fe', 'Re:Ru:Pt:Mo:Co', 
'Re:Ru:Pt:Nb:Ru', 'Re:Ru:Pt:Pt:W', 'Re:V:Mn:Pt:Re', 'Re:V:Ni:Co:Zr', 'Re:V:Re:Ru:Co', 
'Re:V:Re:Zr:Mo', 'Re:V:V:Co:Fe', 'Re:V:Zr:Co:Mn', 'Re:W:Cr:Mn:Re', 'Re:W:Nb:Al:Fe', 
'Re:W:Ni:Re:Nb', 'Re:W:Ru:Fe:Fe', 'Re:W:V:Fe:Zr', 'Re:W:W:Co:Ni', 'Re:W:Zr:Nb:Ni', 
'Re:Zr:Co:Cr:Ni', 'Re:Zr:Co:Zr:Fe', 'Re:Zr:Cr:Co:Co', 'Re:Zr:Mo:Pt:Mn', 'Re:Zr:Nb:Ru:Cr', 
'Re:Zr:Ni:Zr:Pt', 'Re:Zr:Pt:Nb:Re', 'Re:Zr:Pt:Nb:W', 'Re:Zr:Ru:W:W', 'Re:Zr:W:Cr:Zr', 
'Ru:Al:Cr:W:Nb', 'Ru:Co:Fe:Al:Mn', 'Ru:Co:W:Pt:Pt', 'Ru:Cr:Co:W:Ni', 'Ru:Cr:Cr:Cr:Mo', 
'Ru:Cr:Nb:Cr:Mo', 'Ru:Cr:Ni:Nb:Ru', 'Ru:Cr:Ni:V:Pt', 'Ru:Cr:Pt:Mn:Nb', 'Ru:Cr:Ru:Ru:Ru', 
'Ru:Cr:V:Mn:Mn', 'Ru:Cr:V:V:Nb', 'Ru:Cr:Zr:Al:Pt', 'Ru:Cr:Zr:Co:Pt', 'Ru:Fe:Co:W:Mo', 
'Ru:Fe:Mn:Nb:Pt', 'Ru:Fe:Mo:Zr:Co', 'Ru:Fe:Re:Cr:Zr', 'Ru:Fe:Re:Fe:Pt', 'Ru:Mn:Cr:Cr:Al', 
'Ru:Mn:Cr:Mo:Al', 'Ru:Mn:Cr:Ru:W', 'Ru:Mn:Fe:W:Ru', 'Ru:Mo:Co:Zr:Nb', 'Ru:Mo:Mo:Ru:W', 
'Ru:Mo:Nb:Mo:Zr', 'Ru:Mo:Pt:Re:Al', 'Ru:Mo:Re:Fe:Mn', 'Ru:Mo:Re:Mn:Al', 'Ru:Mo:Ru:W:Mo', 
'Ru:Mo:Zr:Co:Nb', 'Ru:Nb:Al:Re:Co', 'Ru:Nb:Fe:Cr:Pt', 'Ru:Nb:Fe:Mo:Fe', 'Ru:Nb:Mo:Fe:Mo', 
'Ru:Nb:Re:Mo:Pt', 'Ru:Nb:V:W:W', 'Ru:Ni:Cr:Mn:Nb', 'Ru:Ni:Mo:Mo:Fe', 'Ru:Ni:Ni:Pt:Mo', 
'Ru:Ni:Ru:Cr:V', 'Ru:Ni:V:Zr:Al', 'Ru:Pt:Al:V:Co', 'Ru:Pt:Cr:Co:Pt', 'Ru:Pt:Ni:Mo:Ru', 
'Ru:Pt:Re:V:Zr', 'Ru:Pt:Ru:W:W', 'Ru:Pt:W:Re:Ni', 'Ru:Pt:Zr:Co:Zr', 'Ru:Re:Cr:Mo:Nb', 
'Ru:Re:Fe:Fe:Zr', 'Ru:Re:Fe:Ru:Nb', 'Ru:Re:Fe:Zr:W', 'Ru:Re:Nb:Co:Cr', 'Ru:Re:Re:Re:Nb', 
'Ru:Ru:Mo:Pt:Cr', 'Ru:Ru:Nb:Mn:V', 'Ru:Ru:Re:W:Co', 'Ru:Ru:Ru:W:Re', 'Ru:Ru:W:Mo:Pt', 
'Ru:Ru:W:Ni:Ni', 'Ru:V:Cr:Mo:Mn', 'Ru:V:W:Re:Zr', 'Ru:V:Zr:Mo:Cr', 'Ru:W:Cr:Cr:Mo', 
'Ru:W:Ru:Fe:Fe', 'Ru:Zr:Cr:Nb:Pt', 'Ru:Zr:Fe:Fe:W', 'Ru:Zr:Fe:Ni:V', 'Ru:Zr:Nb:Fe:Pt', 
'Ru:Zr:Ni:Fe:Ru', 'Ru:Zr:Ni:Re:Nb', 'V:Al:Mo:V:Al', 'V:Al:Nb:Cr:W', 'V:Al:Ni:V:Mo', 
'V:Co:Co:Re:Zr', 'V:Cr:Cr:Co:V', 'V:Cr:Cr:Zr:Cr', 'V:Cr:Mn:Mo:Pt', 'V:Cr:Pt:V:Fe', 
'V:Fe:Co:Re:Al', 'V:Fe:Cr:Pt:W', 'V:Fe:Fe:Al:W', 'V:Fe:Nb:Ru:Mo', 'V:Fe:Re:Ni:Zr', 
'V:Mn:Co:Fe:Co', 'V:Mn:Co:Fe:Nb', 'V:Mn:Mo:Ni:Ni', 'V:Mn:Ru:Pt:Fe', 'V:Mn:V:Ni:V', 
'V:Mo:Ni:Fe:Nb', 'V:Mo:Pt:W:W', 'V:Mo:W:Al:Mo', 'V:Mo:Zr:Fe:Ni', 'V:Nb:Cr:Al:Cr', 
'V:Nb:Mn:Al:Al', 'V:Nb:Re:Cr:Re', 'V:Nb:Ru:V:Ni', 'V:Ni:Al:Pt:Fe', 'V:Ni:Al:W:Co', 
'V:Ni:Ni:Co:V', 'V:Ni:V:V:Ni', 'V:Ni:W:Re:Fe', 'V:Pt:Co:Co:Pt', 'V:Pt:Co:Mn:Re', 
'V:Pt:Co:Ru:V', 'V:Pt:Pt:Fe:V', 'V:Pt:Re:Ni:Ru', 'V:Pt:Zr:Co:Pt', 'V:Re:Al:Cr:Ni', 
'V:Re:Cr:Ru:Pt', 'V:Re:Ru:Mo:Co', 'V:Re:Ru:Zr:W', 'V:Re:Zr:Pt:Ru', 'V:Ru:Co:Co:Pt', 
'V:Ru:Co:Pt:Zr', 'V:Ru:Mn:Mn:Re', 'V:Ru:Nb:Al:Ru', 'V:Ru:Nb:Pt:Fe', 'V:Ru:Ni:Ru:Ru', 
'V:Ru:V:Mn:Ni', 'V:V:Co:Pt:Pt', 'V:V:Cr:W:Re', 'V:V:Mn:Mn:Mn', 'V:V:Mn:V:Ni', 
'V:V:Nb:V:Co', 'V:V:Ni:Fe:Al', 'V:W:Mo:Nb:Zr', 'V:Zr:Mn:W:Nb', 'V:Zr:Ni:Nb:Zr', 
'V:Zr:Ru:Cr:Nb', 'V:Zr:Ru:W:Co', 'V:Zr:Zr:Cr:Al', 'W:Al:Ru:Nb:Cr', 'W:Al:V:Nb:Mo', 
'W:Al:V:Ru:Pt', 'W:Al:Zr:Nb:Ni', 'W:Co:Al:Mn:Cr', 'W:Co:V:Pt:Pt', 'W:Co:W:Ru:Co', 
'W:Co:Zr:Fe:Pt', 'W:Cr:Ni:Ni:Mo', 'W:Cr:Zr:Mn:Fe', 'W:Cr:Zr:V:Ni', 'W:Fe:Ni:Mo:W', 
'W:Fe:Ni:Re:Mo', 'W:Fe:V:Mn:Pt', 'W:Mn:Mn:Mo:Zr', 'W:Mn:Nb:Ru:Ni', 'W:Mn:Pt:Nb:Nb', 
'W:Mn:V:Cr:Fe', 'W:Mn:V:Fe:V', 'W:Mo:Cr:Al:Fe', 'W:Mo:Mo:Pt:Al', 'W:Mo:Nb:W:Mn', 
'W:Mo:Re:Mn:Cr', 'W:Mo:Zr:Re:Ru', 'W:Nb:Fe:Co:Al', 'W:Nb:Ni:V:Fe', 'W:Nb:Re:Ru:Re', 
'W:Nb:Ru:Al:Ni', 'W:Nb:Ru:Cr:V', 'W:Ni:Al:Al:Mo', 'W:Ni:Co:V:Re', 'W:Ni:Cr:Pt:Ni', 
'W:Ni:Cr:Zr:Pt', 'W:Ni:Mn:Al:W', 'W:Ni:Pt:Ni:Ru', 'W:Pt:Cr:Ru:Mo', 'W:Pt:Ni:Ru:Al', 
'W:Pt:Ru:Fe:Ru', 'W:Pt:W:Al:Al', 'W:Re:Cr:Fe:Fe', 'W:Re:Fe:Re:Cr', 'W:Re:Mn:W:Mo', 
'W:Re:Mo:Co:Ni', 'W:Re:Nb:Co:Re', 'W:Re:Pt:Pt:Cr', 'W:Re:Zr:Re:Mn', 'W:Ru:Nb:Zr:Pt', 
'W:V:Co:Fe:Fe', 'W:V:Cr:V:Mn', 'W:V:Fe:V:Mn', 'W:V:Nb:Al:Nb', 'W:V:Ni:V:V', 
'W:V:Ru:Al:Zr', 'W:V:Ru:Re:Mo', 'W:V:Zr:Ru:Ni', 'W:W:Co:Cr:Nb', 'W:W:Co:Mn:W', 
'W:W:Ni:Mo:Ni', 'W:W:Re:Zr:Nb', 'W:W:V:Mo:Ru', 'W:W:V:Zr:Nb', 'W:Zr:Co:Mn:Mn', 
'W:Zr:Nb:Ni:Mo', 'W:Zr:Re:Nb:Ru', 'W:Zr:Ru:Mn:Ni', 'W:Zr:W:Mn:Re', 'W:Zr:W:V:Zr', 
'Zr:Al:Mo:Ru:Mn', 'Zr:Al:Nb:Cr:Pt', 'Zr:Al:Ni:Zr:W', 'Zr:Al:Ru:Pt:Pt', 'Zr:Co:Mo:V:Pt', 
'Zr:Co:Ni:Mo:Pt', 'Zr:Co:Pt:V:Nb', 'Zr:Co:Re:Pt:V', 'Zr:Cr:Mo:Mo:Cr', 'Zr:Cr:Re:Ni:Re', 
'Zr:Cr:Ru:Ru:Pt', 'Zr:Cr:V:Pt:Fe', 'Zr:Cr:Zr:Zr:Pt', 'Zr:Fe:Ni:Mo:Al', 'Zr:Mn:Al:Fe:W', 
'Zr:Mn:Co:Mn:Mn', 'Zr:Mn:Nb:Ru:Ni', 'Zr:Mn:Nb:Ru:Pt', 'Zr:Mn:Ni:Nb:Cr', 'Zr:Mo:Fe:Fe:Mn', 
'Zr:Mo:Pt:Al:Cr', 'Zr:Nb:Mo:Fe:Co', 'Zr:Nb:Re:Cr:Ru', 'Zr:Nb:W:Al:Co', 'Zr:Ni:Mo:Mn:Fe', 
'Zr:Ni:W:Fe:Ru', 'Zr:Pt:Cr:Fe:Cr', 'Zr:Pt:Cr:Nb:Al', 'Zr:Pt:Nb:Co:Fe', 'Zr:Pt:W:W:Pt', 
'Zr:Re:Co:Al:Zr', 'Zr:Re:Mn:W:Cr', 'Zr:Re:Re:V:Mo', 'Zr:Ru:Cr:Pt:Ru', 'Zr:Ru:Fe:Zr:Nb', 
'Zr:Ru:Mn:Cr:Ni', 'Zr:Ru:Mo:Ni:Nb', 'Zr:V:Co:Mo:W', 'Zr:V:Co:Re:W', 'Zr:V:Mn:Re:Al', 
'Zr:V:Ni:Cr:Mo', 'Zr:V:Re:Pt:Cr', 'Zr:W:Mn:Fe:Al', 'Zr:W:Mo:Mo:V', 'Zr:W:Ni:Re:Cr', 
'Zr:W:Re:Re:Co', 'Zr:W:Zr:Al:Ni', 'Zr:Zr:Al:Co:Cr', 'Zr:Zr:Fe:Co:Zr', 'Zr:Zr:Mn:Cr:Pt', 
'Zr:Zr:Re:Co:W', 'Zr:Zr:Ru:Co:Mo', 'Zr:Zr:V:Zr:Al', 'Zr:Zr:W:W:Zr', 'Zr:Zr:Zr:Al:Ru', 
'Zr:Zr:Zr:Co:Mo', 
\end{verbatim}
\end{footnotesize}
} % end of \setstretch

\begin{figure}[htb]
\includegraphics[width=10cm]{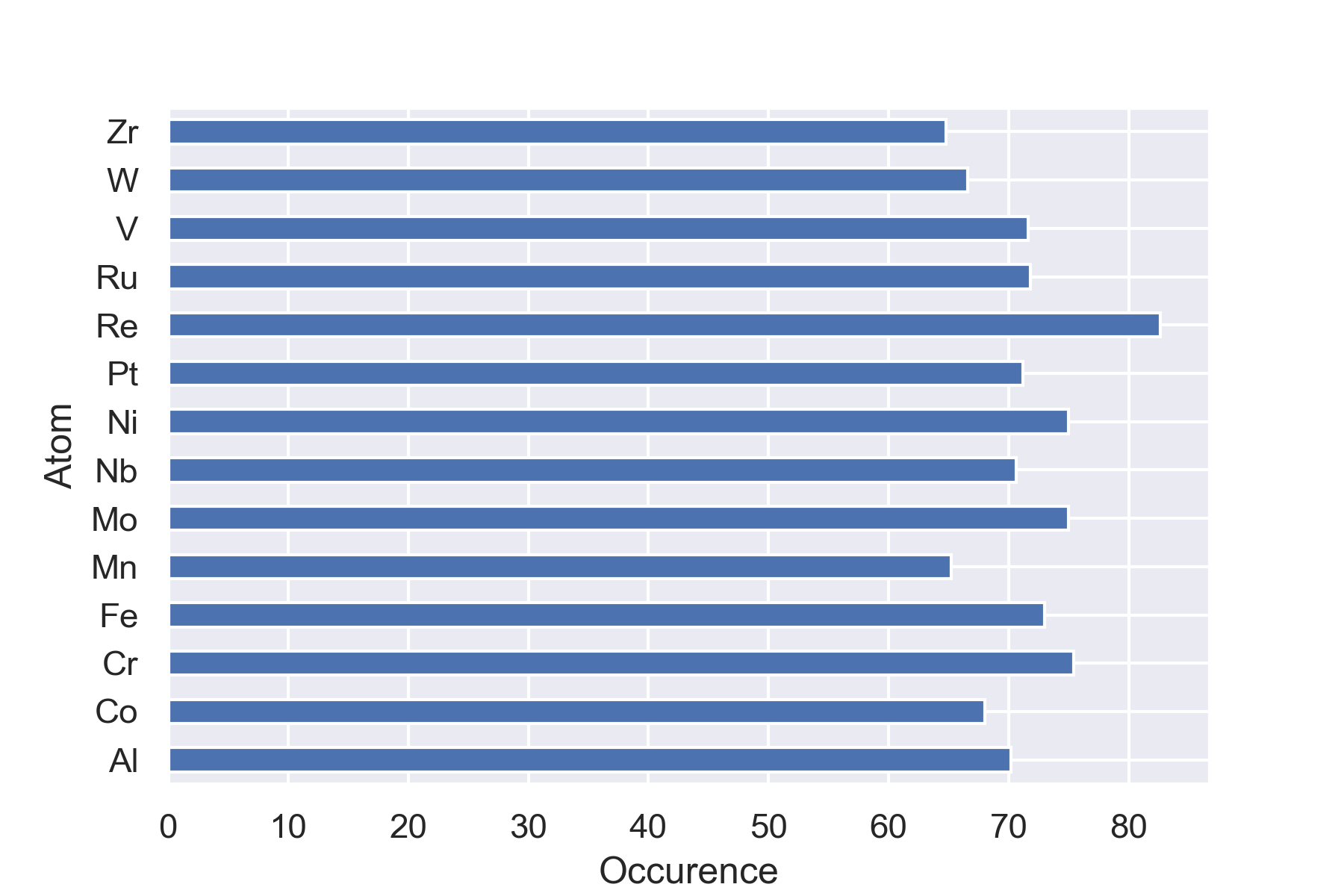}
\caption{\label{f:occu_ele_test} Occurrence of the $n=14$ elements in the testing database (1001 entries).}
\end{figure}

%--------------------------------------------------------------------------
\newpage
\section{Prediction from several simulations of training and testing sets}
\label{a:simu}

\begin{figure}[htb]
\includegraphics[width=5.9cm]{fig3a}% 
\includegraphics[width=5.9cm]{fig3b}% 
\includegraphics[width=5.9cm]{fig3c}\\% 
\includegraphics[width=5.9cm]{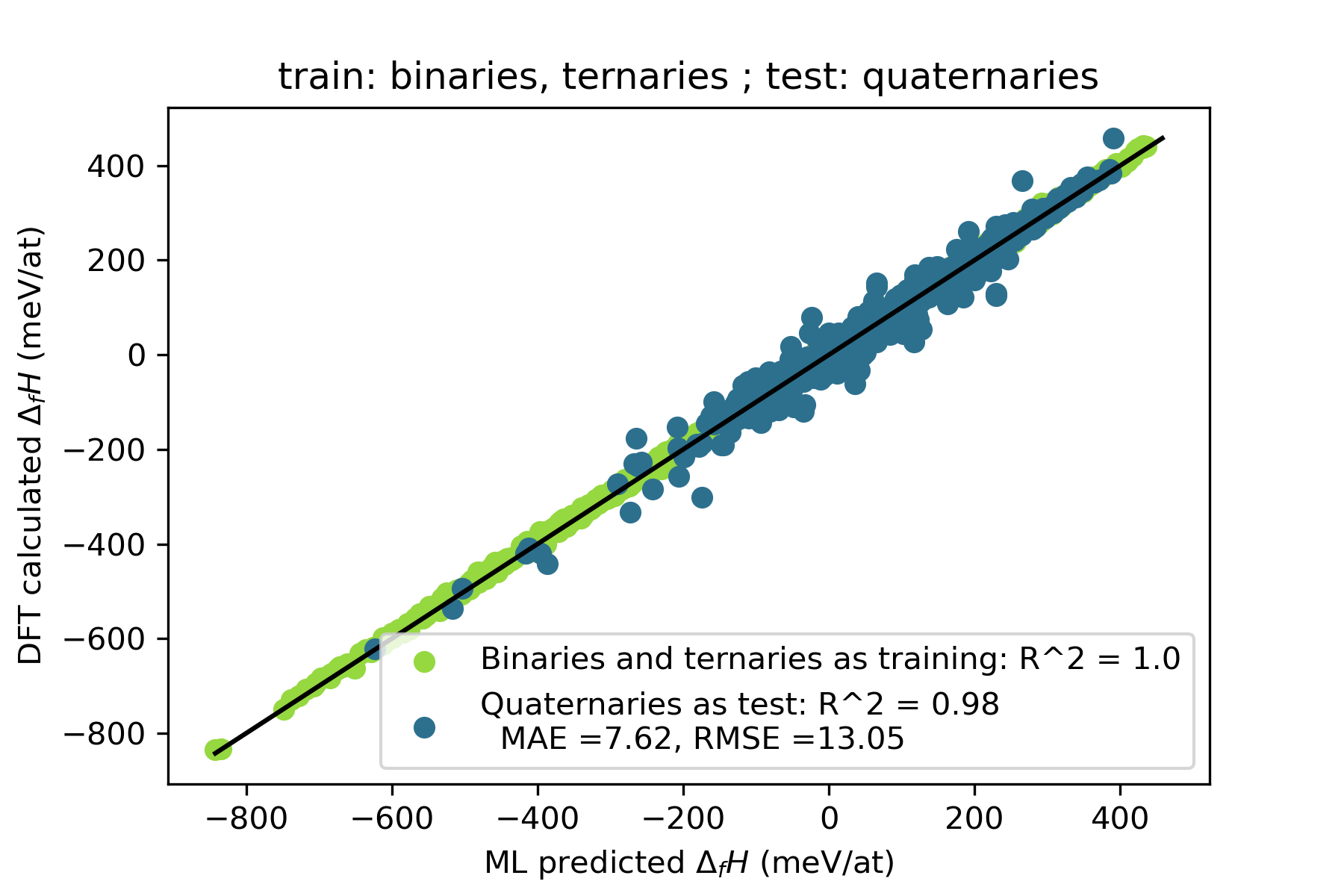}% 
\includegraphics[width=5.9cm]{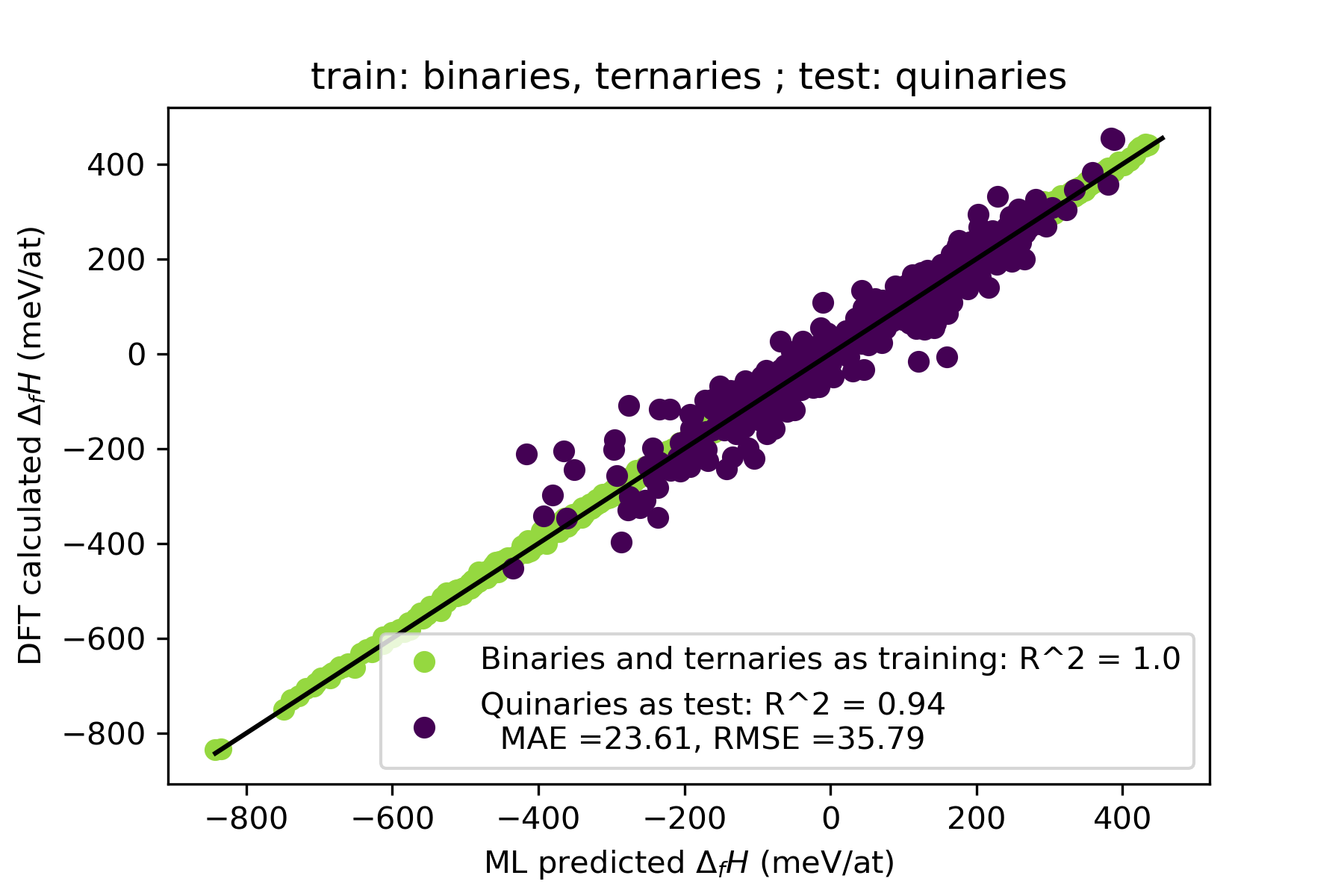}% 
\includegraphics[width=5.9cm]{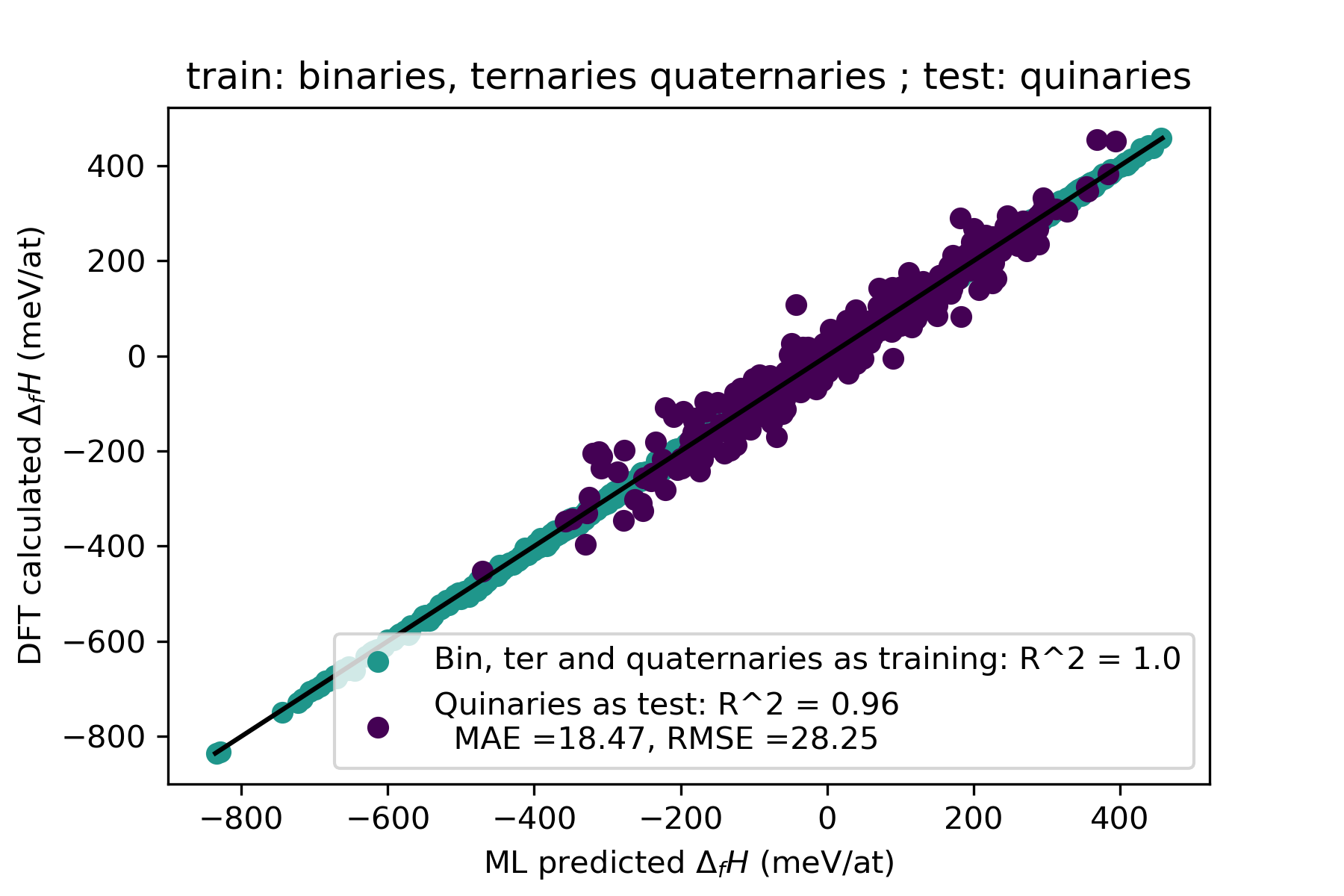}\\% 
\includegraphics[width=5.9cm]{fig3d}%  
\includegraphics[width=5.9cm]{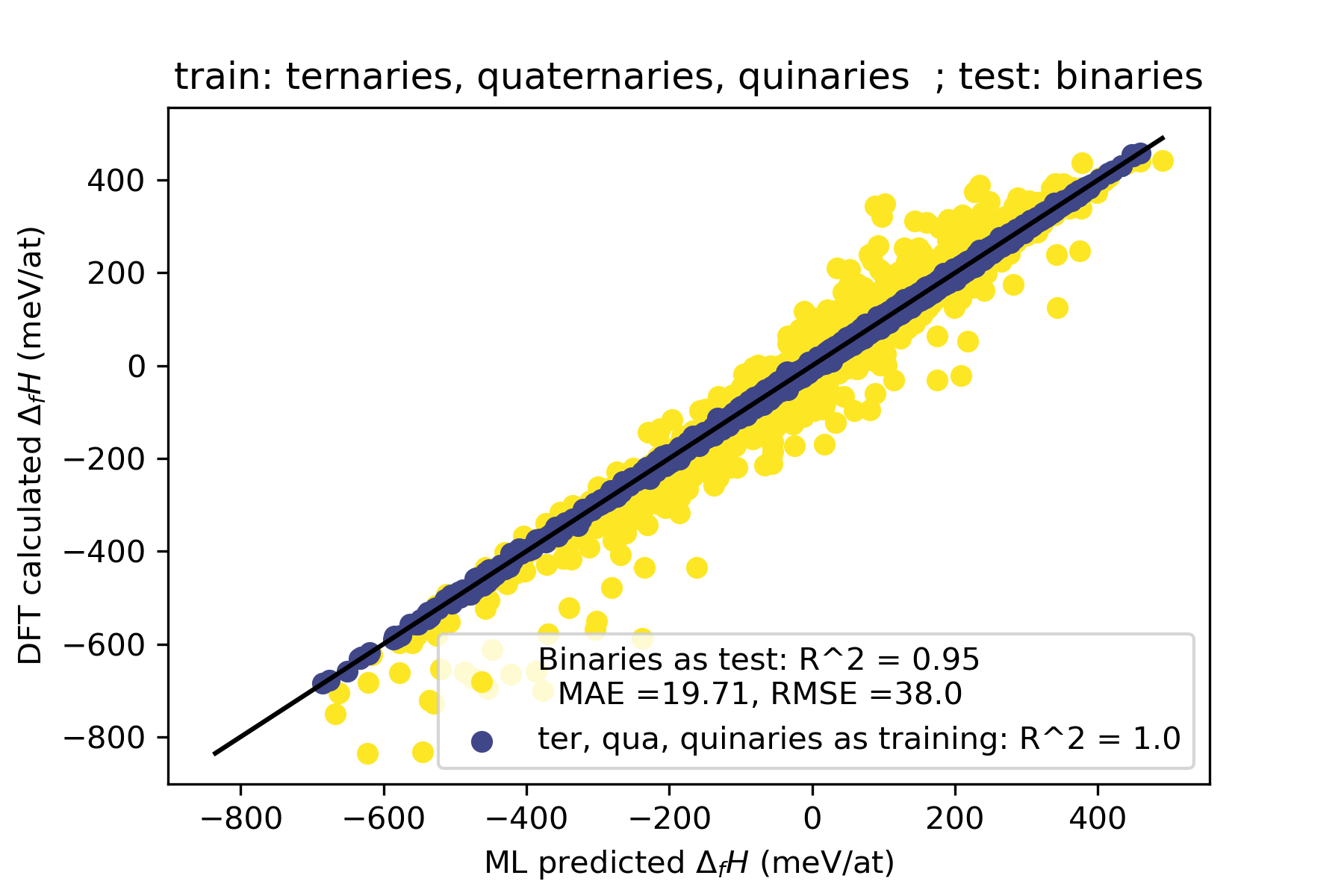}%  
\caption{\label{fig:simu} Simulation of the prediction with MPR from several proportions of training and testing sets including binaries (yellow), ternaries (green), quaternaries (blue) or quinaries (dark purple).
The diagonal line indicates the perfect agreement between predicted and real values.}
\end{figure}

%--------------------------------------------------------------------------
\newpage
\section{Influence of additional featuring}
\label{a:wwu}

\begin{figure}[htb]
\includegraphics[height=6cm]{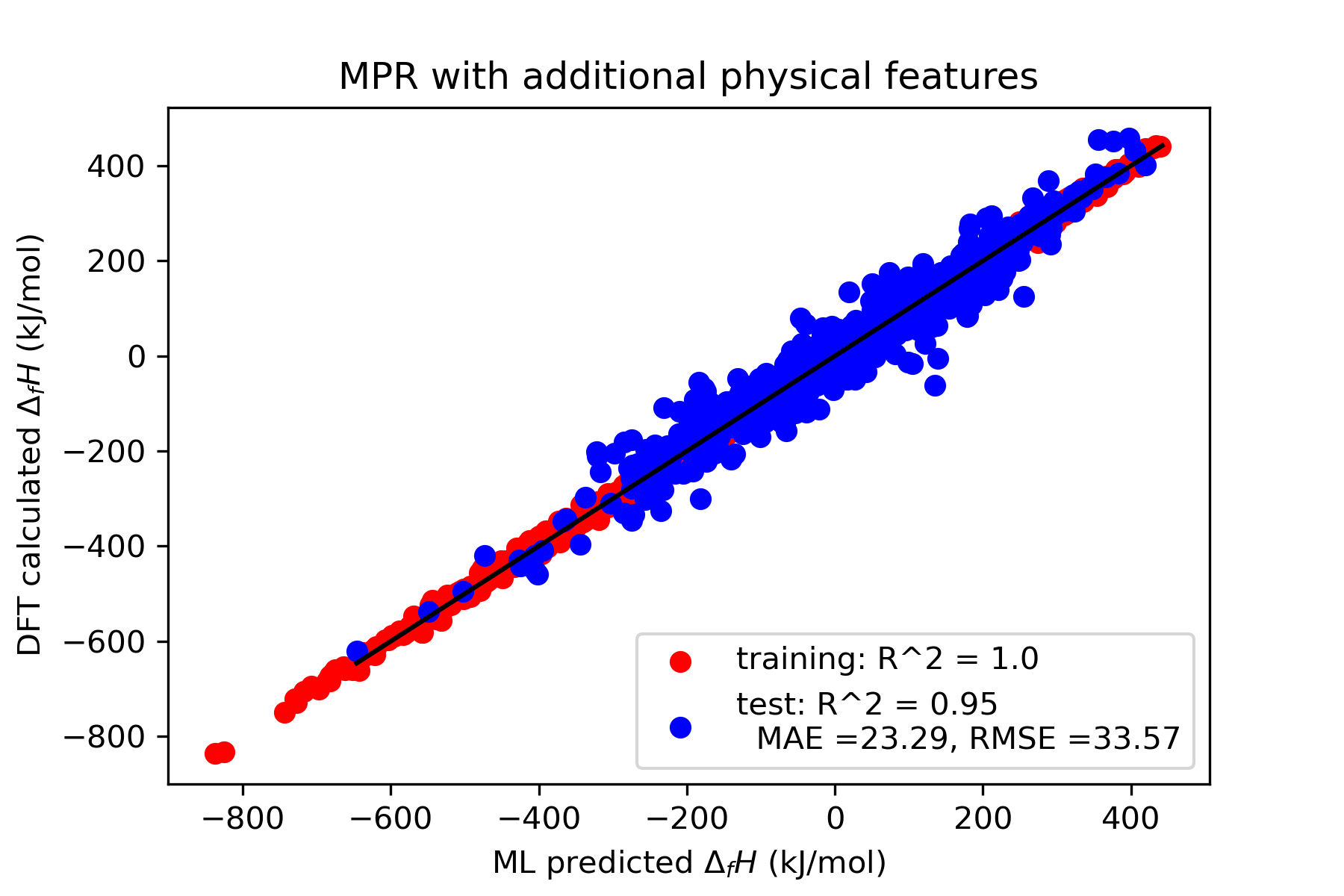}% 
\includegraphics[height=6cm]{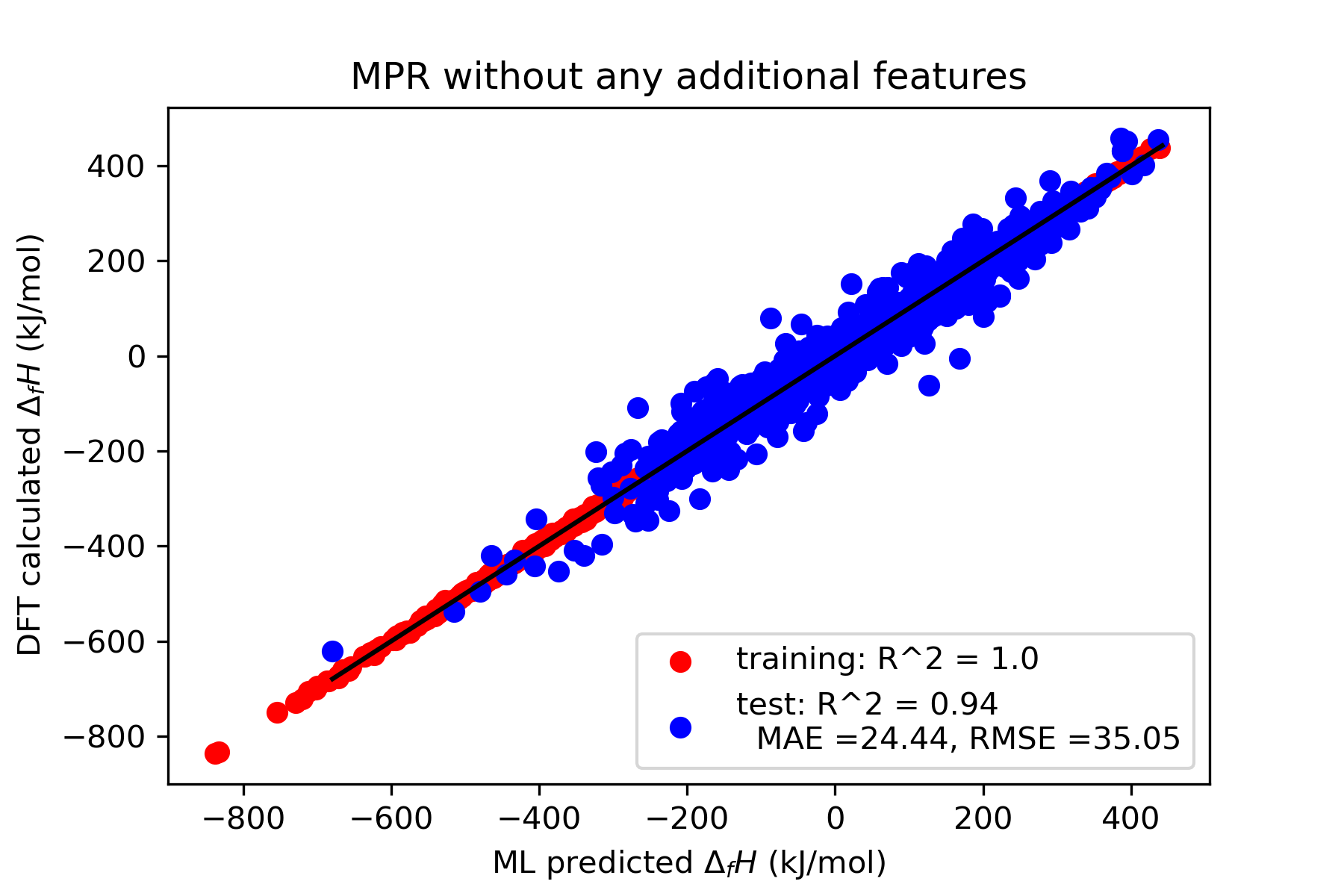}\\ 
\caption{\label{fig:wwu} Prediction with MPR of the randomized set from the learning  database  with (right) and without (left) additional featuring (atomic radius and number of valence electrons). 
The diagonal line indicates the perfect agreement between calculated and predicted values.}
\end{figure}

%--------------------------------------------------------------------------
\section{Results of prediction of every $14^5$ configurations}
\label{a:predi}

All of the data, including heat of formation ($\Delta_f H$), cell parameters ($a$, $c$) and 7 internal parameters ($x^{4f}$, $x^{8i_1}$,	$y^{8i_1}$, $x^{8i_2}$,	$y^{8i_2}$, $x^{8j}$,	$z^{8j}$), generated for the $14^5=$ 537,824\,configurations in this study are available from the authors on request.

%--------------------------------------------------------------------------
\newpage
\section{Prediction of $\sigma$-phase crystal parameters}
\label{a:cell}

\begin{figure}[htb]
\includegraphics[height=6cm]{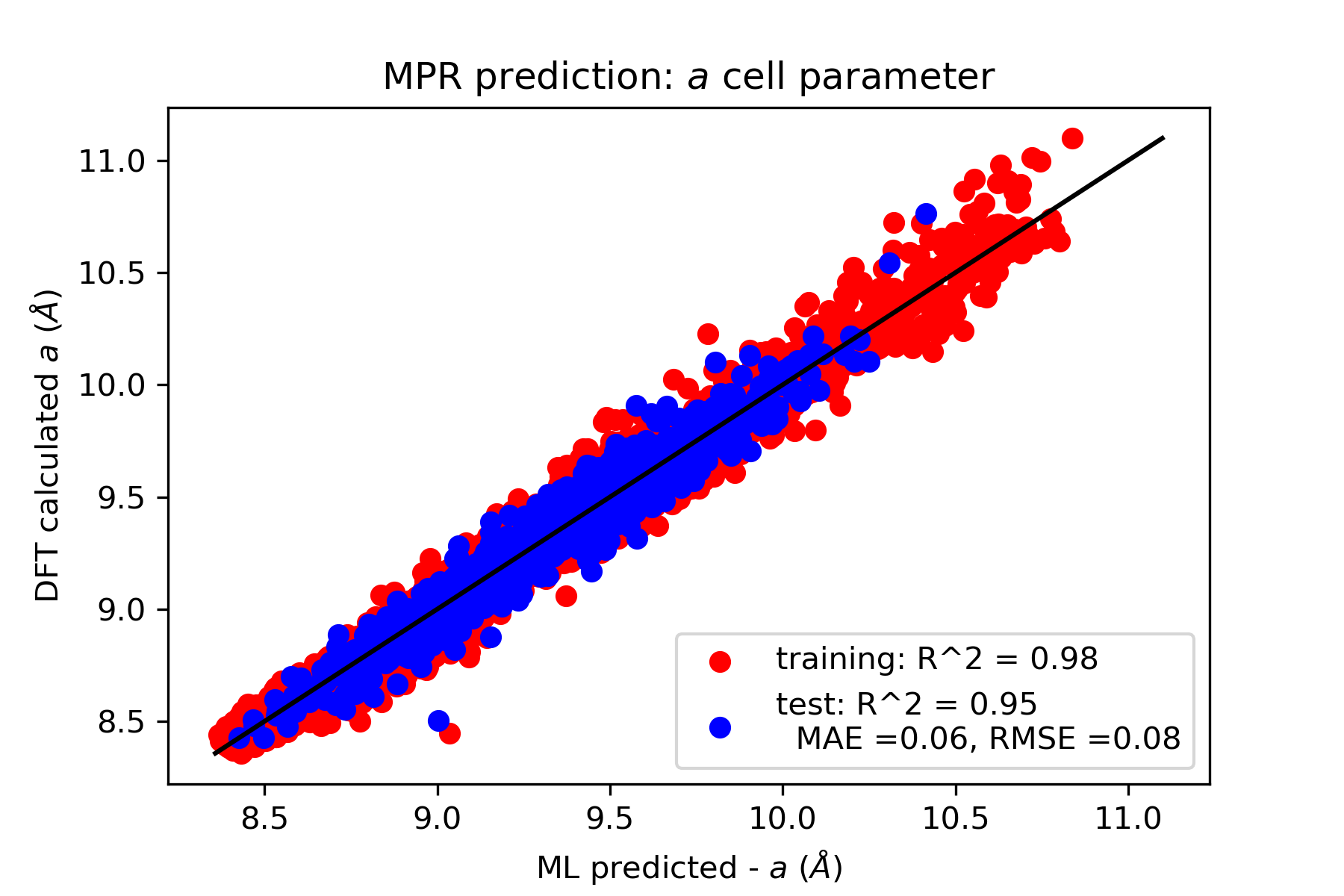}% 
\includegraphics[height=6cm]{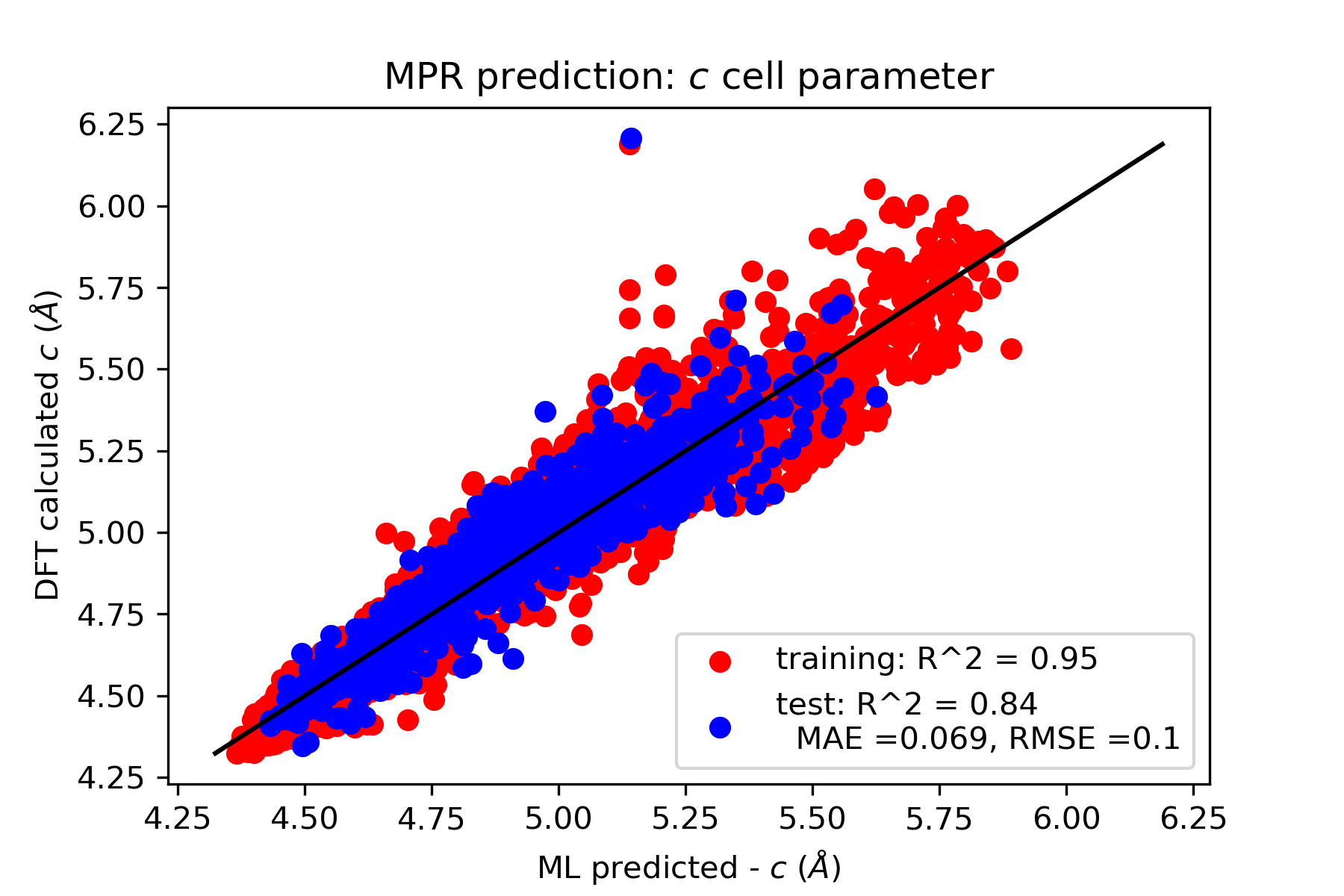}\\ 
\caption{\label{fig:cell} Prediction of both $a$ and $c$ tetragonal cell parameters of randomized 1001 testing configurations from the MPR learning of the training database (9974 data in red).}
\end{figure}

%--------------------------------------------------------------------------
\newpage
\section{DFT calculation methodology}
\label{a:dft}
The Density Functional Theory (DFT) method has been used to perform the electronic structure calculations.\cite{Hoh64,Koh65}
As implemented in the Vienna \textit{Ab initio} Simulation Package (VASP),
the Projector Augmented Wave (PAW) pseudo-potentials method was used.\cite{Kre93,Kre96,Kre99,Bl94}
Exchange and correlations were considered in the Generalized Gradient Approximation (GGA) with the Perdew-Burke-Ernzerhof (PBE) functional.\cite{Per96,Per97}
Convergence tests were performed with respect to the number of plane waves and the \verb"k"-points mesh.
The self-consistent total energy calculations converged to less than 0.01\,meV,  with a 400\,eV cutoff energy and with \verb"k"-points spacing less than 0.06\,\AA$^{-1}$ in each direction, samplings generated by the Monkhorst-Pack procedure (\verb"10x10x10" points in the irreducible wedge of the Brillouin zone).\cite{Mon76}. 
For all the 9974 + 1001 = 10,975 configurations, both the lattice parameters and internal atomic coordinates were fully relaxed by several intermediate steps, within a residual Hellmann-Feynman force of less than 5\,meV/\AA. 
Total energies have been calculated using the linear tetrahedron method with Bl\"ochl corrections~\cite{bl94a}.
The magnetism was not yet considered since routine calculations with spin polarization are not robust but are necessary for a correct energy prediction and will be considered in an future work.
The heat of formation, $\Delta_fH$, of a given $i$:$j$:$k$:$l$:$m$ configuration is given by the difference of its total DFT energy related to the element energies in their stable reference state (SER).
Additional tools, such as Vesta and Phonopy, for crystal representation and for checking relaxed structure symmetry were used.\cite{vesta,Cha11}

\newpage
\tableofcontents

\end{document}